\documentclass[fleqn,twoside]{article}
\usepackage{espcrc2}
\usepackage{epsfig}
\usepackage{latexsym}

\usepackage{graphicx}

\newcommand{\AmS}{{\protect\the\textfont2
  A\kern-.1667em\lower.5ex\hbox{M}\kern-.125emS}}

\title{SUPERSYMMETRY ON THE RUN: LHC AND DARK MATTER}

\author{D. I.~Kazakov\address{BLTP, JINR, Dubna and ITEP, Moscow}}

\begin{document}

\begin{abstract}
Supersymmetry, a
new symmetry that relates bosons and fermions in particle
physics, still escapes observation. Search for SUSY is one of the main aims of the recently launched Large Hadron Collider. The other possible manifestation of SUSY is the Dark Matter in the Universe.  The present lectures contain a brief introduction to supersymmetry in particle physics.
The main notions of supersymmetry  are introduced. The
supersymmetric extension of the Standard Model - the  Minimal
Supersymmetric Standard Model -  is considered in more detail.
Phenomenological features of the MSSM as well as possible
experimental signatures of SUSY at the LHC are described.
The DM problem and its possible SUSY solution is presented.

\vspace{1pc}
\end{abstract}

\maketitle
\renewcommand{\thesubsection}{\thesection.\arabic{subsection}}
\renewcommand{\thesubsubsection}{\thesubsection.\arabic{subsubsection}}
\renewcommand{\theequation}{\thesection.\arabic{equation}}

\section{Introduction: What is supersymmetry}

{\it Supersymmetry} is a {\it boson-fermion} symmetry that is
aimed to unify all forces in Nature including gravity within a
singe framework~\cite{super}-\cite{Books}. Modern views on
supersymmetry in particle physics are based on string paradigm,
though the low energy manifestations of SUSY can be possibly found
at modern colliders and in non-accelerator experiments.

Supersymmetry emerged from the attempts to generalize the
Poincar\'e algebra to mix representations with different
spin~\cite{super}. It happened to be a problematic task due to the
no-go theorems preventing such generalizations~\cite{theorem}. The
way out was found by introducing the so-called graded Lie
algebras, i.e. adding the anti-commutators to the usual
commutators of the Lorentz algebra. Such a generalization,
described below, appeared to be the only possible one within
relativistic field theory.

If $Q$ is a generator of SUSY algebra, then acting on a boson
state it produces a fermion one and vice versa
 $$ \bar Q| boson\!\!> = |fermion\!\!> , \ \
 Q| fermion\!\!> = | boson\!\!>\! \!.$$

Since bosons commute with each other and fermions anticommute, one
immediately finds that SUSY generators should also anticommute,
they must be {\em fermionic}, i.e. they must change the spin by a
half-odd amount and change the statistics. The key element
of SUSY algebra is
\begin{equation}\label{ant}
  \{Q_\alpha, \bar{Q}_{\dot \alpha}\}=2\sigma_{\alpha,\dot
\alpha}^\mu P_\mu,
\end{equation}
where $Q$ and $\bar Q$ are SUSY generators and $P_\mu$ is the
generator of translation, the four-momentum.

In what follows we describe SUSY algebra in more detail and
construct its representations which are needed to build a SUSY
generalization of the Standard Model of fundamental interactions.
Such a generalization is based on a softly broken SUSY quantum
filed theory and contains the SM as a low energy theory.

Supersymmetry promises to solve some  problems of the SM and of
Grand Unified Theories. In what follows we describe supersymmetry
as a nearest option for the new physics on a TeV scale.

\section{Motivation of SUSY in particle physics}
 \setcounter{equation} 0

\subsection{Unification with gravity}

 The {\em general idea} is a unification of all forces
of Nature including quantum gravity. However, the graviton has
spin 2, while the other gauge bosons (photon, gluons, $W$ and $Z$
weak bosons) have spin 1. Therefore, they correspond to different
representations of the Poincar\'e algebra. To mix them one can use
supersymmetry transformations. Starting with the graviton  state
of spin 2 and acting by SUSY generators we get the following chain
of states:
 $$spin \ 2
\rightarrow spin \ \frac 32 \rightarrow spin \ 1
\rightarrow spin \ \frac 12 \rightarrow spin \ 0 .$$
 Thus, a partial unification of matter (fermions) with forces (bosons)
naturally arises from an attempt to unify gravity with other
interactions.

 Taking infinitesimal transformations
$\delta_\epsilon = \epsilon^\alpha Q_\alpha, \ \bar{\delta}_{\bar
\epsilon} = \bar{Q}_{\dot \alpha}{\bar \epsilon}^{\dot \alpha},$
and using eq.(\ref{ant}) one gets
 \begin{equation}
\{\delta_\epsilon,\bar{\delta}_{\bar \epsilon} \}
 =2(\epsilon \sigma^\mu \bar \epsilon )P_\mu ,
 \label{com}
 \end{equation}
where $\epsilon$ is a transformation parameter. Choosing
$\epsilon$ to be local, i.e. a function of a space-time point
$\epsilon = \epsilon(x)$, one finds from eq.(\ref{com}) that an
anticommutator of two SUSY transformations is a local coordinate
translation. And a theory which is invariant under local
coordinate transformation is General Relativity. Thus, making SUSY
local, one naturally obtains General Relativity, or a theory of
gravity, or supergravity~\cite{Rev}.

\subsection{Unification  of  gauge couplings}

According to the  Grand Unification  {\em hypothesis}, gauge
symmetry increases with energy~\cite{GUT}. All known interactions
are different branches of a unique interaction associated with a
simple gauge group. The unification
 (or splitting) occurs at high energy.
To reach this goal one has to consider how the couplings change
with energy. This is described by the renormalization group
equations. In the SM the strong and weak couplings associated with
non-Abelian gauge groups decrease with energy, while the
electromagnetic one associated with the Abelian group on the
contrary increases. Thus, it becomes possible that at some energy
scale they become equal.

After the precise measurement of the $SU(3)\times SU(2) \times
U(1)$ coupling constants, it has become possible to check the
unification numerically. The three coupling constants to be
compared are
 \begin{eqnarray}
\alpha_1&=&(5/3)g^{\prime2}/(4\pi)=5\alpha/(3\cos^2\theta_W),
\nonumber \\ \alpha_2&=& g^2/(4\pi)=\alpha/\sin^2\theta_W, \\
\alpha_3&=& g_s^2/(4\pi) \nonumber
 \end{eqnarray}
where $g',~g$ and $g_s$ are the usual $U(1)$, $SU(2)$ and $SU(3)$
coupling constants and $\alpha$ is the fine structure constant.
The factor of 5/3 in the definition of $\alpha_1$ has been
included for  proper normalization of the generators.

In the modified minimal subtraction ($\overline{MS}$) scheme, the
world averaged values of the coup\-lings at the Z$^0$ energy are
obtained from a fit to the LEP and Tevatron data~\cite{SM}:
 \begin{eqnarray}
  \label{worave}
  \alpha^{-1}(M_Z)             & = & 128.978\pm 0.027   \nonumber\\
  \sin^2\theta_{\overline{MS}} & = & 0.23146\pm 0.00017\\
  \alpha_s                     & = & 0.1184\pm 0.0031, \nonumber
 \end{eqnarray}
that gives
 \begin{eqnarray} \alpha_1(M_Z)&=&0.017 \nonumber,\\
 \alpha_2(M_Z)&=&0.034, \\
\ \alpha_3(M_Z)&=&0.118\pm 0.003.\nonumber
 \end{eqnarray}
 Assuming that the SM is valid up to the unification
scale, one can then use the known RG equations for the three
couplings. In the leading order they are:
 \begin{equation}
\frac{d\tilde{\alpha}_i}{dt} =  b_i\tilde{\alpha}_i^2, \ \ \ \
\tilde{\alpha}_i=\frac{\alpha_i}{4\pi}, \ \ \ \ \
t=log(\frac{Q^2}{\mu^2}), \label{alpha}
 \end{equation}
where for the SM the coefficients  are $b_i=(41/10, -19/6, -7)$.

\begin{figure}
\begin{center}
  \leavevmode\hspace*{-1.0cm}
  \epsfxsize=9cm \epsfysize=5cm
 \epsffile{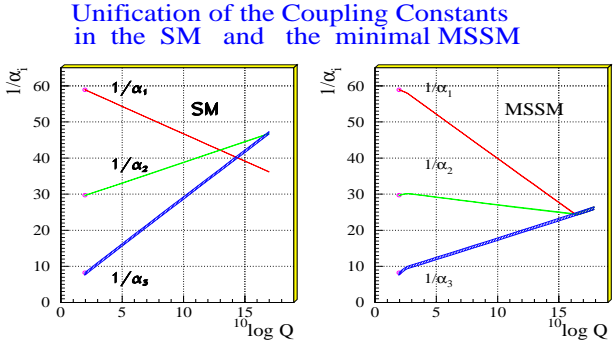}
 \end{center}
\vspace{-1cm}
 \caption{Evolution of the inverse of the three coupling constants
in the Standard Model (left) and in the supersymmetric extension
of the SM (MSSM) (right).}\label{unif}
 \end{figure}

The solution to eq.(\ref{alpha}) is very simple
 \begin{equation}
\frac{1}{\tilde{\alpha}_i(Q^2)} = \frac{1}{\tilde{\alpha}_i(\mu^2)}-
 b_i log(\frac{Q^2}{\mu^2}). \label{alphasol}
 \end{equation}
The result is demonstrated in  Fig.\ref{unif} showing the
evolution of the inverse of the couplings as a function of the
logarithm of energy. In this presentation, the evolution becomes a
straight line in first order. The second order corrections are
small and do not cause any visible deviation from a straight line.
Fig.\ref{unif} clearly demonstrates that within the SM the
coupling constant unification at a single point is impossible. It
is excluded by more than 8 standard deviations. This result means
that the unification can only be obtained if new physics enters
between the electroweak and the Planck scales.
%

In the SUSY case, the slopes of the RG evolution curves are
modified. The coefficients $b_i$ in eq.(\ref{alpha}) now are
$b_i=(33/5, 1, -3)$. The SUSY particles are assumed to effectively
contribute  to the running of the coupling constants only for
energies above the typical SUSY mass scale. It turns out that
within the SUSY model a perfect unification can be obtained as is
shown in Fig.\ref{unif}.  From the fit requiring unification one
finds for the break point $M_{SUSY}$ and the unification point
$M_{GUT}$~\cite{ABF}
 \begin{eqnarray}
M_{SUSY} &= & 10^{3.4\pm 0.9\pm 0.4} \ GeV , \nonumber\\ M_{GUT}
&= & 10^{15.8\pm 0.3\pm 0.1} \ GeV , \label{MSUSY}\\
\alpha^{-1}_{GUT} &= & 26.3 \pm 1.9 \pm 1.0 , \nonumber
 \end{eqnarray} The first error originates from
the uncertainty in the coupling constant, while the second one is
due to the uncertainty in the mass splittings between the SUSY
particles.

This observation was considered as the first "evidence" for
supersymmetry, especially since $M_{SUSY}$  was found in the range
preferred by the fine-tuning arguments.

\subsection{Solution of the hierarchy problem}

The appearance of two different scales $V \gg v$ in a GUT theory,
namely, $M_W$ and $M_{GUT}$, leads to a very  serious problem
which is called the {\em hierarchy problem}. There are two aspects
of this problem.

The first one is the very existence of the hierarchy. To get the
desired spontaneous symmetry breaking pattern, one needs
 \begin{equation}
 \begin{array}{cc}
m_H  \sim  v  \sim  10^2 \ \ \mbox{GeV} \\ m_{\Sigma} \sim
V  \sim  10^{16} \ \ \mbox{GeV}
 \end{array} \ \
\frac{m_H}{m_{\Sigma}} \sim 10^{-14}   \ll 1 , \label{hier}
 \end{equation}
where $H$ and $\Sigma$ are the Higgs fields responsible for the
spontaneous breaking of the $SU(2)$ and the GUT groups,
respectively. The question arises of how to get so small number in
a natural way.

\begin{figure}[ht]\hspace*{-0.5cm}
 \leavevmode
  \epsfxsize=8cm
 \epsffile{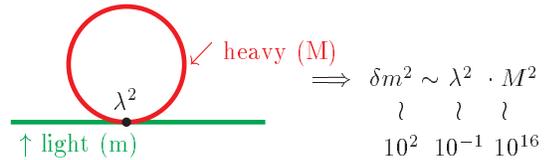}
 \vspace{-1.7cm}
 \caption{Radiative correction to the light Higgs boson mass}\label{fig:hierar}
 \end{figure}

The second aspect of the hierarchy problem is connected with the
preservation of a given hierarchy. Even if we choose the hierarchy
like eq.(\ref{hier}) the radiative corrections will destroy it! To
see this, consider the radiative correction to the light Higgs
mass given by the  Feynman diagram shown in Fig.\ref{fig:hierar}.
\vspace*{-1.5cm} 
This correction, proportional to the mass squared of the heavy
particle, obviously, spoils the hierarchy if it is not cancelled.
This very accurate cancellation with a precision $\sim 10^{-14}$
needs a fine tuning of the coupling constants.

\begin{figure}[htb]\hspace*{-0.4cm}
 \leavevmode
  \epsfxsize=8cm
 \epsffile{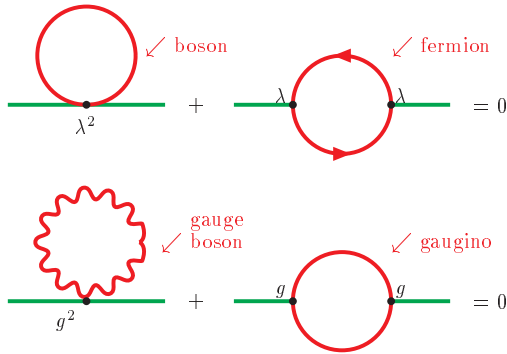}
\vspace{-1.5cm}
 \caption{Cancellation of quadratic terms (divergencies)}\label{fig:cancel}
 \end{figure}

The only known way of achieving this kind of cancellation of
quadratic terms (also known as the cancellation of the quadratic
divergencies) is supersymmetry. Moreover, SUSY automatically
cancels quadratic corrections in all orders of PT. This is due to
the contributions of superpartners of  ordinary particles. The
contribution from boson loops cancels those from the fermion ones
because of an additional factor (-1) coming from Fermi statistics,
as shown in Fig.\ref{fig:cancel}.
One can see here  two types of contribution. The first line is the
contribution of the heavy Higgs boson and its superpartner. The
strength of interaction is given by the Yukawa coupling $\lambda$.
The second line represents the gauge interaction proportional to
the gauge coupling constant $g$ with the contribution from the
heavy gauge boson and heavy gaugino.

In both the cases the cancellation of quadratic terms takes place.
This cancellation is true up to the SUSY breaking scale,
$M_{SUSY}$, which  should not be very large ($\leq$ 1 TeV) to make
the fine-tuning natural. Indeed, let us take the Higgs boson mass.
Requiring for consistency of perturbation theory that the
radiative corrections to the Higgs  boson mass do not  exceed the
mass itself gives
 \begin{equation}
\delta M_h^2 \sim g^2 M^2_{SUSY} \sim M_h^2.\label{del}
 \end{equation}
 So, if $M_h \sim 10^2$ GeV and $g \sim 10^{-1}$, one
needs $M_{SUSY} \sim 10^3$ GeV in order that the relation
(\ref{del}) is valid. Thus, we again get the same rough estimate
of $M_{SUSY} \sim $ 1 TeV as from the gauge coupling unification
above.

 That is why it is usually said that supersymmetry solves the
hierarchy problem.  We show below how SUSY can also explain the
origin of the hierarchy.

\subsection{Astrophysics and Cosmology}

\begin{figure}[ht]\vspace{0.3cm}
  \leavevmode
 \epsfxsize=6cm \epsfysize=5.5cm
 \hspace*{0.3cm}\epsffile{solar.eps}

  \epsfxsize=6.0cm
  \hspace*{0.3cm}
  \epsfysize=5.6cm
 \epsffile{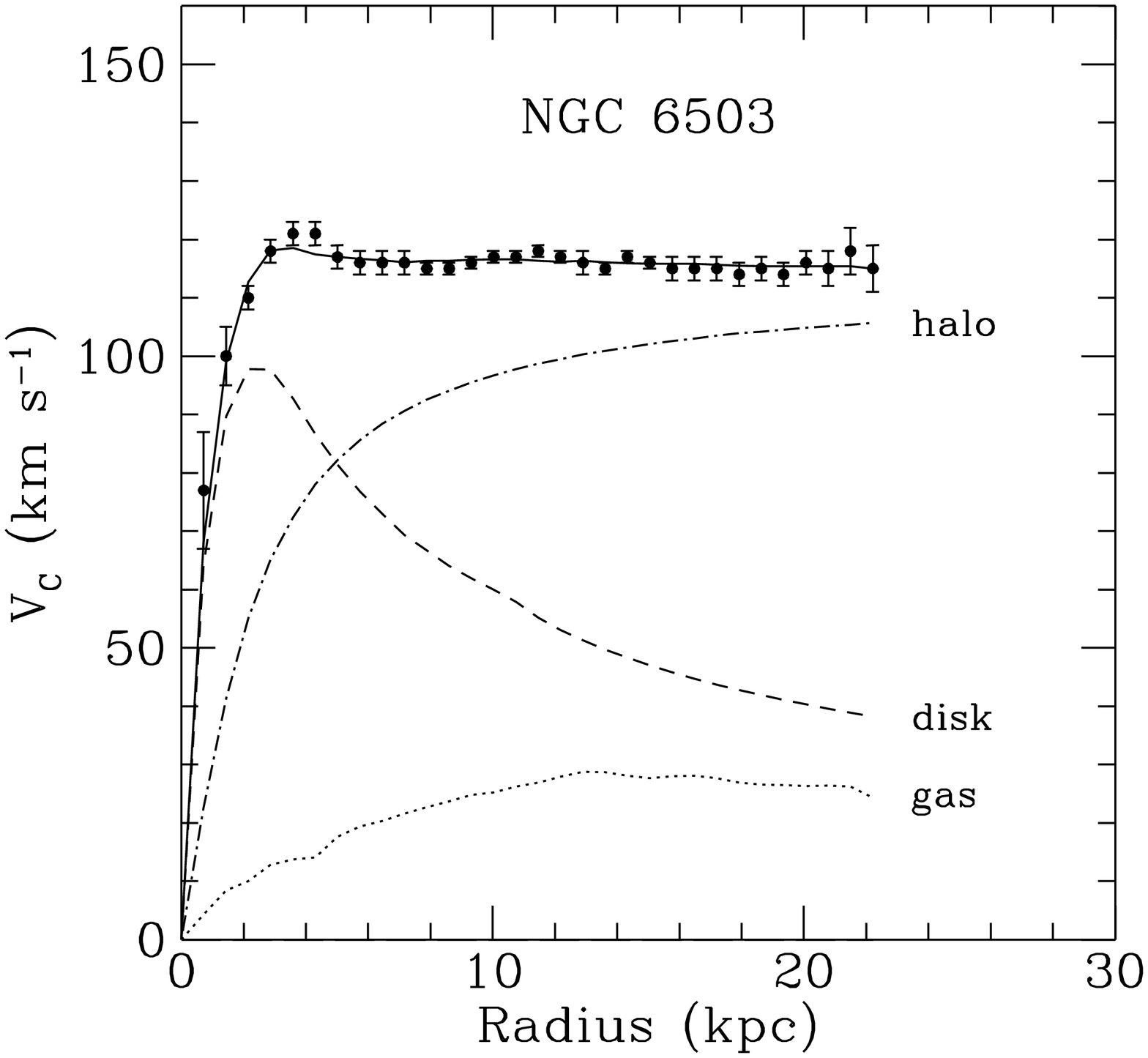}\vspace{-0.5cm}
 \caption{
Roration curves for the solar system and galaxy}\label{gal}
\end{figure}

The shining matter is not the only one in the Universe.
Considerable amount consists of the so-called dark matter. The
direct evidence for the presence of the dark matter are the
rotation curves of galaxies~\cite{rotcurve} (see Fig.\ref{gal}). To explain these
curves one has to assume the existence of galactic halo made of
non-shining matter which takes part in gravitational interaction.
There are two possible
types of the dark matter: the hot one, consisting of light
relativistic particles and the cold one, consisting of massive
weakly interacting particles (WIMPs)~\cite{WIMP}. The hot dark matter might
consist of neutrinos, however, this has problems with galaxy
formation. As for the cold dark matter, it has no candidates
within the SM. At the same time, SUSY provides an excellent
candidate for the cold dark matter, namely neutralino, the
lightest superparticle~\cite{abun}.

\subsection{Beyond GUTs: superstring}

Another motivation for supersymmetry follows from even more
radical changes of basic ideas related to the ultimate goal of
construction of consistent unified theory of everything. At the
moment the only viable conception is the superstring
theory~\cite{string}. In the superstring theory, strings are
considered as fundamental objects, closed or open, and are
nonlocal in nature. Ordinary particles are considered as string
excitation modes. String interactions, which are local, generate
proper interactions of usual particles, including gravitational
ones.

To be consistent, the string theory should be conformal invariant
in D-dimensional target space and have a stable vacuum. The first
requirement is valid in classical theory but may be violated by
quantum anomalies. Cancellation of quantum anomalies takes place
when space-time dimension of a target space equals to a critical
one which is $D_c=26$ for bosonic string and $D_c=10$ for a
fermionic one.

The second requirement is that the massless  string excitations
(the particles of the SM)  are stable. This assumes the absence of
tachyons, the states with imaginary mass, which can be guaranteed
only in supersymmetric string theories!

\section{Basics of supersymmetry}
 \setcounter{equation} 0

\subsection{Algebra of SUSY}

Combined with the usual Poincar\'e and internal symmetry algebra
the Super-Poincar\'e Lie algebra contains additional SUSY
generators $Q_{\alpha}^i$ and $\bar Q_{\dot
\alpha}^i$~\cite{WessB}
 \begin{eqnarray}  &&\hspace*{-0.7cm}{[} P_{\mu},P_{\nu}{]}  =   0, \nonumber \\  &&\hspace*{-0.7cm} {[} P_{\mu},M_{\rho\sigma}{]} =
i(g_{\mu \rho}P_{\sigma} -
 g_{\mu \sigma}P_{\rho}), \nonumber \\ &&\hspace*{-0.7cm}{[} M_{\mu \nu} , M_{\rho \sigma} {]}  \! = \!   i(\!g_{\nu \rho}M_{\mu
\sigma}\!\! -\!\! g_{\nu \sigma}M_{\mu \rho}\!\! -\!\! g_{\mu \rho}M_{\nu \sigma}
\!\!+\!\! g_{ \mu \sigma}M_{\nu \rho}\!),\nonumber \\ &&\hspace*{-0.6cm} {[} B_r , B_s {]}    =    iC_{rs}^t B_{t}, \nonumber\\ &&\hspace*{-0.7cm} {[}
B_r , P_{\mu} {]}    =    {[} B_r ,M_{\mu \sigma} {]} = 0, \label{group} \\ &&\hspace*{-0.7cm} {[} Q_{\alpha}^i , P_{\mu} {]} =  {[} \bar Q_{\dot \alpha}^i ,
P_{\mu} {]} = 0, \nonumber\\ &&\hspace*{-0.7cm} {[} Q_{\alpha}^i , M_{\mu \nu} {]} \! = \!  \frac{1}{2} (\!\sigma_{\mu \nu}\!)_{\alpha}
^{\beta}Q_{\beta}^i ,\! \ {[}\bar Q_{\dot \alpha}^i ,M_{\mu
\nu}{]}\! =\!  \frac{-1}{2} \bar Q_{\dot \beta}^i (\!\bar \sigma_{\mu
\nu}\!)_{\dot \alpha} ^{\dot \beta} ,\nonumber \\  &&  \hspace*{-0.7cm} {[} Q_{\alpha}^i , B_r {]}  =  (b_r)_{j}^i Q_{\alpha}^j , \ \ \ {[}
 \bar Q_{\dot \alpha}^i , B_r {]} = - \bar Q_{\dot \alpha}^j (b_r)_j^i , \nonumber\\ &&\hspace*{-0.7cm} \{ Q_{\alpha}^i , \bar Q_{\dot \beta}^j \}
  =   2 \delta^{ij} (\sigma ^{\mu})_{\alpha \dot \beta }P_{\mu} ,\nonumber\\ &&\hspace*{-0.7cm} \{ Q_{\alpha}^i , Q_{\beta}^j \}  =   2 \epsilon_
{\alpha \beta}Z^{ij} ,
\ \ \ Z_{ij} = a_{ij}^r b_r , \ \ Z^{ij} = Z_{ij}^+ ,\nonumber \\ &&\hspace*{-0.7cm} \{ \bar Q_{\dot \alpha}^i , \bar Q_{\dot \beta}^j \}  =  - 2 \epsilon
_{\dot \alpha \dot \beta}Z^{ij} , \ \ \ {[}Z_{ij} , anything {]} =
0 , \nonumber\\ &&\hspace*{-0.7cm} \alpha , \dot \alpha  =  1,2 \ \ \ \ i,j = 1,2, \ldots , N .\nonumber
 \end{eqnarray}
Here $P_{\mu}$ and $M_{\mu \nu}$  are four-momentum and angular
momentum operators, respectively, $B_r$ are the internal symmetry
generators, $Q^i$ and $\bar Q^i$ are the spinorial SUSY generators
and $Z_{ij}$ are the so-called central charges; $\alpha , \dot
\alpha, \beta , \dot \beta $ are the spinorial indices. In the
simplest case one has one spinor generator $Q_\alpha$ (and the
conjugated one $\bar Q_{\dot{\alpha}}$) that corresponds to an
ordinary or N=1 supersymmetry. When $N>1$ one has an extended
supersymmetry.

A natural question arises: how many SUSY generators are possible,
i.e. what is the value of $N$? To answer this question, consider
massless states. Let us start with the ground state labeled by
energy and helicity, i.e. projection of a spin on the direction of
momenta, and let it be annihilated by $Q_i$
 $${\rm Vacuum} =|E,\lambda >, \ \ \ \ \ \ \  Q_i|E,\lambda >=0. $$
Then  one and more  particle states can be constructed with the
help of a creation operators as $$\begin{array}{lll}
\mbox{\underline{State}} & \mbox{\underline{Expression}}  &
\hspace{-1.5cm}\# \ \mbox{ \underline{of States}} \\  \\
\mbox{vacuum} & |E,\lambda
\!>
& 1 \\
 1\!-\!\mbox{particle} & \bar Q_i|E,\lambda\!>= |E,\lambda\! +\!1/2\!>_i
& N \\
 2\!-\!\mbox{particle} & \bar Q_i\bar Q_j|E,\lambda\!>= |E,\lambda\! +\!1\!>_{ij}
&\hspace{-0.5cm}\frac{N(N-1)}{2} \\ ... & ... & ... \\
N\!-\!\mbox{particle} & \bar Q_1... \bar
Q_N|E,\lambda\!>= |E,\lambda\! +\!\frac N2\!\!> & 1
\end{array} $$
Total $\#$ of states: $\displaystyle
\sum_{k=0}^{N}\left(\begin{array}{c} N\\ k
\end{array}\right) =2^N=2^{N-1}$ bosons + $2^{N-1}$
fermions. The energy $E$ is not changed, since according to
(\ref{group}) the operators $ \bar Q_i$ commute with the
Hamiltonian.

Thus, one has a sequence of bosonic and fermionic states and the
total number of bosons equals  that of fermions. This is a generic
property of any supersymmetric theory. However, in CPT invariant
theories the number of states is doubled, since CPT transformation
changes the sign of helicity. Hence, in CPT invariant theories,
one has to add the states with opposite helicity to the above
mentioned ones.

Consider some examples. Let us take $N=1$ and $\lambda= 0$. Then
one has the following set of states:
 $$\begin{array}{lllllc}
 N=1 & \lambda=0 &&&&\\
\mbox{helicity}&0\ \frac 12& & \mbox{helicity}&0\ -\frac 12 \\
  & & \stackrel{CPT}{\Longrightarrow}   & & \\
  \# \ \mbox{of states}& 1\ 1&&\# \ \mbox{of states} &1\ \ \  \ 1
\end{array}$$
Hence, a complete $N=1$ multiplet is
 $$ \begin{array}{llccc} N=1 \ \ &
\mbox{helicity}&-1/2&0&1/2  \\ &
 \# \ \mbox{of states}&\ 1& 2&1
\end{array}$$ which contains one complex scalar and
one spinor with two helicity states.

This is an example of the so-called self-conjugated multiplet.
There are also self-conjugated multiplets with $N>1$ corresponding
to extended supersymmetry. Two particular examples are the $N=4$
super Yang-Mills multiplet and the $N=8$ supergravity multiplet
$$N=4 \ \ \
 \mbox{SUSY YM} \  \ \  \lambda=-1 $$
 $$\begin{array}{cccccccccc}
\mbox{helicity}&& &-1&-1/2&0&1/2&1 &&  \\
\# \ \mbox{of states}&&&\ \ 1&\ 4&6 & 4 & 1 &&\\
\end{array} $$
$$N=8  \ \ \  \mbox{SUGRA} \ \ \   \lambda=-2  $$
$$\begin{array}{ccccccccc}
-2&-3/2&-1&-1/2&0&1/2&1&3/2&2 \\
 1&8&28 & 56 & 70 &56&28&  8& 1
\end{array} $$ One can see that the multiplets of extended
supersymmetry are very rich and contain a vast number of
particles.

The constraint on the number of SUSY generators comes from a
requirement of consistency of the corresponding QFT.  The number
of supersymmetries and the maximal spin of the particle in the
multiplet are related by
 $$N \leq 4S,$$
where $S$ is the maximal spin. Since the theories with spin
greater than 1 are non-renormalizable and the theories with spin
greater than 5/2 have no consistent coupling to gravity, this
imposes a constraint on the number of SUSY generators
$$\begin{array}{cl} N \leq 4 & \ \ \ \ \mbox{for renormalizable
theories (YM),} \\ N \leq 8 & \ \ \ \ \mbox{for (super)gravity}.
\end{array}$$

 In what follows, we shall consider simple supersymmetry, or $N=1$
supersymmetry, contrary to extended supersymmetries with $N > 1$.
In this case, one has the following types of supermultiplets which are used in the
construction of SUSY generalization of the SM
$$\begin{array}{ll} \hspace*{1.2cm} (\phi,\ \ \psi) & \hspace*{1cm} (\lambda, \ \ A_\mu)\\
Spin=0, \ Spin=1/2 & Spin=1/2, \ Spin=1 \\
\ \ scalar\  \ \ \ \ \ chiral & majorana\ \ \ \   vector  \\
\hspace*{1.7cm} fermion & fermion
\end{array}$$
each of them contains two physical states, one boson and one fermion. They are called
chiral and vector multiplets, respectively.
Construction the generalization of the SM one has to put all the particles into these multiplets.
For instance, quarks should go into chiral multiplet and photon  into vector multiplet.

\subsection{Superspace and supermultiplets}

An elegant formulation of supersymmetry transformations and
invariants can be achieved in the framework of superspace
\cite{sspace}.
 Superspace differs from the ordinary Euclidean (Minkowski)
space by adding of two new coordinates, $\theta_{\alpha}$ and
$\bar \theta_{\dot \alpha}$, which are Grassmannian,
 i.e. anti\-com\-muting, variables
 $$\{ \theta_{\alpha}, \theta_{\beta} \} = 0 , \ \ \{\bar
\theta_{\dot \alpha}, \bar \theta_{\dot \beta} \} = 0, \ \
\theta_{\alpha}^2 = 0,\ \ \bar \theta_{\dot \alpha}^2=0, $$
$$\alpha,\beta, \dot\alpha, \dot\beta =1,2.$$
 Thus, we go from space to superspace
$$\begin{array}{cc} Space & \ \Rightarrow \ \ Superspace \\
x_{\mu} & \ \ \ \ \ \ \ x_{\mu}, \theta_{\alpha} , \bar \theta_
{\dot \alpha} \end{array}$$
 A SUSY group element can be constructed in
superspace in the same way as an ordinary translation in the usual
space
\begin{equation}
 G(x,\theta ,\bar \theta ) = e^{\displaystyle
i(-x^{\mu}P_{\mu} + \theta Q + \bar \theta \bar Q)}.\label{st}
\end{equation}
 It leads to a supertranslation in superspace
 \begin{equation}
 \begin{array}{ccl}
x_{\mu} & \rightarrow & x_{\mu} + i\theta \sigma_{\mu} \bar
\varepsilon
 - i\varepsilon \sigma_{\mu} \bar \theta, \\
\theta & \rightarrow & \theta + \varepsilon , \ \ \bar \theta
\rightarrow  \bar \theta + \bar \varepsilon ,\label{sutr}
\end{array}
 \end{equation}
where $\varepsilon $ and $\bar \varepsilon $ are Grassmannian
transformation parameters. From eq.(\ref{sutr}) one can easily
obtain the representation for the supercharges (\ref{group})
acting on the superspace
 \begin{equation}
Q_\alpha \!=\!\frac{\partial }{\partial \theta_\alpha
}\!-\!i\sigma^\mu_{\alpha \dot \alpha}\bar{\theta}^{\dot \alpha
}\partial_\mu , \  \bar{Q}_{\dot \alpha}\! =\!\frac{-\partial
}{\partial \bar{\theta}_{\dot \alpha}
}\!+\!i\theta_\alpha\sigma^\mu_{\alpha \dot \alpha}\partial_\mu .
\label{q}
 \end{equation}

Working in superspace all the fields become functions of not only the space-time
 point $x_\mu$ but also the Grassmanian coordinates $\theta$, i.e. they become superfields.
 The superfield contains inside the whole supermultiplet. We will not describe the superfields here
 and refer the reader to existing literature. What is important for us is that this formalizm is straightforward
  and allows one to construct a SUSY generalization of any theory.

\section{SUSY generalization of the Standard Model. The MSSM}
 \setcounter{equation} 0

As has been already mentioned, in SUSY theories the number of
bosonic degrees of freedom equals that of fermionic. At the same
time,  in the SM one has 28 bosonic and 90 fermionic degrees of
freedom (with massless neutrino, otherwise 96). So the SM is to a
great extent non-supersymmetric. Trying to add some new particles
to supersymmetrize the SM, one should take into account the
following observations:

$\bullet$  There are no fermions with quantum numbers of the gauge
bosons;

$\bullet$ Higgs fields have nonzero v.e.v.s; hence they cannot be
 superpartners of quarks and leptons since
this would induce  spontaneous violation of baryon and lepton
numbers;

 $\bullet$ One needs at least two complex chiral Higgs multiplets to
give masses to Up and Down quarks.

The latter is due to the form of a superpotential and chirality of
matter superfields. Indeed, the superpotential should be invariant
under the $SU(3)\times SU(2)\times U(1)$ gauge group. If one looks
at the Yukawa interaction in the Standard Model, one finds that it
is indeed $U(1)$ invariant since the sum of hypercharges in each
vertex equals zero. For the up quarks  this is achieved by taking
the conjugated Higgs doublet $\tilde{H}=i\tau_2H^\dagger$ instead
of $H$. However, in SUSY $H$ is a chiral superfield and hence a
superpotential, which is constructed out of  chiral fields, can
contain only $H$ but not $\tilde H$ which is an antichiral
superfield.

Another reason for the second  Higgs doublet is related to chiral
anomalies. It is known that chiral anomalies spoil the gauge
invariance and, hence, the renormalizability of the theory. They
are canceled in the SM between quarks and leptons in each
generation~\cite{Peskin}
$$\begin{array}{l}TrY^3=3\ (\frac{1}{27}+\frac{1}{27}-\frac{64}{27}+\frac{8}{27})-1-1+8=0\\
\hspace*{0.8cm} color\ \  u_L \ \ \ d_L \ \ \ \  u_R \ \ d_R \hspace{0.5cm} \nu_L\ \ e_L \ \ e_R
\end{array}$$

 However, if one introduces a chiral Higgs superfield,
it contains higgsinos, which are chiral fermions, and contain
anomalies. To cancel them one has to add the second Higgs doublet
with the opposite hypercharge. Therefore, the Higgs sector in SUSY
models is inevitably enlarged, it contains an even number of
doublets.

 {\em Conclusion}: In SUSY models supersymmetry
associates {\em known} bosons with {\em new} fermi\-ons and {\em
known} fermi\-ons with {\em new} bosons.

\subsection{The field content}

Consider the particle content of the Minimal Supersymmetric
Standard Model \cite{HT,MSSM}. According to the previous discussion,
in the minimal version  we double the number of particles
(introducing a superpartner to each particle) and add another
Higgs doublet (with its superpartner).

\begin{figure}[ht]
\begin{center}\vspace{-0.5cm}
 \leavevmode
  \epsfxsize=6.5cm
 \epsffile{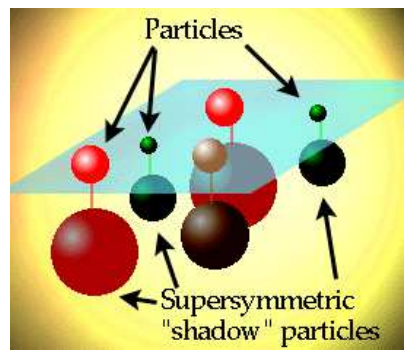}
\end{center}\vspace{-1.0cm} \caption{The shadow world of SUSY particles
}\label{fig:shadow}
 \end{figure}

Thus, the characteristic feature of any supersymmetric
generalization of the SM is the presence of superpartners (see
Fig.\ref{fig:shadow})~\cite{shadow}.  If supersymmetry is exact,
superpartners of ordinary particles should have the same masses
and have to be observed. The absence of them at modern energies is
believed to be explained by the fact that their masses are very
heavy, that means that supersymmetry should be broken. Hence,  if
the energy of accelerators is high enough, the superpartners will
be created.

The particle content of the MSSM then appears as shown in the table.
\begin{table*}
\begin{center}
{\bf Particle Content of the MSSM}
\end{center}\vglue 0.2cm

\begin{center}
\renewcommand{\tabcolsep}{0.03cm}
\begin{tabular}{|lllccc|}\hline
Superfield & \ \ \ \ \ \ \ Bosons & \ \ \ \ \ \ \ Fermions &
$SU(3)$& $SU(2$ & $U_Y(1)$ \\ \hline \hline Gauge  &&&&& \\
${\bf G^a}$   & gluon \ \ \ \ \ \ \ \ \ \ \ \ \ \ \  $g^a$ &
gluino$ \ \ \ \ \ \ \ \ \ \ \ \ \tilde{g}^a$ & 8 & 0 & 0 \\ ${\bf
V^k}$ & Weak \ \ \ $W^k$ \ $(W^\pm, Z)$ & wino, zino \
$\tilde{w}^k$ \ $(\tilde{w}^\pm, \tilde{z})$ & 1 & 3& 0 \\ ${\bf
V'}$   & Hypercharge  \ \ \ $B\ (\gamma)$ & bino \ \ \ \ \ \ \ \ \
\ \ $\tilde{b}(\tilde{\gamma })$ & 1 & 1& 0 \\ \hline \hline Matter &&&&&
\\ $\begin{array}{c} {\bf L_i} \\ {\bf E_i}\end{array}$ & sleptons
\ $\left\{
\begin{array}{l} \tilde{L}_i=(\tilde{\nu},\tilde e)_L \\ \tilde{E}_i =\tilde
e_R \end{array} \right. $ & leptons \ $\left\{ \begin{array}{l}
L_i=(\nu,e)_L
\\ E_i=e_R^c \end{array} \right.$ & $\begin{array}{l} 1 \\ 1 \end{array} $  &
$\begin{array}{l} 2 \\ 1 \end{array} $ & $\begin{array}{r} -1 \\ 2
\end{array} $ \\ $\begin{array}{c} {\bf Q_i} \\ {\bf U_i} \\ {\bf D_i}
\end{array}$ & squarks \ $\left\{ \begin{array}{l}
\tilde{Q}_i=(\tilde{u},\tilde d)_L \\ \tilde{U}_i =\tilde u_R \\
\tilde{D}_i =\tilde d_R\end{array}  \right. $ & quarks \ $\left\{
\begin{array}{l} Q_i=(u,d)_L \\ U_i=u_R^c \\ D_i=d_R^c \end{array}
\right.$ & $\begin{array}{l} 3
\\ 3^* \\ 3^* \end{array} $  & $\begin{array}{l} 2 \\ 1 \\ 1 \end{array} $ &
$\begin{array}{r} 1/3 \\ -4/3 \\ 2/3 \end{array} $\\  \hline  \hline Higgs
&&&&& \\ $\begin{array}{c} {\bf H_1} \\ {\bf H_2}\end{array}$ &
Higgses \ $\left\{
\begin{array}{l} H_1 \\ H_2 \end{array}  \right. $ & higgsinos \ $\left\{
 \begin{array}{l} \tilde{H}_1 \\ \tilde{H}_2 \end{array} \right.$ &
$\begin{array}{l} 1 \\ 1 \end{array} $  & $\begin{array}{l} 2 \\ 2
\end{array} $ &
$\begin{array}{r} -1 \\ 1
\end{array} $
 \\ \hline \hline
\end{tabular}
\end{center}
\end{table*}
Hereafter, tilde denotes a superpartner of an ordinary particle.

The presence of an extra Higgs doublet in SUSY model is a novel
feature of the theory. In the MSSM  one has two doublets with the
quantum numbers (1,2,-1) and (1,2,1), respectively:
 \begin{eqnarray*}
H_1&=&\left(\begin{array}{c} H^0_1 \\ H_1^- \end{array}\right) =
\left( \begin{array}{c} v_1 +\frac{
S_1+iP_1}{\sqrt{2}} \\ H^-_1 \end{array}\right),\\
  H_2&=&\left(
\begin{array}{c} H^+_2 \\ H_2^0 \end{array}  \right) = \left(
\begin{array}{c} H^+_2 \\ v_2 +\frac{
S_2+iP_2}{\sqrt{2}}
\end{array} \right),
\end{eqnarray*}
where  $v_i$ are the vacuum expectation values of the neutral
components.

Hence, one has 8=4+4=5+3 degrees of freedom. As in the case of the
SM, 3 degrees of freedom can be gauged away, and one is left with
5 physical states compared to 1 in the SM. Thus, in the MSSM, as
actually in any of two Higgs doublet models,
 one has  five  physical Higgs bosons: two CP-even neutral,
one CP-odd neutral  and two charged. We consider the mass
eigenstates below.

\subsection{Lagrangian of the MSSM}

To construct a SUSY Lagrangian one has to follow the following three steps:
\begin{itemize}
\item $1^{st}$ step:  Take your favorite Lagrangian written in terms of fields
\item $2^{nd}$ step: Replace the fields ($\phi,\psi, A_\mu$) by superfields $\Phi, V$
\item  $3^{rd}$ step:  Replace the Action by superAction
$$A=\int d^4x {\cal L}(x) \ \ \Rightarrow \ \ A=\int d^4x\ d^4\theta {\cal L}(x,\theta,\bar \theta)$$
\end{itemize}
At the last step one has to perform the integration over the Grassmannian variables. The
rules of integration are very easy~\cite{ber}:
$$\int d\theta_\alpha =0, \ \ \int\theta_\alpha d\theta_\beta= \delta_{\alpha,\beta}.$$

Now we can construct the Lagrangian of the MSSM. It consists of two parts; the
first part is the SUSY generalization of the Standard Model, while the
second one represents the SUSY breaking as mentioned above.
 \begin{equation}
 {\cal L}={\cal L}_{SUSY}+{\cal L}_{Breaking},
 \end{equation}
where
 \begin{equation}
 {\cal L}_{SUSY}={\cal L}_{Gauge}+{\cal L}_{Yukawa}.
 \end{equation}

 We will not describe the gauge part since it is essentially the gauge invariant kinetic terms
 but rather concentrate on Yukawa terms. They are given by the so-called superpotential
 which is nothing else but the usual Yukawa terms with the fields replaced by superfields as explained above.
 \begin{eqnarray} &&
{\cal L}_{Yukawa}   = \epsilon_{ij}(y^U_{ab}Q_a^j U^c_bH_2^i +
y^D_{ab}Q_a^jD^c_bH_1^i \nonumber  \\ &&
       +  y^L_{ab}L_a^jE^c_bH_1^i + \mu H_1^iH_2^j), \label{R}\end{eqnarray}
where $i,j=1,2$ are the $SU(2)$ and $a,b=1,2,3$ are the
generation indices; colour indices are suppressed. This part of
the Lagrangian almost exactly repeats that of the SM. The only difference is the last term which describes
the Higgs mixing. It is absent in the SM since there is only one
Higgs field there.

However, one can write down the other Yukawa terms
  \begin{eqnarray} && {\cal L}_{Yukawa}  =  \epsilon_{ij}(\lambda^L_{abd}L_a^i L_b^jE_d^c +
\lambda^{L\prime}_{abd}L_a^iQ_b^jD_d^c\nonumber\\ && +\mu'_aL^i_aH_2^j)
+\lambda^B_{abd}U_a^cD_b^cD_d^c. \label{NR}
 \end{eqnarray}
These terms are absent in the SM. The reason is very simple: one
can not replace the superfields in eq.(\ref{NR}) by the ordinary
fields like in eq.(\ref{R}) because of the Lorentz invariance.
These terms have a different property, they violate either lepton
(the first 3 terms  in eq.(\ref{NR})) or baryon number (the last
term). Since both effects are not observed in Nature, these terms
must be suppressed or excluded. One can avoid such terms
introducing a  special symmetry called $R$-symmetry\cite{r-symmetry}. This is the
global $U(1)_R$ invariance
 \begin{equation}
U(1)_R: \ \ \theta \to e^{i\alpha} \theta ,\ \  \Phi \to
e^{in\alpha}\Phi , \label{RS}
 \end{equation}
which is reduced  to the discrete group $Z_2$, called the
$R$-parity. The $R$-parity quantum number
 is given by
$R=(-1)^{3(B-L)+2S}$ 
for particles with spin $S$. Thus, all the ordinary particles have
the $R$-parity quantum number equal to $R=+1$, while all the
superpartners have $R$-parity quantum number equal to $R=-1$.
The  first part of  the Yukawa Lagrangian is R-symmetric, while the  second part is R-nonsymmetric.
The $R$-parity obviously forbids the  terms. However, it may
well be that these terms are present, though experimental limits
on the couplings are very severe
$$\lambda^L_{abc}, \ \ \lambda^{L\prime}_{abc} < 10^{-4}, \ \ \ \
\ \lambda^B_{abc} < 10^{-9}.$$

\subsection{Properties of interactions}

If one assumes that the $R$-parity is preserved, then the
 interactions of superpartners are essentially the same as in the
SM, but two of three particles involved into an interaction at any
vertex
are replaced by superpartners. The reason for it is the
$R$-parity. Conservation of the $R$-parity has two consequences

$\bullet$ the superpartners are created in pairs;

$\bullet$  the lightest superparticle (LSP) is stable. Usually it
is  photino $\tilde \gamma $, the superpartner of a photon with
some admixture of neutral higgsino.

\begin{figure}[ht]\vspace{-0.5cm}
 \leavevmode
  \epsfxsize=7cm \hspace*{-1.0cm}
 \epsffile{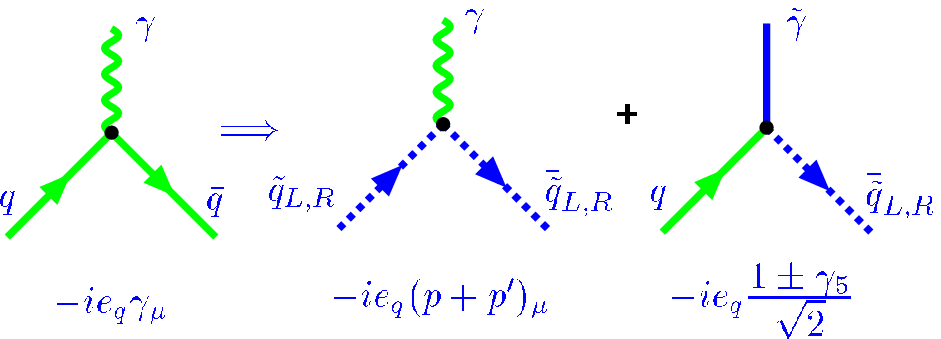}
 \leavevmode
  \epsfxsize=7.5cm \hspace*{0.0cm}
 \epsffile{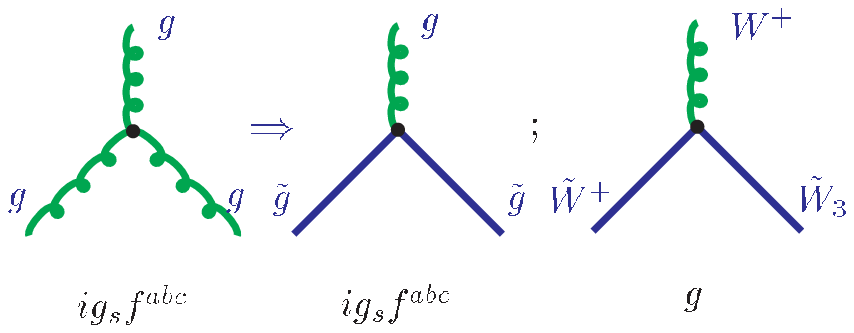}
\vspace{0.3cm}
 \leavevmode
  \epsfxsize=7.5cm \hspace*{-0.0cm}
 \epsffile{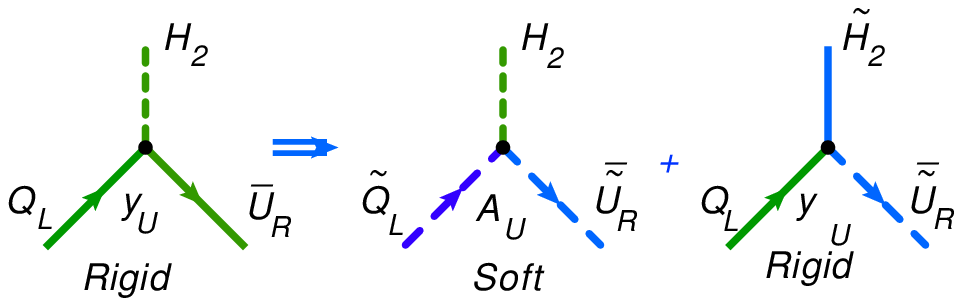}
\vspace{-1.0cm}
 \caption{Gauge-matter interaction, Gauge self-interaction and Yukawa-type interaction}
 \label{yukint}
 \end{figure}

Typical vertices are  shown in Figs.\ref{yukint}. The tilde above
a letter denotes the corresponding superpartner. Note that the
coupling is the same in all the vertices involving superpartners.

\subsection{Creation and decay of superpartners}

\begin{figure}[ht]\vspace{-0.0cm}\hspace*{-0.3cm}
 \leavevmode
  \epsfxsize=8cm
 \epsffile{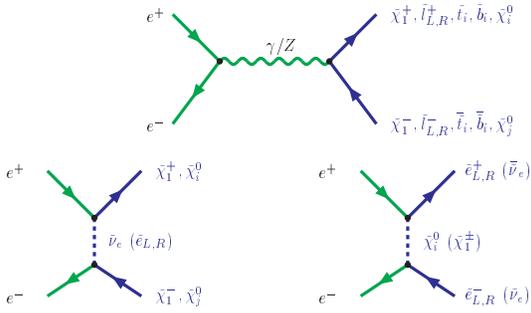}
\vspace{-0.3cm}
 \caption{Creation of superpartners at electron-positron  colliders}\label{creation}
 \end{figure}

The above-mentioned rule   together with the Feynman rules for the
SM enables one to draw diagrams describing creation of
superpartners. One of the most promising processes is the $e^+e^-$
annihilation (see Fig.\ref{creation}).
 \begin{figure}[h]
\leavevmode \hspace*{-0.2cm}
  \epsfxsize=7.5cm \epsfysize=5.8cm
   \epsffile{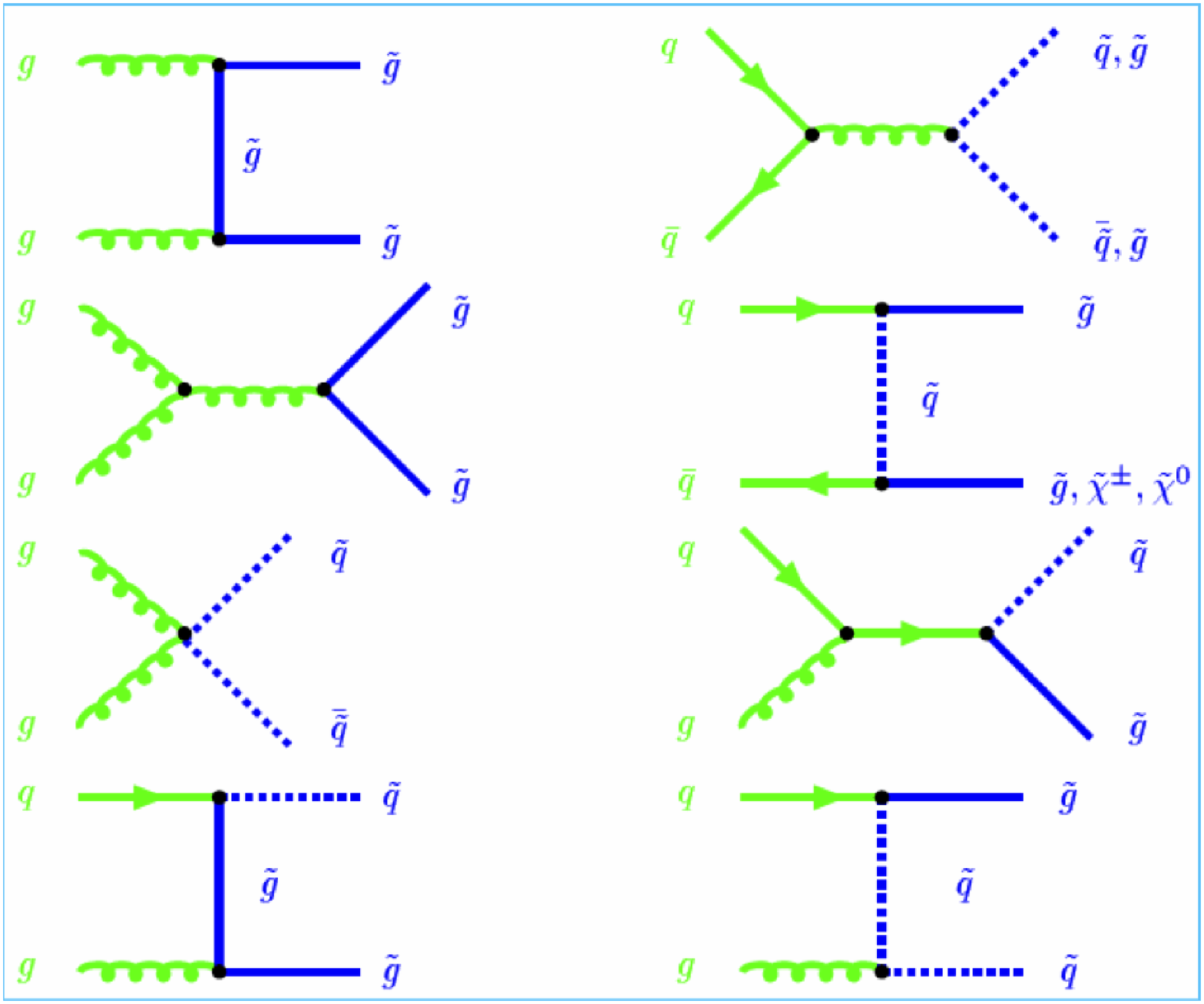}
\vspace{-0.3cm} \caption{Gluon fusion, $q\bar q$ scattering,
quark-gluon scattering}\label{fusion}
\end{figure}

The usual kinematic restriction is given by the c.m. energy
$m^{max}_{sparticle} \leq \frac{\sqrt s}{2}.\ $ Similar processes
take place at hadron colliders with electrons and positrons being
replaced by quarks and gluons.

Experimental signatures at hadron colliders are similar to those
at $e^+e^-$ machines; however, here one has much wider
possibilities. Besides the usual annihilation channel, one has numerous processes of gluon
fusion, quark-antiquark and quark-gluon scattering (see
Fig.\ref{fusion}).

Creation of superpartners can be  accompanied by creation of
ordinary particles as well. We consider various experimental
signatures  below. They crucially
depend on SUSY breaking pattern and on the mass spectrum of
superpartners.

\begin{figure}[htb]\vspace{-0.1cm}
 \leavevmode \hspace*{0.5cm}
  \epsfxsize=6.5cm \epsfysize=4.8cm
 \epsffile{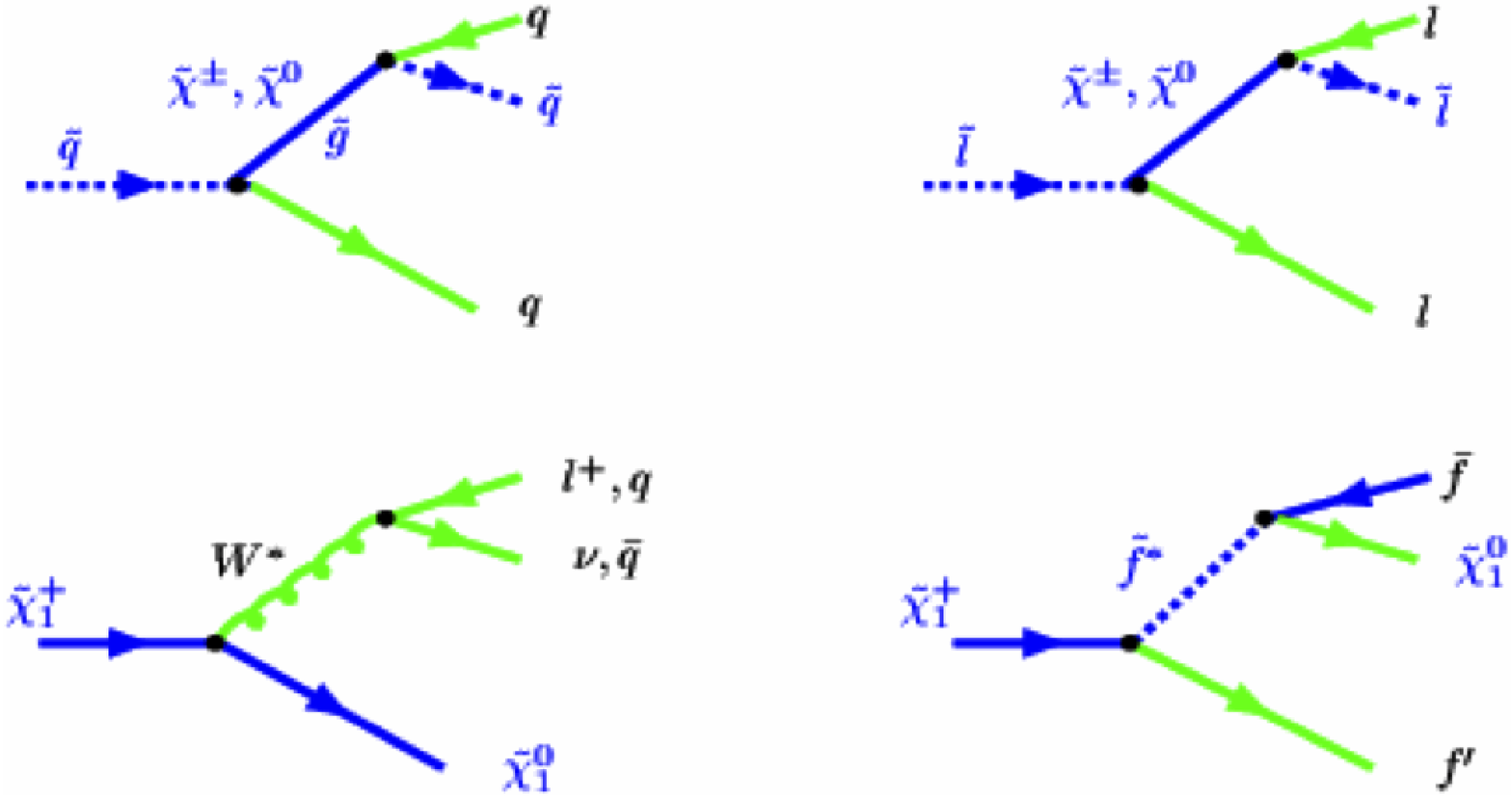}
\vspace{-0.5cm}
 \caption{Decay of superpartners}\label{decay}
 \end{figure}

The decay properties of superpartners also depend on their masses. For
the quark and lepton superpartners the main processes are shown in
Fig.\ref{decay}.

\section{Breaking of SUSY in the MSSM}
 \setcounter{equation} 0

 Usually it is assumed that supersymmetry is broken spontaneously  via the v.e.v.s of some fields.  However, in  the case of supersymmetry one can not use the scalar fields like the Higgs field, but  rather the auxiliary fields present in  any SUSY  multiplet.
 There are two basic mechanisms of spontaneous SUSY breaking: the Fayet-Iliopoulos (or D-type) mechanism~\cite{Fayet}  based on the  $D$  auxiliary field from  a vector multiplet and the O'Raifeartaigh  (or F-type) mechanism~\cite{O'R} based on  the $F$  auxiliary field from  a chiral multiplet. Unfortunately, one can not explicitly use these mechanisms within the MSSM
since none of the fields of the MSSM can develop non-zero v.e.v.
 without spoiling the gauge invariance. Therefore,  a
 spontaneous SUSY breaking should take place via
some other fields.

The most common scenario for producing
low-energy supersymmetry breaking is called the {\em hidden
sector} one~\cite{hidden}. According to this scenario, there exist
two sectors: the usual matter belongs to the "visible" one, while
the second, "hidden" sector, contains fields which lead to
breaking of supersymmetry. These two sectors interact with each
other by exchange of some fields called {\em messengers}, which
mediate SUSY breaking from the hidden to the visible sector. There
might be various types of messenger fields: gravity, gauge, etc.
The hidden sector is the weakest part of the MSSM. It contains a
lot of ambiguities and leads to uncertainties of the MSSM
predictions considered below.

So far there are known four main mechanisms to mediate SUSY breaking from a
hidden to a visible sector:\vspace{-0.2cm}
\begin{itemize}
\item Gravity mediation (SUGRA)~\cite{gravmed};\\[-0.7cm]
\item Gauge mediation~\cite{gaugemed};\\[-0.7cm]
\item Anomaly mediation~\cite{anommed};\\[-0.7cm]
\item Gaugino mediation~\cite{gauginomed}.
\end{itemize}

All four mechanisms of soft SUSY breaking are different in details
but are  common in results. Predictions for the sparticle spectrum
depend on the mechanism of SUSY breaking.
In what follows, to calculate the mass spectrum of superpartners,
we need an explicit form of SUSY breaking terms. For the
 MSSM and without  the $R$-parity violation one has in general
\begin{eqnarray}
&&\hspace{-0.4cm}-{\cal L}_{Breaking} = \label{soft}\\
&&\hspace{-0.4cm} =  \sum_{i}^{}m^2_{0i}|\varphi_i|^2+\left( \frac 12
\sum_{\alpha}^{}M_\alpha \tilde \lambda_\alpha\tilde \lambda_\alpha+
BH_1H_2\right.\nonumber\\
 & &\hspace{-0.4cm}+ \left. A^U_{ab}\tilde Q_a\tilde U^c_bH_2+A^D_{ab}\tilde Q_a
 \tilde D^c_bH_1+ A^L_{ab}\tilde L_a\tilde E^c_bH_1
  \right) , \nonumber
  \end{eqnarray}
where we have suppressed the $SU(2)$ indices.  Here $\varphi_i$
are all scalar fields, $\tilde \lambda_\alpha $  are the gaugino
fields, $\tilde Q, \tilde U, \tilde D$ and $\tilde L, \tilde E$
are the squark and slepton  fields, respectively, and $H_{1,2}$
are the SU(2) doublet Higgs fields.

Eq.(\ref{soft}) contains a vast number of free parameters which
spoils the prediction power of the model. To reduce their number,
we adopt the so-called {\em universality} hypothesis, i.e., we
assume the universality or equality of various soft parameters at
a high energy scale, namely, we put  all the spin 0 particle
masses to be equal to the universal value $m_0$, all the spin 1/2
particle (gaugino) masses to be equal to $m_{1/2}$ and all the
cubic and quadratic terms, proportional to $A$ and $B$, to repeat
the structure of the Yukawa superpotential (\ref{R}). This is an
additional requirement motivated by the supergravity mechanism of
SUSY breaking. Universality is not a necessary requirement and one
may consider nonuniversal soft terms as well. However, it will not
change the qualitative picture presented below; so for simplicity,
in what follows we consider the universal boundary conditions. In
this case, eq.(\ref{soft}) takes the form
\begin{eqnarray}
&&\hspace{-0.5cm}-{\cal L}_{Breaking}= \label{soft2} \\
&&\hspace{-0.5cm}=m_0^2\sum_{i}^{}|\varphi_i|^2+\left( \frac{ m_{1/2}}{2}
\sum_{\alpha}^{} \tilde \lambda_\alpha\tilde \lambda_\alpha\right.+  B[\mu H_1H_2] \nonumber\\
 &  & \left.\hspace{-0.5cm}  +\ A[y^U_{ab}\tilde Q_a\tilde U^c_bH_2+y^D_{ab}\tilde Q_a
 \tilde D^c_bH_1+ y^L_{ab}\tilde L_a\tilde E^c_bH_1]
\right).\nonumber
\end{eqnarray}

Thus, we are left with five free parameters, namely, $m_0,m_{1/2},A,B$ and $\mu$
versus two parameters of the SM coming from the Higgs potential, $m^2$ and $\lambda$.
In supersymmetry the Higgs potential is not arbitrary but is calculated from the Yuakawa and gauge terms
as we shall see below.

The soft terms explicitly break supersymmetry. As will be shown
later, they lead to the mass spectrum of superpartners different
from that of  ordinary particles. Remind that the masses of quarks
and leptons remain zero until $SU(2)$ invariance is spontaneously
broken.

\subsection{The soft terms and the mass formulas}

There are two main sources of the mass terms in the Lagrangian:
the $D$ terms and soft ones. With given values of
$m_0,m_{1/2},\mu,Y_t,Y_b,Y_\tau, A$, and $B$ one can construct the
mass matrices for all the particles. Knowing them at the GUT
scale, one can solve the corresponding RG equations, thus linking
the values at the GUT and electroweak scales. Substituting these
parameters into the mass matrices, one can predict the mass
spectrum of superpartners \cite{spectrum,BEK}.

\subsubsection{Gaugino-higgsino mass terms}

The mass matrix for  gauginos, the superpartners of the gauge
bosons, and for higgsinos, the superpartners of the Higgs bosons,
is nondiagonal, thus leading to their mixing. The mass terms look
like
\begin{eqnarray}
&&{\cal L}_{Gaugino-Higgsino}= \\
=&& -\frac{1}{2}M_3\bar{\lambda}_a\lambda_a
 -\frac{1}{2}\bar{\chi}M^{(0)}\chi -(\bar{\psi}M^{(c)}\psi + h.c.),\nonumber
\end{eqnarray}
where $\lambda_a , a=1,2,\ldots ,8,$ are the Majorana gluino
fields and
\begin{equation}
\chi = \left(\begin{array}{c}\tilde{B}^0 \\ \tilde{W}^3 \\
\tilde{H}^0_1 \\ \tilde{H}^0_2
\end{array}\right), \ \ \ \psi = \left( \begin{array}{c}
\tilde{W}^{+} \\ \tilde{H}^{+}
\end{array}\right)
\end{equation}
are, respectively, the Majorana neutralino and Dirac chargino
fields.

The neutralino mass matrix is
$$
M^{(0)}\!\!=\!\!\left(\!
\renewcommand{\tabcolsep}{0.04cm}
\begin{tabular}{cccc}
$M_1$ & $0$ & -$M_Zc_\beta s_W$ & $M_Zs_\beta s_W$ \\
$0$ &
$M_2$ & $M_Zc_\beta c_W$   & -$M_Zs_\beta c_W$  \\
-$M_Zc_\beta s_W$ & $M_Zc_\beta c_W$  & $0$ & -$\mu$ \\
$M_Zs_\beta s_W$ & -$M_Zs_\beta c_W$  & -$\mu$ & $0$
\end{tabular}\! \right),\label{neut)}
$$
where $\tan\beta = v_2/v_1$ is the ratio of two Higgs v.e.v.s and
$\sin_W= \sin\theta_W$ is the usual sinus of the weak mixing
angle. The physical neutralino masses  $M_{\tilde{\chi}_i^0}$ are
obtained as eigenvalues of this matrix after diagonalization.

For charginos one has
\begin{equation}
M^{(c)}=\left(
\begin{array}{cc}
M_2 & \sqrt{2}M_W\sin\beta \\ \sqrt{2}M_W\cos\beta & \mu
\end{array} \right).\label{char}
\end{equation}
This matrix has two chargino eigenstates
$\tilde{\chi}_{1,2}^{\pm}$ with mass eigenvalues
\begin{eqnarray}
M^2_{1,2}&=&\frac{1}{2}\left[M^2_2+\mu^2+2M^2_W \mp \right.\\ &&
\left.\hspace*{-1.8cm} \sqrt{(M^2_2\!\!-\!\!\mu^2)^2\!\!+\!\!4M^4_Wc^2_{2\beta}
\!\!+\!\!4M^2_W(\!M^2_2\!+\!\mu^2\!\!+\!\!2M_2\mu s_{2\beta}\! )}\right].\nonumber
\end{eqnarray}

\subsubsection{Squark and slepton masses}

Non-negligible Yukawa couplings cause a mixing between the
electroweak eigenstates and the mass eigenstates of the third
generation particles.  The mixing matrices for
$\tilde{m}^{2}_t,\tilde{m}^{2}_b$ and $\tilde{m}^{2}_\tau$ are
\begin{equation} \label{stopmat}
\left(\begin{array}{cc} \tilde m_{tL}^2& m_t(A_t-\mu\cot \beta )
\\ m_t(A_t-\mu\cot \beta ) & \tilde m_{tR}^2 \end{array}  \right),
\nonumber
\end{equation}
\begin{equation} \label{sbotmat}
\left(\begin{array}{cc} \tilde  m_{bL}^2& m_b(A_b-\mu\tan \beta )
\\ m_b(A_b-\mu\tan \beta ) & \tilde  m_{bR}^2 \end{array}
\right), \nonumber
\end{equation}
\begin{equation} \label{staumat} \left(\begin{array}{cc}
\tilde  m_{\tau L}^2& m_{\tau}(A_{\tau}-\mu\tan \beta ) \\
m_{\tau}(A_{\tau}-\mu\tan \beta ) & \tilde m_{\tau R}^2
\end{array}  \right)               \nonumber
\end{equation}
with
\begin{eqnarray*}
  \tilde m_{tL}^2&=&\tilde{m}_Q^2+m_t^2+\frac{1}{6}(4M_W^2-M_Z^2)\cos
  2\beta ,\\
  \tilde m_{tR}^2&=&\tilde{m}_U^2+m_t^2-\frac{2}{3}(M_W^2-M_Z^2)\cos
  2\beta ,\\
  \tilde m_{bL}^2&=&\tilde{m}_Q^2+m_b^2-\frac{1}{6}(2M_W^2+M_Z^2)\cos
  2\beta ,\\
  \tilde m_{bR}^2&=&\tilde{m}_D^2+m_b^2+\frac{1}{3}(M_W^2-M_Z^2)\cos
  2\beta ,\\
 \tilde m_{\tau L}^2&=&\tilde{m}_L^2+m_{\tau}^2-\frac{1}{2}(2M_W^2-M_Z^2)\cos
2\beta ,\\ \tilde m_{\tau
R}^2&=&\tilde{m}_E^2+m_{\tau}^2+(M_W^2-M_Z^2)\cos
  2\beta
\end{eqnarray*}
and the  mass eigenstates are  the eigenvalues of these mass matrices. For
the light generations the mixing is negligible.

The first terms here ($\tilde{m}^2$) are the soft ones, which are
calculated using the RG equations starting from their values at
the GUT (Planck) scale. The second ones are the usual masses of
quarks and leptons and the last ones are the $D$ terms of the
potential.

\subsection{The Higgs potential}

As has already been mentioned, the Higgs potential in the MSSM is
totally defined by superpotential (and the soft terms). Due to the
structure of ${\cal L}_{Yukawa}$ the Higgs self-interaction is given by the
$D$-terms while the $F$-terms contribute only to the mass matrix.
The tree level potential is
\begin{eqnarray}
V_{tree}&=&m^2_1|H_1|^2+m^2_2|H_2|^2\!-\!m^2_3(H_1H_2\!+\!h.c.)
\label{Higpot} \nonumber\\ &&\hspace*{-1.5cm}+
\frac{g^2+g^{'2}}{8}(|H_1|^2-|H_2|^2)^2 +
\frac{g^2}{2}|H_1^+H_2|^2,
\end{eqnarray}
where $m_1^2=m^2_{H_1}+\mu^2, m_2^2=m^2_{H_2}+\mu^2$. At the GUT
scale $m_1^2=m^2_2=m_0^2+\mu^2_0, \ m^2_3=-B\mu_0$. Notice that
the Higgs self-interaction coupling in eq.(\ref{Higpot}) is  fixed
and  defined by the gauge interactions as opposed to the SM.

The potential (\ref{Higpot}), in accordance with supersymmetry, is
positive definite and stable. It has no nontrivial minimum
different from zero.  Indeed, let us write the minimization
condition for  the potential (\ref{Higpot})
\begin{eqnarray}&&\hspace{-0.8cm} \frac 12\frac{\delta V}{\delta
H_1}\!=\!m_1^2v_1\! -\!m^2_3v_2\!+\! \frac{g^2\!\!+\!\!g'^2}4(v_1^2\!\!-\!\!v_2^2)v_1\!=\!0,
\label{min1}
\\&&\hspace{-0.8cm} \frac 12\frac{\delta V}{\delta H_2}\!=\!m_2^2v_2\!-\!m^2_3v_1\!+\! \frac{
g^2\!\!+\!\!g'^2}4(v_1^2\!\!-\!\!v_2^2)v_2\!=\!0, \label{min2}
\end{eqnarray}
where we have  introduced the notation $$<H_1>\equiv v_1= v
\cos\beta , \ \  <H_2>\equiv v_2= v \sin\beta, $$
$$ v^2=
v_1^2+v_2^2,\ \  \tan\beta \equiv \frac{v_2}{v_1}.$$ Solution of
eqs.(\ref{min1}),(\ref{min2}) can be expressed in terms of $v^2$
and $\sin 2\beta$
\begin{equation} v^2\!=\!\frac{\displaystyle  4(m^2_1\!-\!m^2_2\tan^2 \beta
)}{\displaystyle (g^2\!+\! g'^2)(\tan^2\beta\! -\!1)}, \sin2\beta
=\frac{\displaystyle 2m^2_3}{\displaystyle m^2_1\!+\!m^2_2}.
\label{min}
\end{equation}
One can easily see from eq.(\ref{min}) that if
$m_1^2=m_2^2=m_0^2+\mu_0^2$, $v^2$ happens to be negative, i.e.
the minimum does not exist.  In fact, real positive solutions to
eqs.(\ref{min1}),(\ref{min2}) exist only if the following
conditions are satisfied:
\begin{equation}
m_1^2+m_2^2 > 2 m_3^2, \ \ \ \  m_1^2m_2^2 < m_3^4 , \label{cond}
\end{equation}
which is not the case at the GUT scale. This means that
spontaneous breaking of the $SU(2)$  gauge invariance, which is
needed in the SM to give masses for all the particles, does not
take place in the MSSM.

This strong statement is valid, however, only at the GUT scale.
Indeed, going down with energy, the parameters of the potential
(\ref{Higpot}) are renormalized.  They become the ``running''
parameters with the energy scale dependence given by the RG
equations.

\subsection{Radiative electroweak symmetry breaking}

The running of the Higgs masses leads to the remarkable phenomenon known as
{\em radiative electroweak symmetry breaking}. Indeed, one can see
in Fig.\ref{16}  that $m_2^2$ (or both $m_1^2$ and $m_2^2$)
decreases when going down from the GUT scale to the $M_Z$ scale
and  can even become negative. As a result, at some value of
$Q^2$  the conditions (\ref{cond}) are satisfied, so that the
nontrivial minimum appears. This triggers spontaneous breaking of
the $SU(2)$ gauge invariance. The vacuum expectations of the Higgs
fields acquire nonzero values and provide masses to quarks,
leptons and $SU(2)$ gauge bosons, and additional masses to their
superpartners.

In this way one also obtains the explanation of why the two scales
are so much different. Due to the logarithmic running of the
parameters, one needs  a long "running time" to get $m_2^2$ (or
both $m_1^2$ and $m_2^2$) to be negative when starting from a
positive value of the order of $M_{SUSY}\sim 10^2 \div 10^3$ GeV
at the GUT scale.

\subsection{The mass spectrum}

\begin{figure}[hbt]\vspace{-0.5cm}
\begin{center}
\leavevmode \epsfxsize=5cm \epsffile{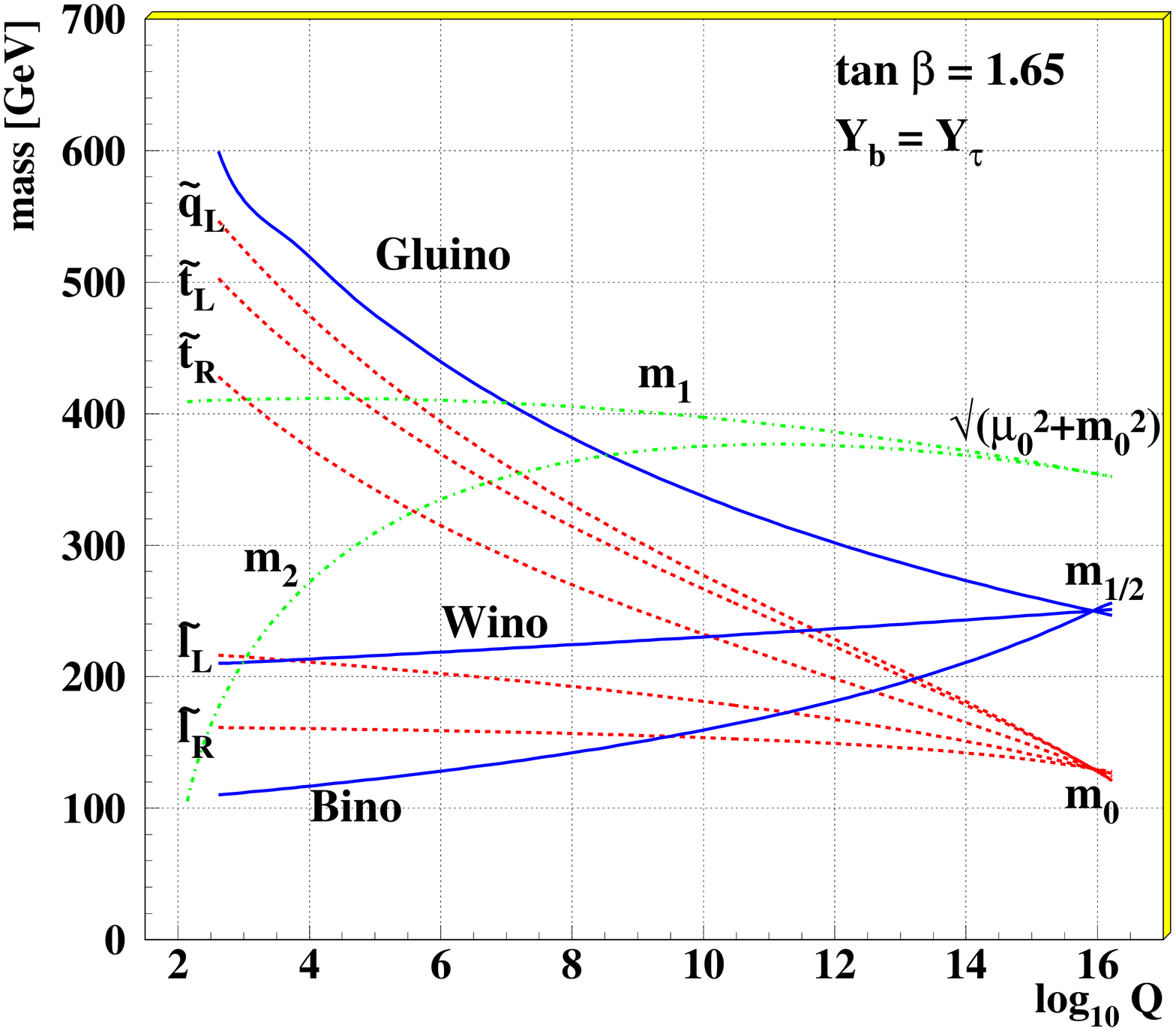}

 \leavevmode
       \epsfxsize=5cm
\epsffile{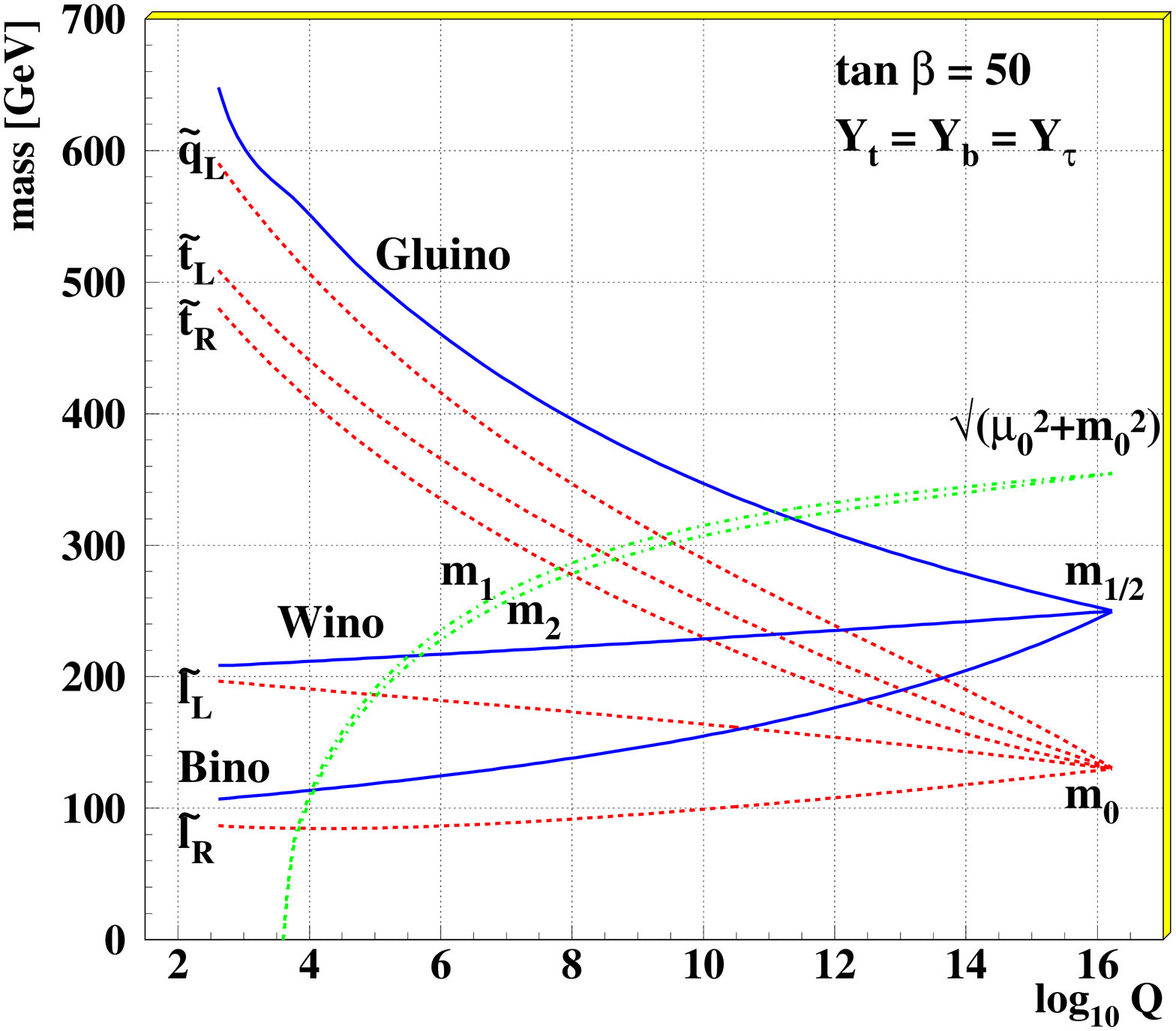}
\end{center}
\vspace{-1.5cm} \caption{An example of
evolution of sparticle masses and soft supersymmetry breaking
parameters $m_1^2=m^2_{H_1}+\mu^2$ and $m_2^2=m^2_{H_2}+\mu^2$ for
low (left) and high (right) values of $\tan\beta$ } \label{16}
\end{figure}

The mass spectrum is defined by low energy parameters.
To calculate the low energy values of the soft terms, we use the
corresponding RG equations~\cite{Ibanez}.
Having all the RG equations, one can now find the RG flow for the
soft terms.  Taking the initial values of the soft masses at the
GUT scale in the interval between $10^2\div 10^3$ GeV consistent
with the SUSY scale suggested by unification of the gauge
couplings (\ref{MSUSY}) leads to the  RG flow of the soft terms
shown in Fig.\ref{16}.~\cite{spectrum,BEK}
%
%

One should mention the following general features common to any
choice of initial conditions:

 i) The gaugino masses follow the running of the gauge couplings
 and split at low energies. The gluino mass is running faster
 than the others  and is usually the heaviest due to the strong interaction.

 ii) The squark and slepton masses also split at low energies, the
 stops (and sbottoms) being the lightest due to relatively big Yukawa couplings
 of the third generation.

  iii) The Higgs masses (or at least one of them) are running down
 very quickly and may even become negative.

\begin{figure}[ht]\vspace{-0.0cm}
\begin{center}
 \leavevmode
  \epsfxsize=5.cm
 \epsffile{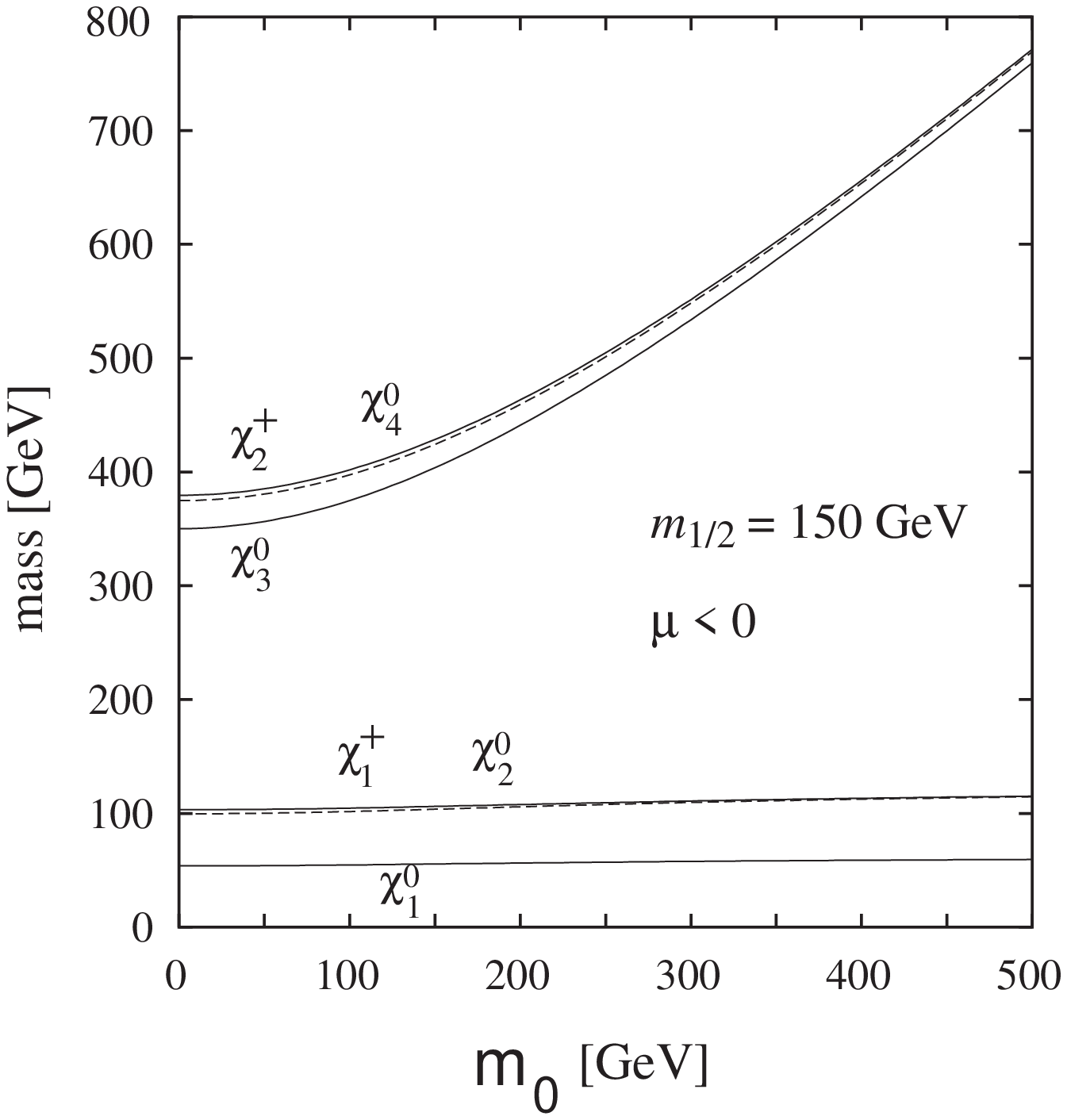}
 \epsfxsize=5.cm\epsffile{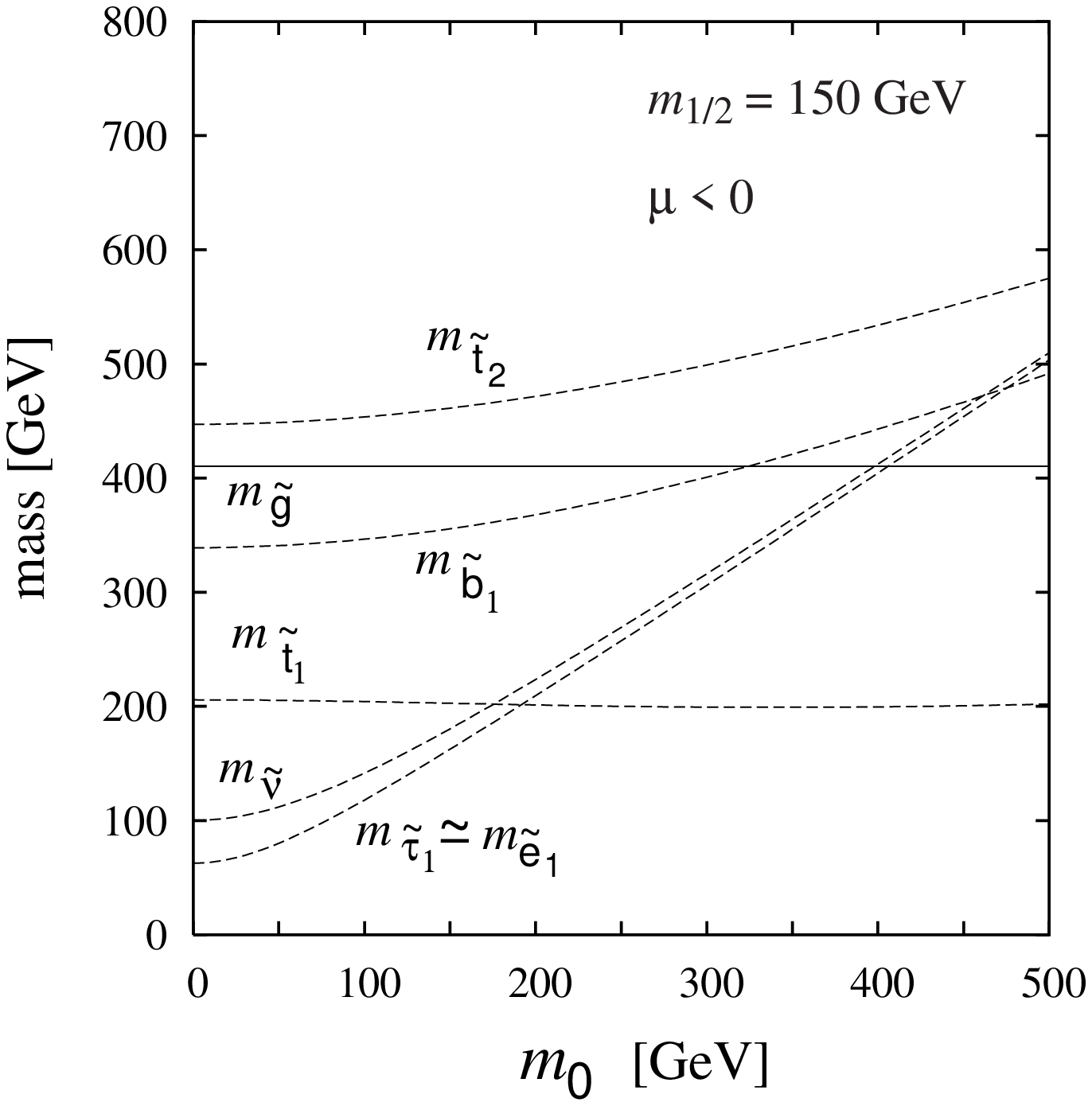}
\end{center}
\vspace{-0.5cm}
\caption{The masses of sparticles as functions of the initial
value $m_0$} \label{fig:barger}
\end{figure}

Typical dependence of the mass spectra on the initial conditions
($m_0$) is also shown in Fig.\ref{fig:barger} ~\cite{Barger,bbog}. For
a given value of $m_{1/2}$ the masses of the lightest particles
are practically independent of $m_0$, while the heavier ones
increase with it monotonically. One can see that the lightest
neutralinos and charginos as well as the stop squark may be rather
light.

Provided conditions (\ref{cond}) are satisfied, one can also calculate the masses of the Higgs bosons.
 The mass matrices
at the tree level are \\ CP-odd components
 $P_1$ and $P_2$ :
\begin{equation}
{\cal M}^{odd} = \left.\frac{\partial^2 V}{\partial P_i \partial
P_j} \right |_{H_i=v_i} \hspace{-0.6cm}= \left(\!\! \begin{array}{cc}  \tan\beta &1
\\1& \cot\beta \end{array}\!\!\right) m_3^2,
\end{equation}
 CP-even neutral components $S_1$ and $S_2$:
\begin{eqnarray}
{\cal M}^{ev} &=& \left.\frac{\partial^2 V}{\partial S_i \partial
S_j} \right|_{H_i=v_i} \hspace{-0.5cm} =\left(\!
\begin{array}{cc} \tan\beta & -1
\\-1& \cot\beta \end{array}\!\right) m_3^2\nonumber \\
& +&\left( \begin{array}{cc}
\cot\beta & -1 \\-1&\tan\beta \end{array}\right) M_Z
\frac{\sin2\beta}{2},
\end{eqnarray}
Charged components $H^-$ and $H^+$:
\begin{eqnarray}
{\cal M}^{ch} &=&\left.\frac{\partial^2 V}{\partial H^+_i
\partial H^-_j} \right|_{H_i=v_i} \\&& \hspace{-0.7cm} = \left(\!\!\!\! \begin{array}{cc}
\tan\beta &1 \\1&\!\!\!\! \cot\beta \end{array}\!\!\!\!\right)
 (m_3^2+M_W\frac{\sin2\beta}{2}).\nonumber
\end{eqnarray} Diagonalizing the mass matrices, one gets the mass
eigenstates:
 $$\hspace{-0.3cm}\begin{array}{l} \left\{\!\!
\begin{array}{l} G^0 \ = -\cos\beta P_1+\sin \beta P_2 ,
\ Goldst \ boson  \to Z_0, \\ A \ = \sin\beta P_1+\cos \beta P_2 ,  \
Neutral \ CP odd \ Higgs, \end{array}\right.\\  \\ \left\{\!\!
\begin{array}{l} G^+\! \!= \!-\!\cos\beta (H^-_1)^*\!+\!\sin \beta H^+_2 , \
Goldst \ boson\!  \to\! W^+, \\ H^+ = \sin\beta (H^-_1)^*+\cos \beta
H^+_2 , \ Charged \ Higgs, \end{array}\right.\\ \\ \left\{\!\!
\begin{array}{l} h \ = -\sin\alpha S_1+\cos\alpha S_2 ,
\ SM \ CP\ even \ Higgs, \\ H \ = \cos\alpha S_1+\sin\alpha S_2 ,
\ Extra \ heavy \ Higgs , \end{array}\right.
\end{array}$$ where the mixing angle $\alpha $ is
given by $$ \tan 2\alpha = \tan 2\beta
\left(\frac{m^2_A+M^2_Z}{m^2_A-M^2_Z}\right).$$
The physical Higgs
bosons acquire the following masses \cite{MSSM}:
 \begin{eqnarray} \mbox{CP-odd
neutral Higgs} \ A: && \hspace{-0.5cm}m^2_A = m^2_1\!+\!m^2_2,
\\ \mbox{Charge Higgses}  \ H^{\pm}: && \hspace{-0.5cm} m^2_{H^{\pm}}=m^2_A\!+\!M^2_W ,
\nonumber  \end{eqnarray}
CP-even neutral Higgses \ \ H, h:
\begin{eqnarray}&&\hspace{-0.7cm} m^2_{H,h}\!=\!
\frac{1}{2}\!\left[\!m^2_A\!\!+\!\!M^2_Z \!\pm\!
\sqrt{(m^2_A\!\!+\!\!M_Z^2)^2\!\!-\!\!4m^2_AM_Z^2c^2_{2\beta}}\!\right],\nonumber\\
&&
\end{eqnarray}
where, as usual,
 $$ M^2_W=\frac{g^2}{2}v^2, \ \
M^2_Z=\frac{g^2+g'^2}{2}v^2 .$$ This leads to the once celebrated
SUSY mass relations
\begin{eqnarray}&& m_{H^{\pm}} \geq M_W, \
m_h \leq m_A \leq M_H,\nonumber \\ &&m_h \leq M_Z |\cos 2\beta| \leq
M_Z, \label{bound}\\ && m_h^2+m_H^2=m_A^2+M_Z^2\nonumber
\end{eqnarray}

Thus, the lightest neutral Higgs boson happens to be lighter than
the $Z$ boson, which clearly distinguishes it from the SM one.
Though we do not know the mass of the Higgs boson in the SM, there
are several indirect constraints leading to the lower boundary of
$m_h^{SM} \geq 135 $ GeV. After including the radiative
corrections,  the mass of the lightest Higgs boson in the MSSM,
$m_h$, reads
\begin{equation}
m_h^2\!=\!M_Z^2 \cos^2(2\beta)\! +\!\frac{3g^2m_t^4}{16\pi^2M_W^2}
\log\frac{\tilde m^2_{t_1}\tilde m^2_{t_2}}{m_t^4} +...
\end{equation}
which leads to about 40 GeV increase~\cite{CW+we}. The second loops correction is negative but small~\cite{feynhiggs}.

\section{Constrained MSSM}
 \setcounter{equation} 0

\subsection{Parameter space of the MSSM}

The Minimal Supersymmetric Standard Model has the following free
parameters: i) three gauge couplings $\alpha_i$; ii) three
matrices of the Yukawa couplings $y^i_{ab}$, where $i = L,U,D$;
iii) the Higgs field mixing parameter  $\mu $; iv) the soft
supersymmetry breaking parameters. Compared to the SM there is an
additional Higgs mixing parameter, but the Higgs self-coupling,
which is arbitrary in the SM,  is fixed by supersymmetry. The main
uncertainty comes from the unknown soft terms.

With the universality hypothesis one is left with the following
set of 5 free parameters defining the mass scales
 $$ \mu, \ m_0, \ m_{1/2}, \ A \ \mbox{and}\
 B \leftrightarrow \tan\beta = \frac{v_2}{v_1}. $$
While choosing parameters and making predictions, one has two
possible ways to proceed:

 i) take the low-energy parameters like superparticle masses
  $\tilde{m}_{t1},\tilde{m}_{t2}, m_A$,
$\tan\beta$, mixings $X_{stop},\mu$, etc. as input  and calculate
cross-sections as functions of these parameters.

 ii) take the high-energy parameters like the above mentioned 5
soft parameters as input, run the RG equations and find the
low-energy values. Now the calculations can be carried out in
terms of the initial parameters. A  typical range of these parameters
is
$$ 100\ GeV \leq m_0,m_{1/2}, \mu \leq 1-2 \ TeV, $$
$$  -3m_0 \leq A_0 \leq 3m_0, \ \ \   1\leq \tan\beta \leq 70.$$
 The experimental constraints are
sufficient to determine these parameters, albeit with large
uncertainties.

\subsection{The choice of constraints}

When subjecting constraints on the MSSM, perhaps, the most
remarkable fact is that all of them can be fulfilled
simultaneously. In our analysis we impose the following
constraints on the parameter space of the MSSM:

$\bullet$ Gauge coupling constant unification; \\ This is one of
the most  restrictive constraints, which we have discussed in Sect
2. It fixes the scale of SUSY breaking of an order of 1 TeV.

$\bullet$ $M_Z$ from electroweak symmetry breaking;\\
Radiative EW symmetry breaking  (see eq.(\ref{min})) defines the
mass of the Z-boson
$$\frac{M_Z^2}{2}\!=\!\frac{m_1^2\!-\!m_2^2\tan^2\beta}{\tan^2\beta\!-\!1}\!=\!
\!-\!\mu^2\!+\!\frac{m_{H_1}^2\!-\!m_{H_2}^2\tan^2\beta}{\tan^2\beta\!-\!1}.$$
This condition determines the value of $\mu$ for given
values of $m_0$ and $m_{1/2}$.

$\bullet$ Precision measurement of decay rates;\\
We take the branching ratio $BR(b\to s \gamma)$ which has been
measured by the CLEO~\cite{CLEO} collaboration and later by
ALEPH~\cite{ALBSG} and the branching ration $BR(B_s\to \mu^+\mu^-)$
measured recently by  CDF collaboration~\cite{bmu}. Susy contributions
should not destroy the agreement with the SM and in some cases can improve it.
This requirement imposes severe
restrictions on the parameter space, especially for the case of
large $\tan\beta$.

$\bullet$ Anomalous magnetic moment of muon;\\
Recent measurement of the anomalous magnetic moment indicates
small deviation from the SM of the order of 2 $\sigma$. The
deficiency may be easily filled with SUSY contribution.

$\bullet$ The lightest superparticle (LSP) should be neutral, otherwhise
we would have charged clouds of stable particles in the Universe which is not observed.

$\bullet$ Experimental lower limits on SUSY masses; \\ SUSY
particles have not been found so far and  the searches at LEP
 impose  the lower limit on the charged lepton and chargino
masses of about  half of the centre of mass energy~\cite{LEPSUSY}.
The lower limit on the neutralino masses  is smaller. There exist
also limits on squark and gluino masses from the Tevatron
collider~\cite{TEVSUSY}.  These limits restrict the  minimal
values for the SUSY mass parameters.

$\bullet$ Dark Matter constraint; \\
Recent very precise astrophysical data restrict the amount of the
Dark matter in the Universe up to 23\%.  Assuming $h_0\sim0.7$ one
finds that the contribution of each relic particle species $\chi$
has to obey $\Omega_\chi h^2_0 \sim 0.13\pm 0.03.$. This serves as
a very severe bound on SUSY parameters~\cite{relictst}.

Having in mind the above mentioned constraints one can  find the
most probable region of the parameter space by minimizing the
$\chi^2$ function~\cite{BEK}.  Since the parameter space is 5 dimensional
one can not plot it explicitly and is bounded to use various projections.
We will accept  the following strategy:
We first choose the value of the
Higgs mixing parameter $\mu$ from the requirement of radiative EW
symmetry breaking and then take the set of values of $\tan\beta$. Parameter
$A$ happens to be less important and we will fix it typically like $A_0=0$.
Then we are left with two parameters $m_0$ and $m_{1/2}$ and we present the
restrictions coming from various constraints in the $m_0,m_{1/2}$ plane.
 \subsection{The excluded regions of parameter space}

\begin{figure*}[ht]\vspace{-0.5cm}
\begin{center}
 \leavevmode
  \epsfxsize=5.cm
 \epsffile{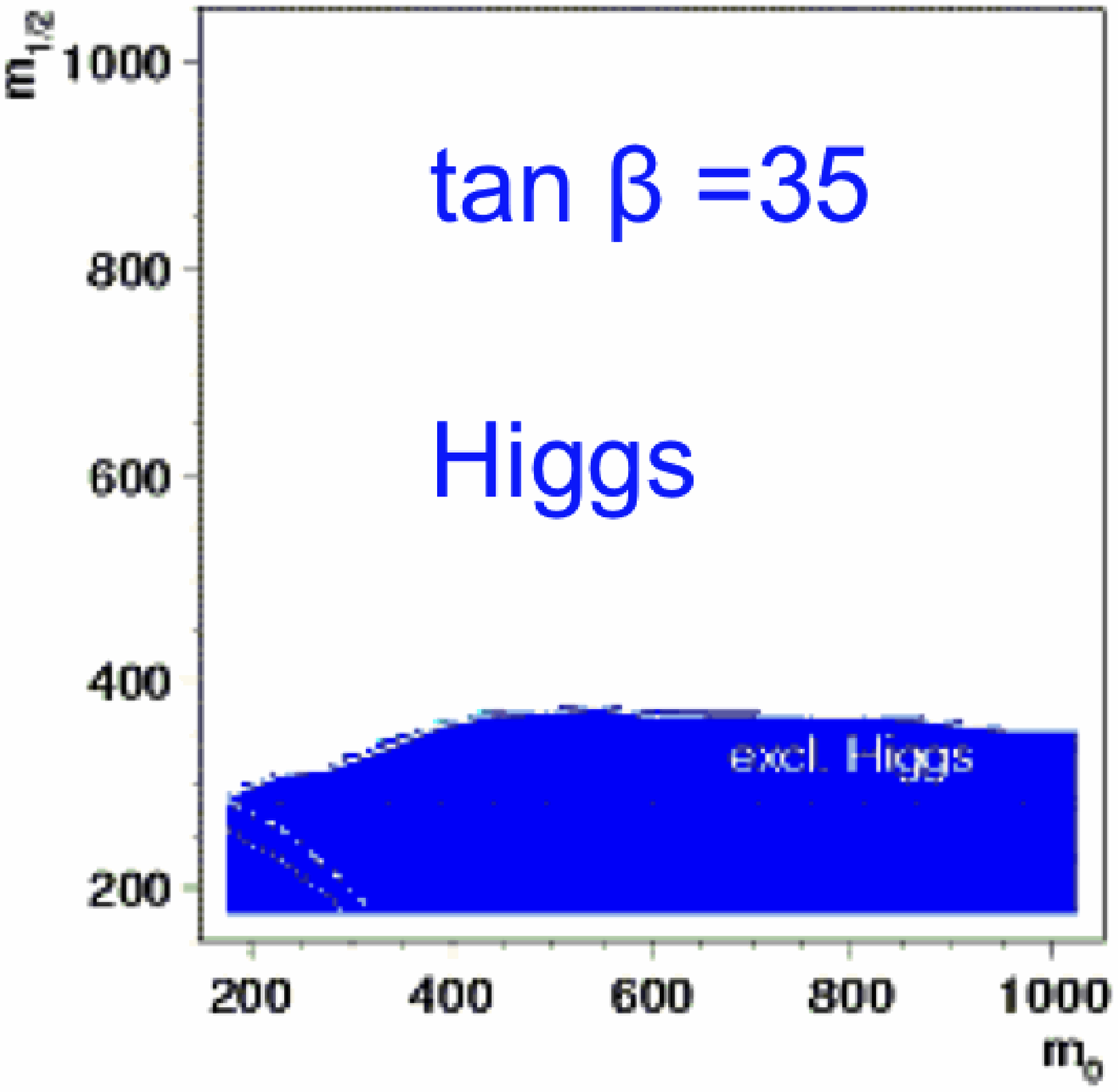}  \epsfxsize=5.cm
 \epsffile{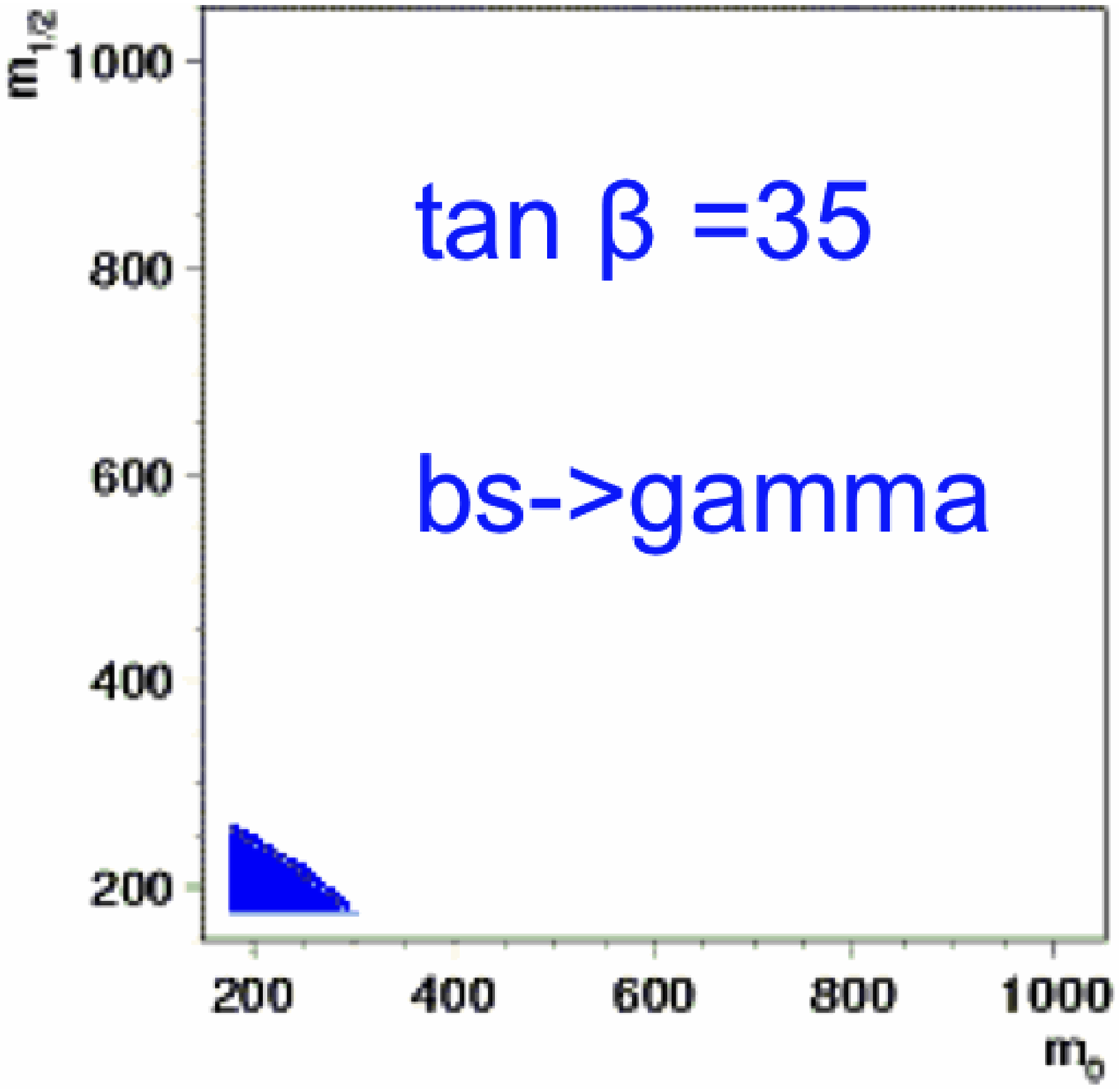}
   \epsfxsize=5.cm
 \epsffile{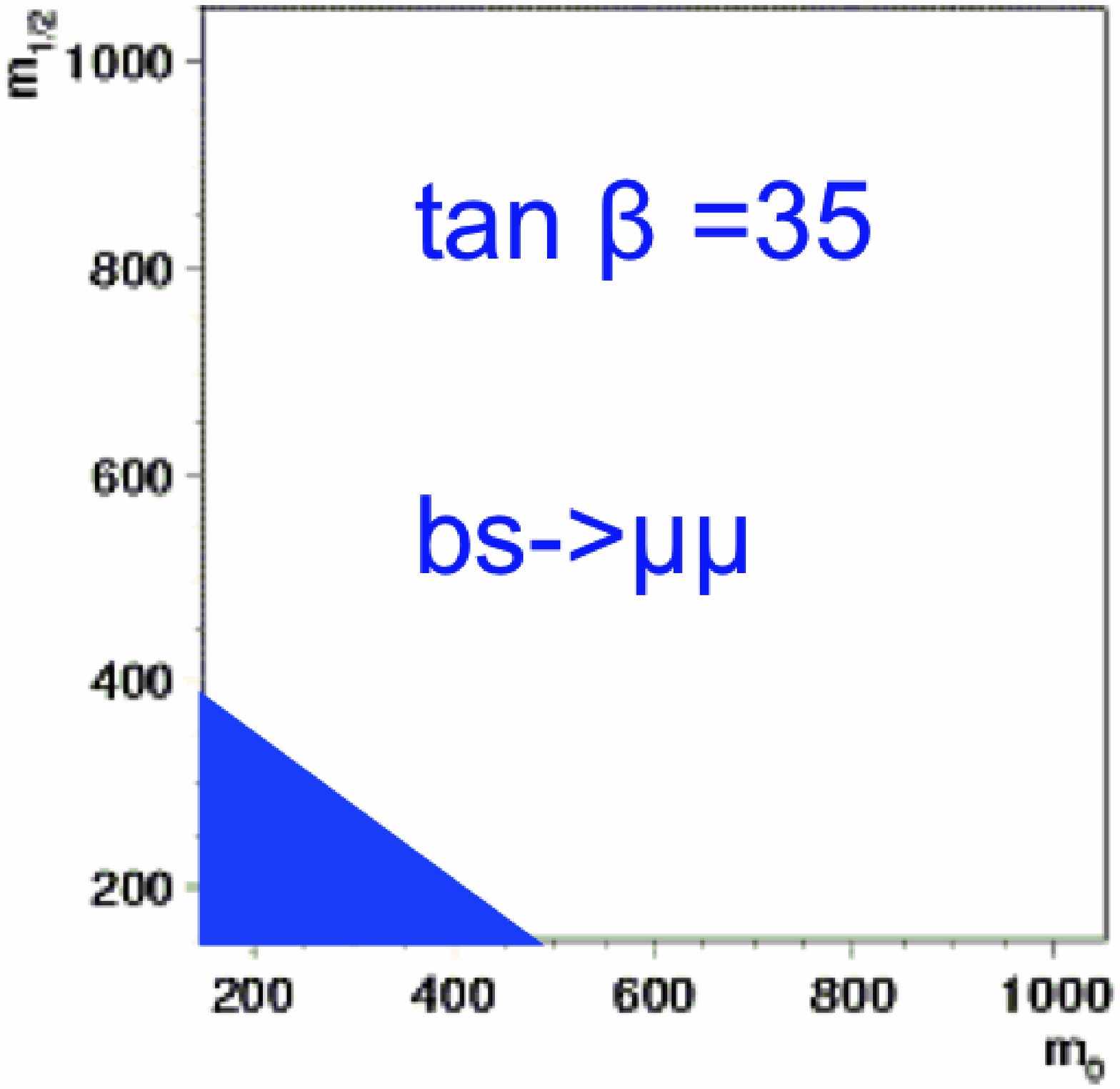}  \epsfxsize=5.cm
 \epsffile{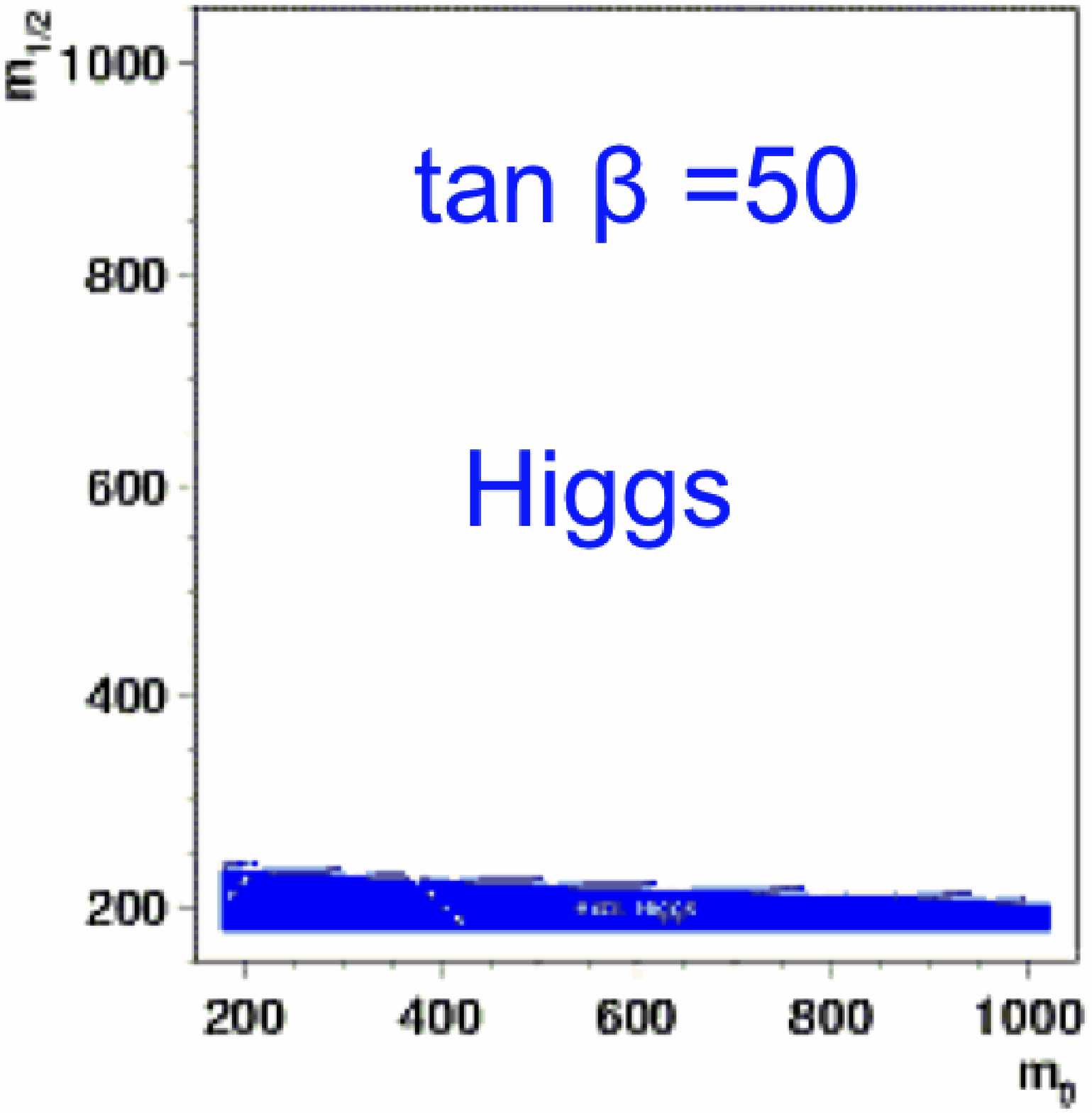}
   \epsfxsize=5.cm
 \epsffile{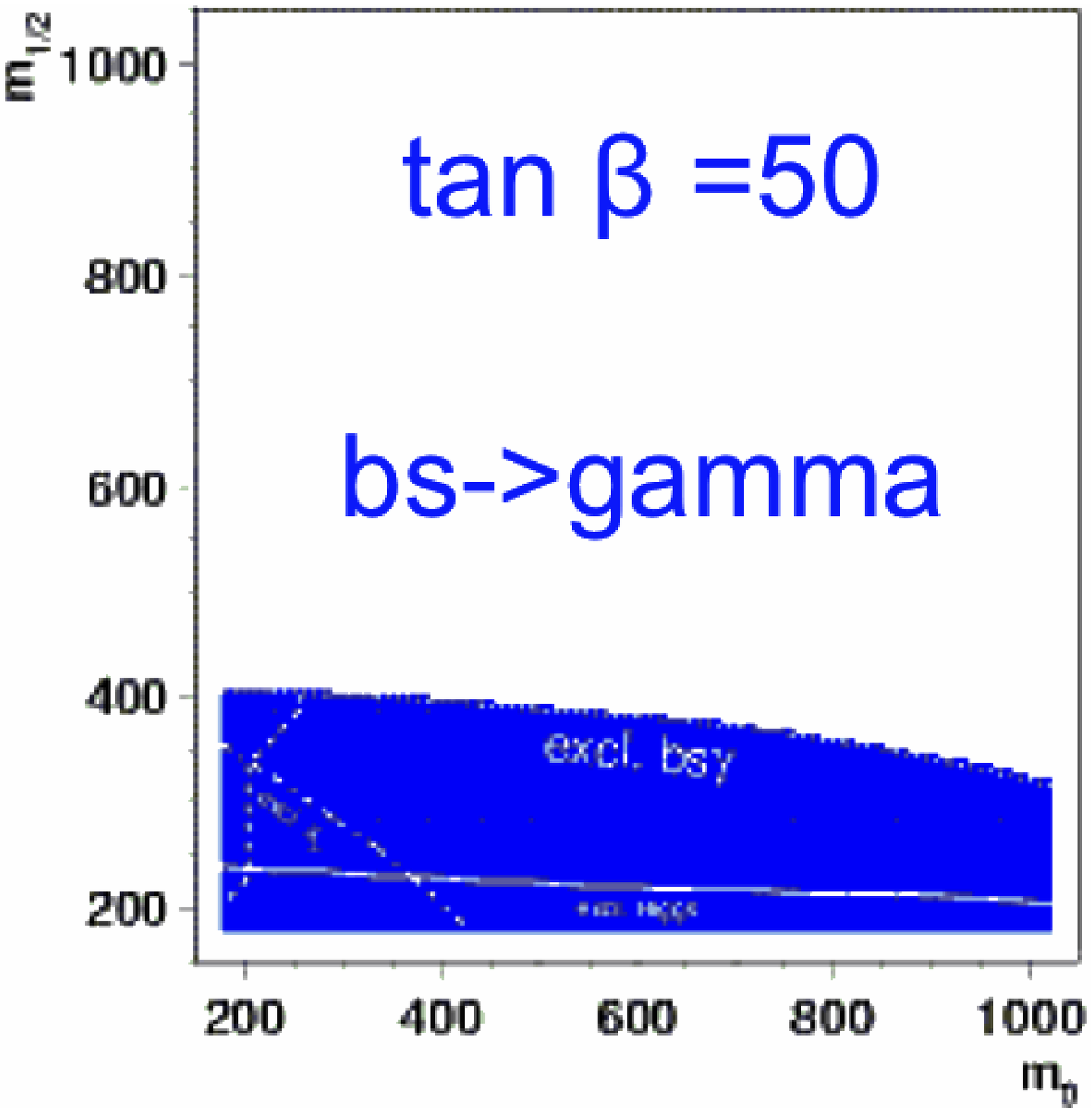}  \epsfxsize=5.cm
 \epsffile{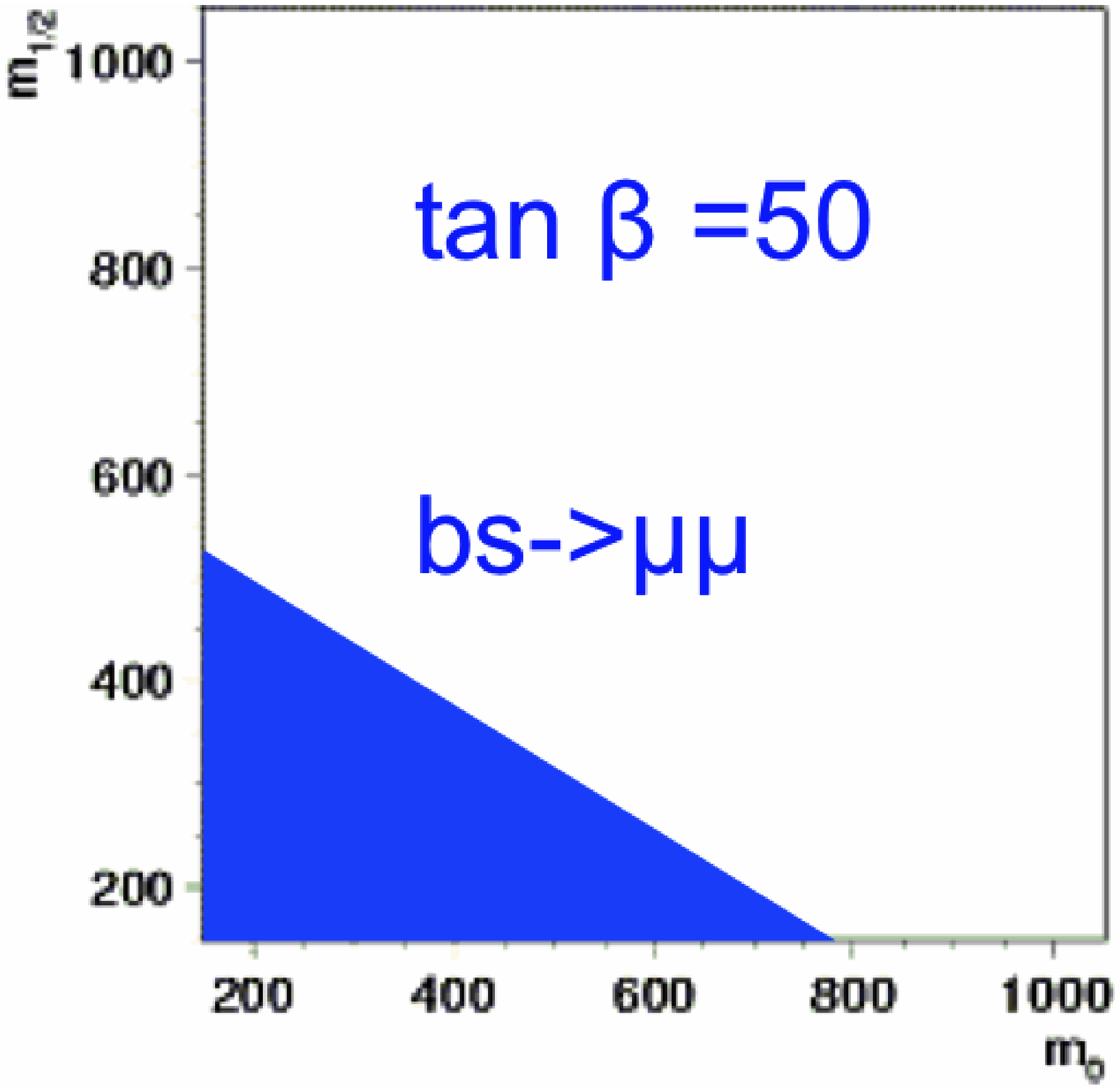}
   \epsfxsize=5.cm
 \epsffile{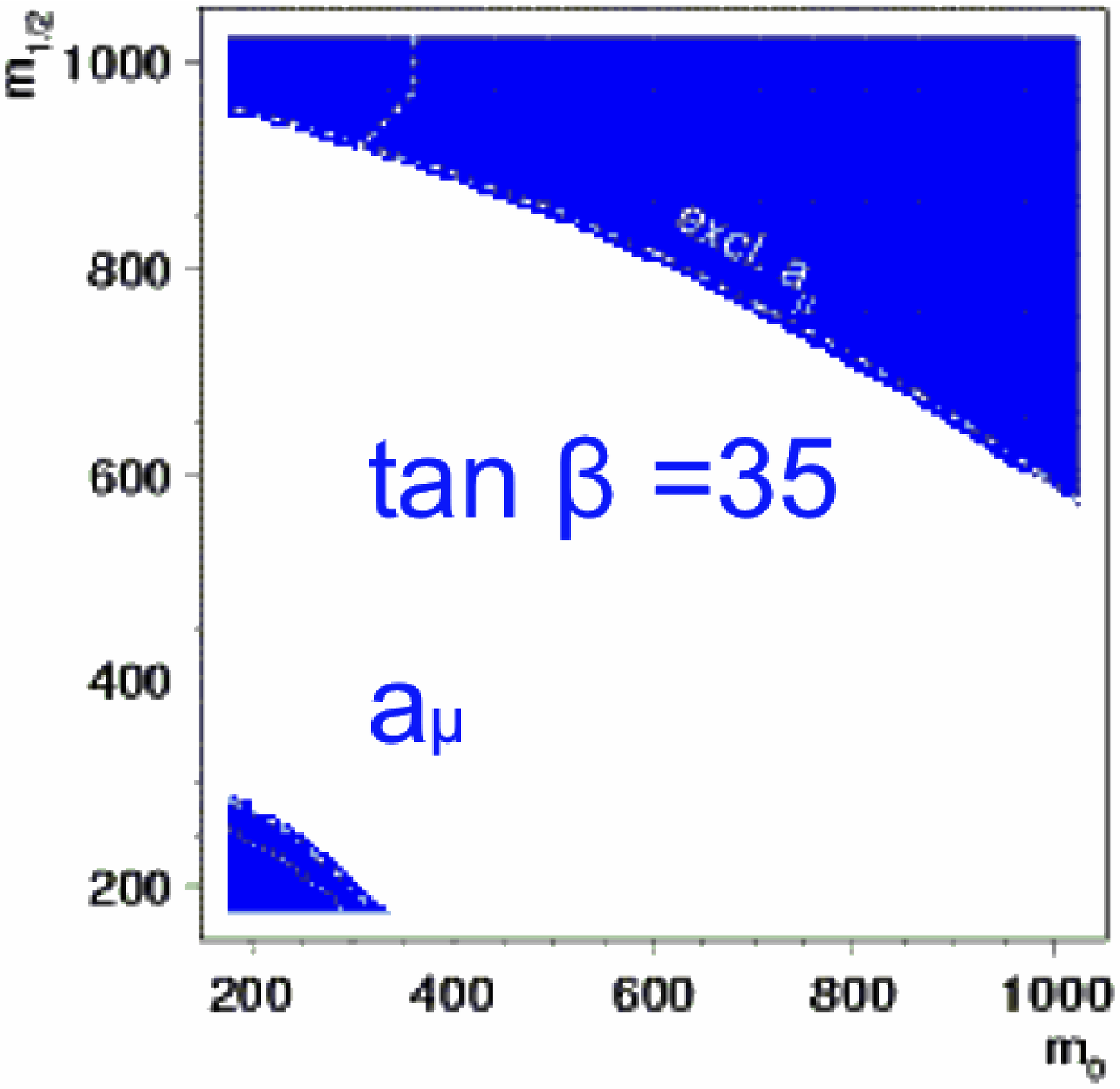}  \epsfxsize=5.cm
 \epsffile{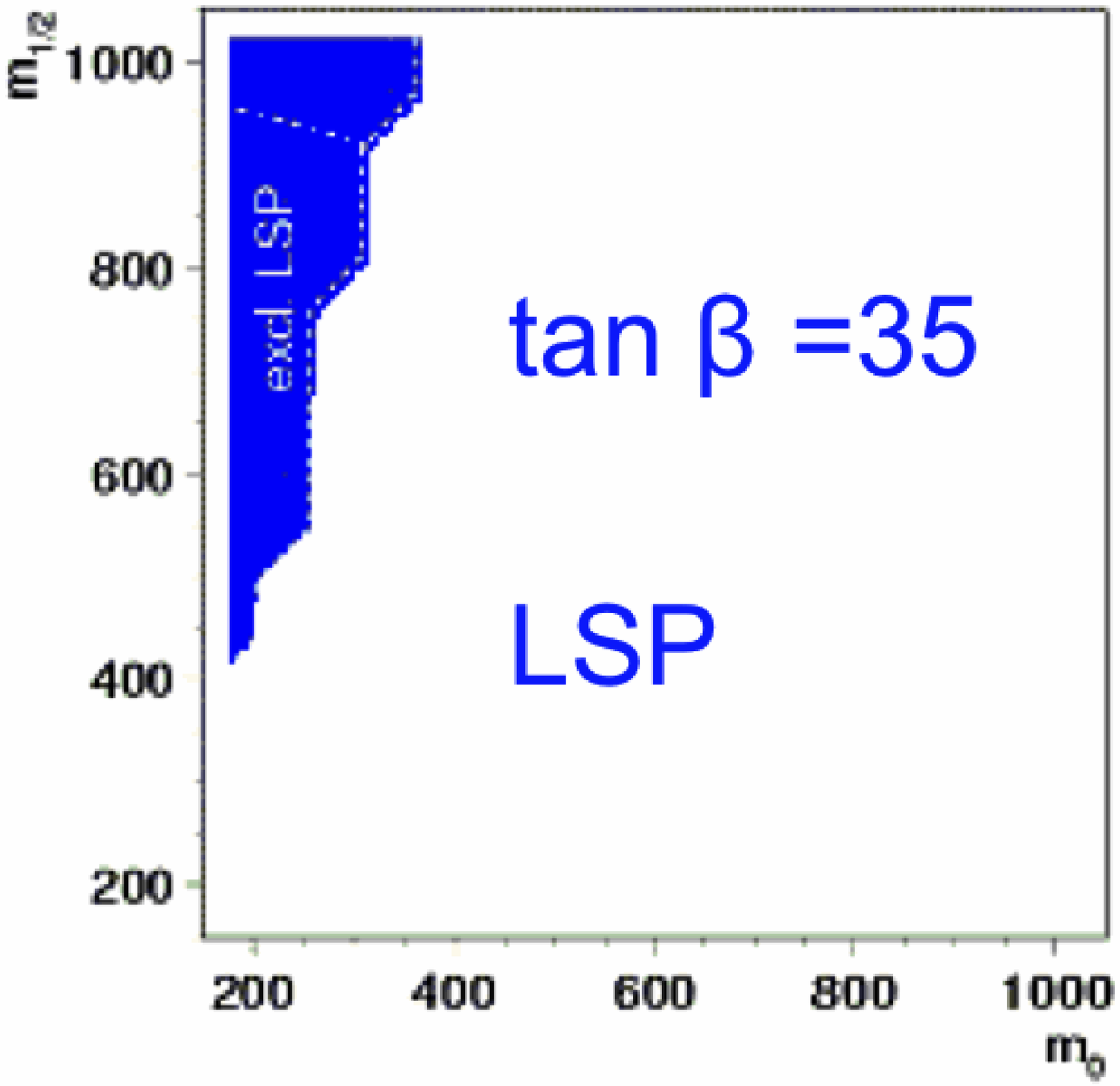}
    \epsfxsize=5.cm \epsffile{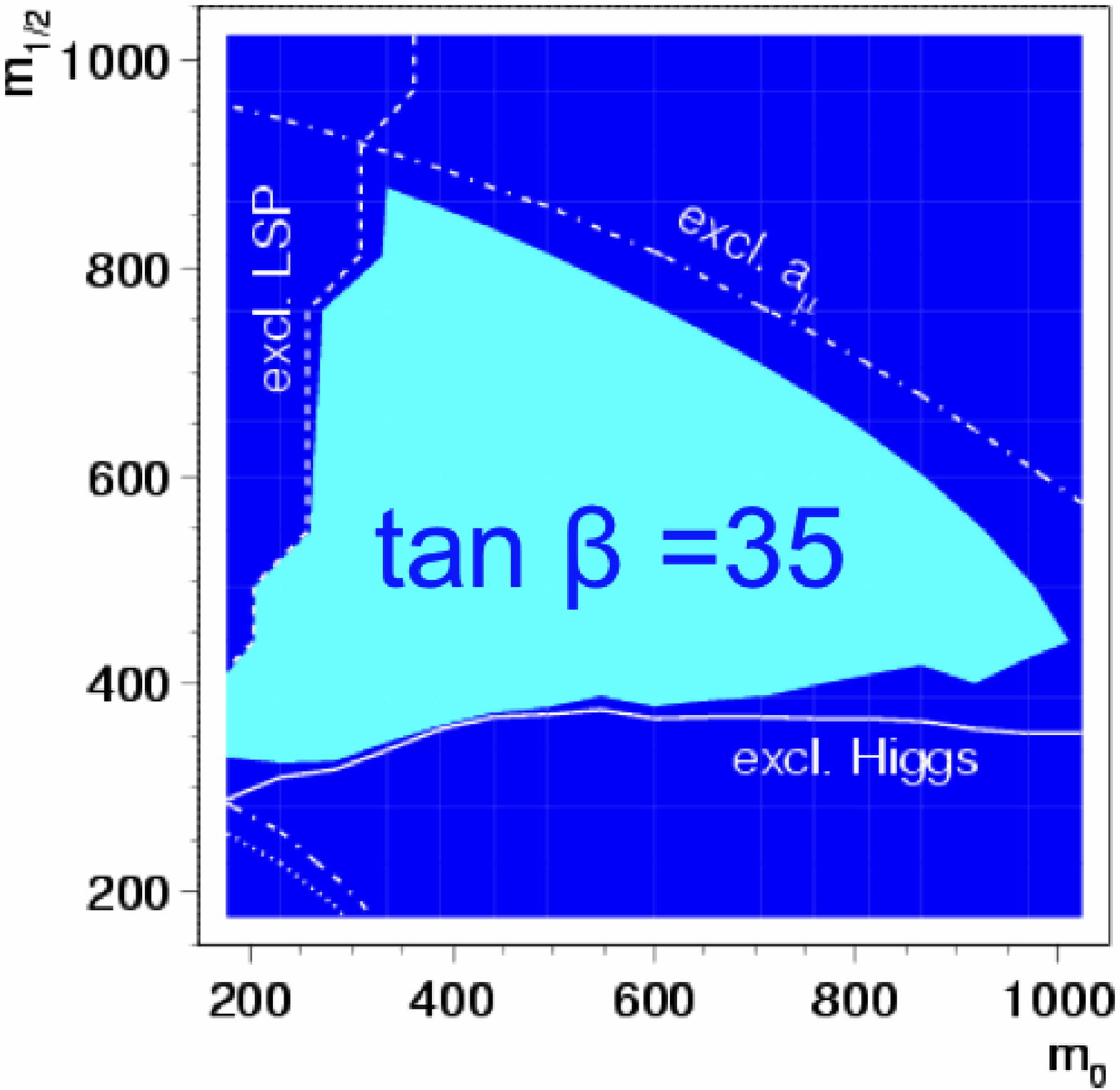}    \epsfxsize=5.cm
   \epsffile{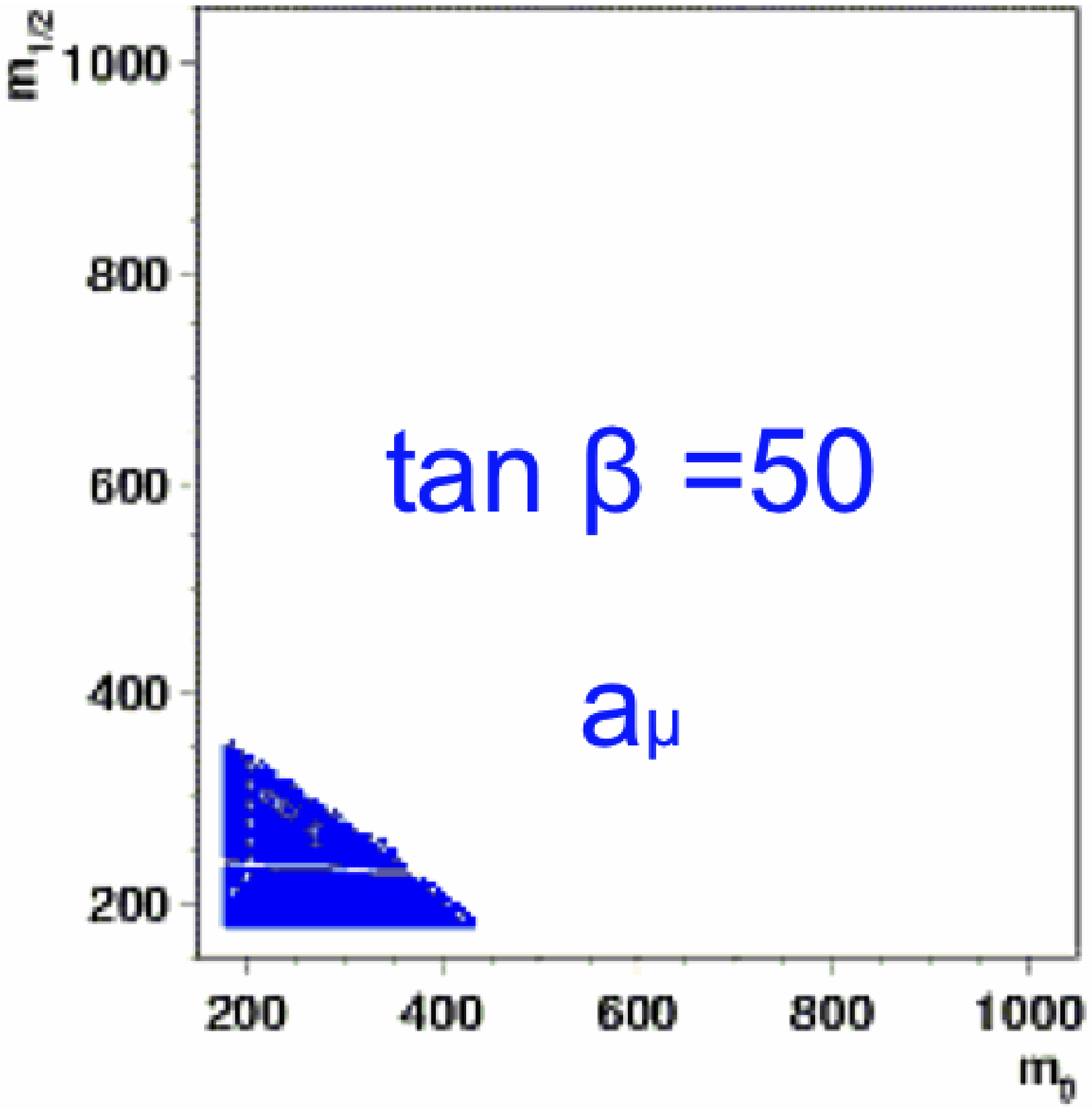}
   \epsfxsize=5.cm
   \epsffile{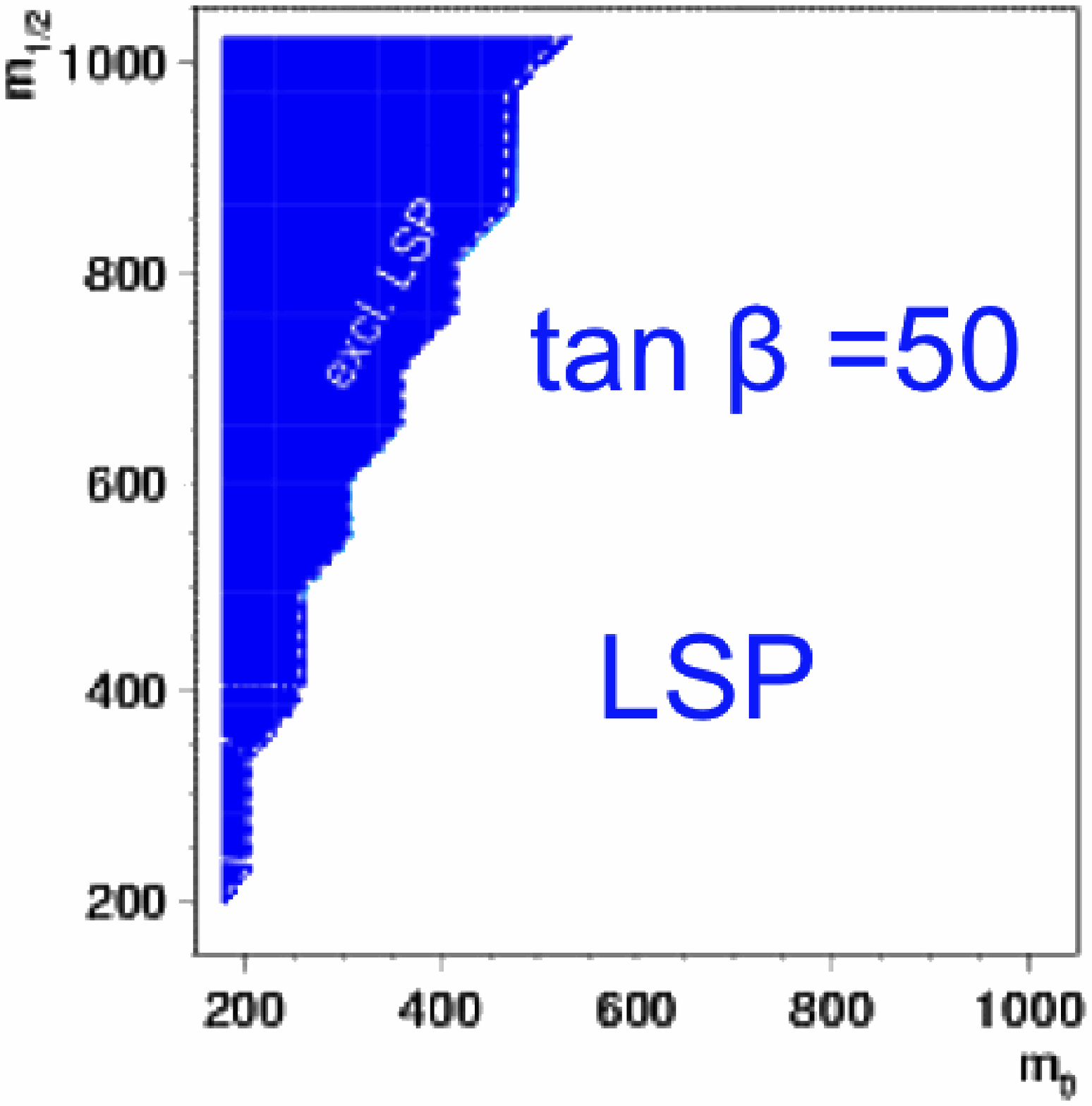}  \epsfxsize=5.cm
 \epsffile{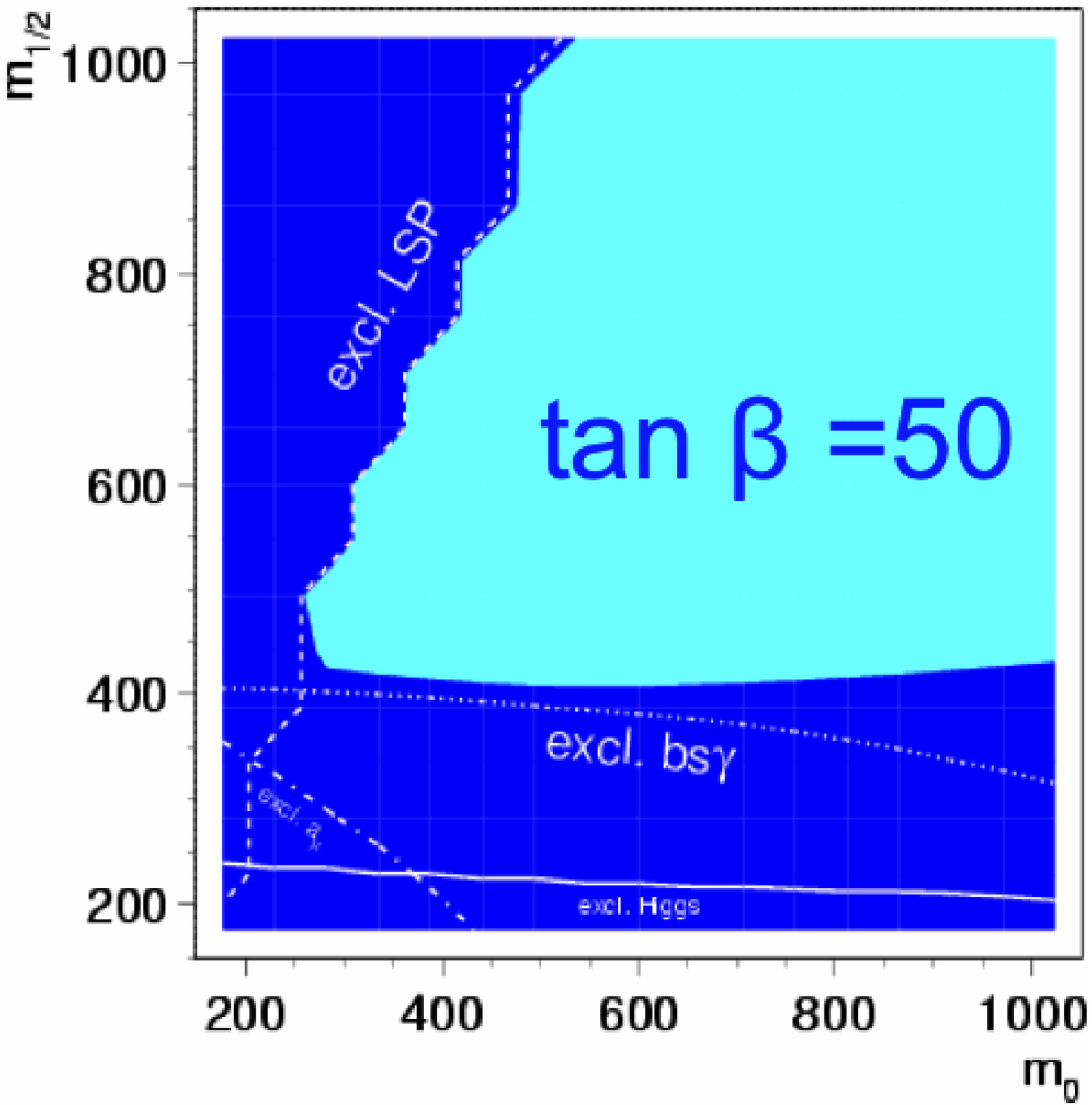}   \end{center}
\vspace{-1.5cm}
\caption{Regions excluded by various constraints for $tan\beta=35$ and $50$ shown in blue( dark). The last plots show the combination of all constraints leaving the allowed region of parameter space} \label{constr}
\end{figure*}

$\bullet$ We start with the Higgs mass constraint. Experimental lower limit on the Higgs  mass  from LEP2:
$m_h \geq 114.7$ GeV cuts the part of the parameter space as shown in Fig.\ref{constr}.

\begin{figure}[ht]\vspace{-1cm}
\begin{center}
 \leavevmode
  \epsfysize=3.cm\epsfxsize=5.cm  \epsffile{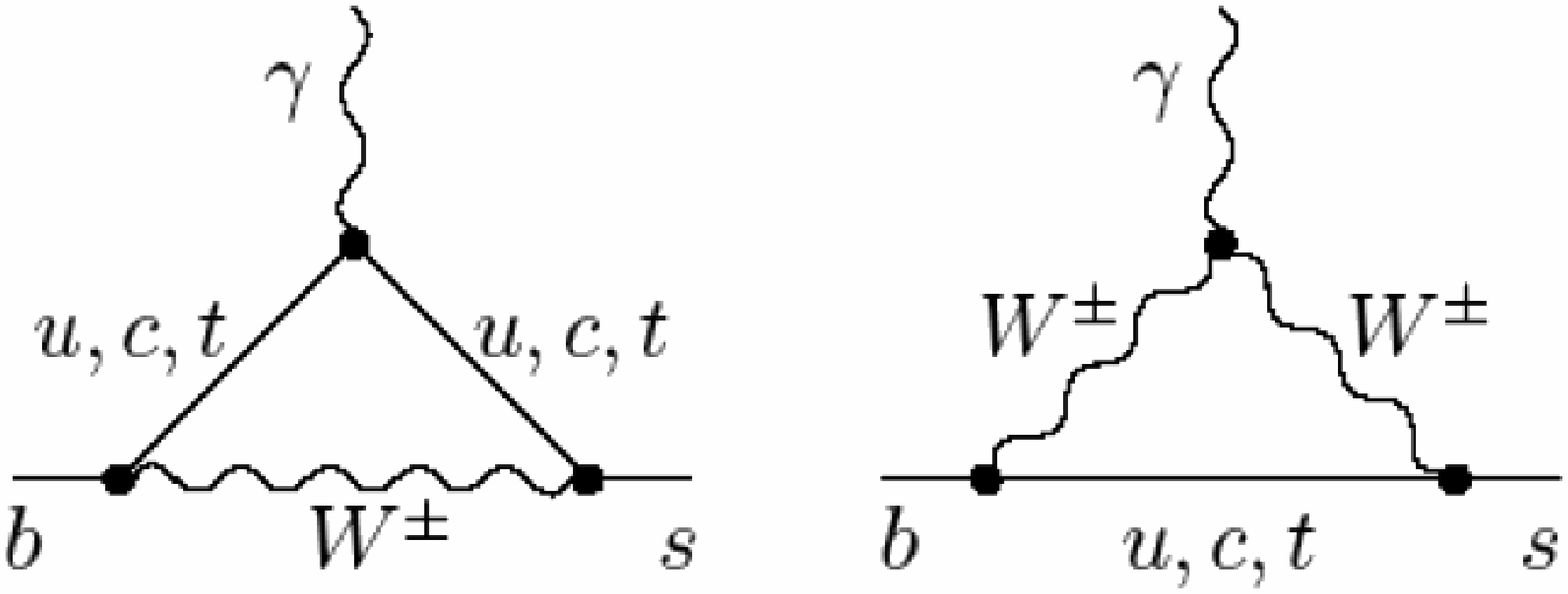}
 \epsfysize=3.cm\epsfxsize=7.cm  \epsffile{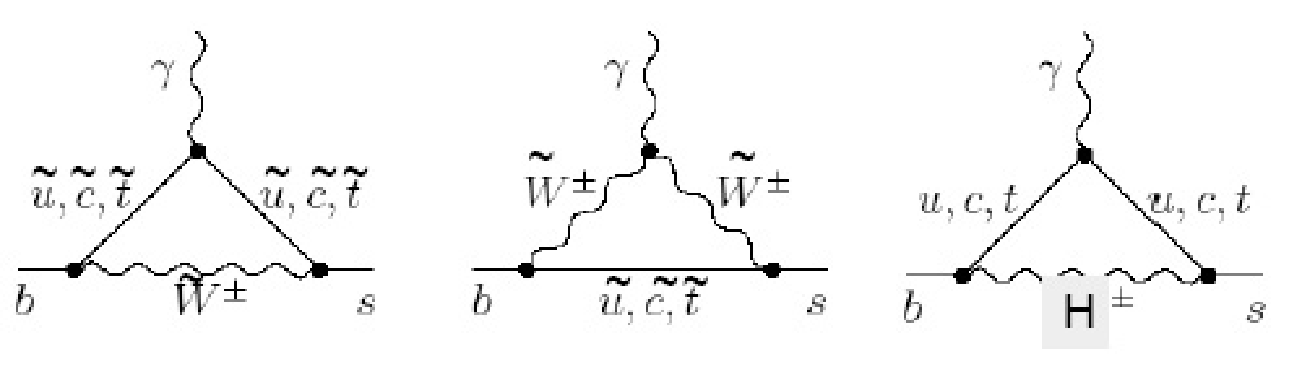}   
 \end{center}
\vspace{-1.cm}
\caption{The diagrams contributing to $b\to s\gamma$ decay in the SM and in the MSSM.} \label{bsg}
\end{figure}

$\bullet$ The next two constrains are related to rare decays where SUSY may contribute. The first one is
$b\to s\gamma$ decay  which in the SM given by the diagrams shown on top of Fig.\ref{bsg}  and leads to
$$BR^{SM}(b\to s\gamma)=(3.28\pm0.33)\cdot 10^{-4}$$ while experiment gives ~\cite{CLEO,ALBSG}
$$BR^{EX}(b\to s\gamma)=(3.43\pm0.36)\cdot 10^{-4}.$$ These two values almost coincide but still leave some room for SUSY. SUSY contribution comes from the diagrams shown in the bottom of Fig.\ref{bsg}  and is enhanced by $\tan\beta$\cite{bsgsusy}
$$BR^{SUSY}\!(b\to s\gamma\!) \propto \mu A_t m_b \tan\beta f(\tilde m^2_{t_1},\tilde m^2_{t_1},m_{\chi^\pm})
$$ The obtained constraints are shown in Fig.\ref{constr}.

\begin{figure}[ht]\vspace{-0.5cm}
\begin{center}
 \leavevmode
  \epsfysize=3.3cm \epsfxsize=5.cm \epsffile{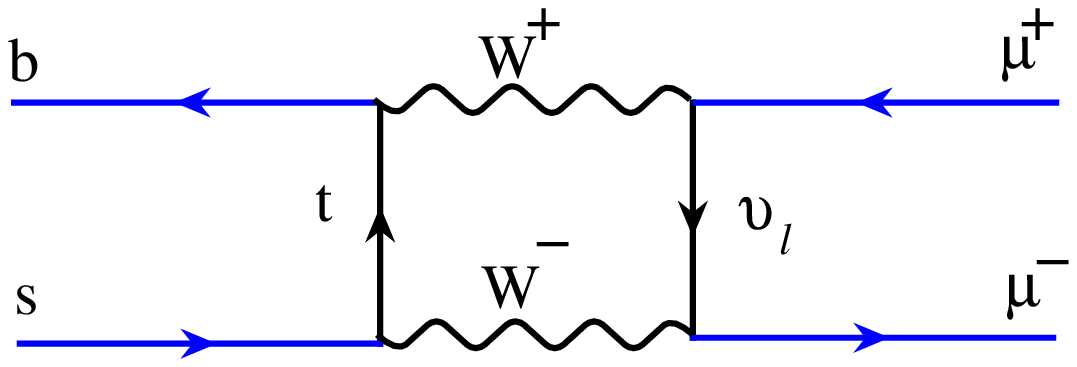}  \vspace{-1.0cm}

    \epsfysize=3.3cm \epsfxsize=5.cm \epsffile{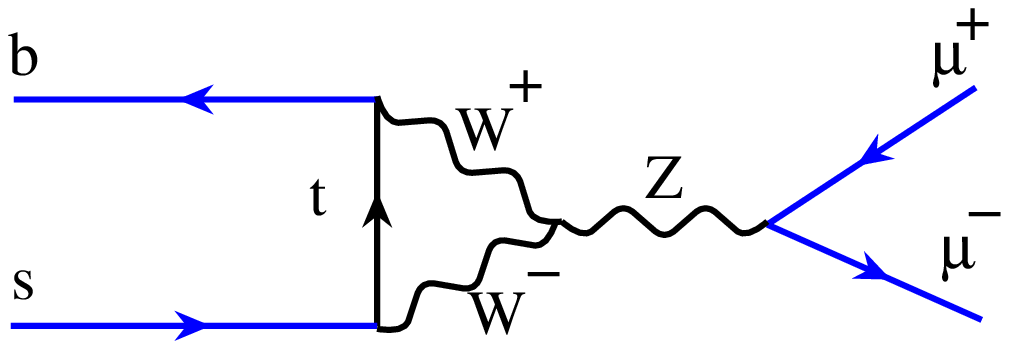}  \vspace{-0.5cm}

    \epsfysize=2.3cm \epsfxsize=4.cm \epsffile{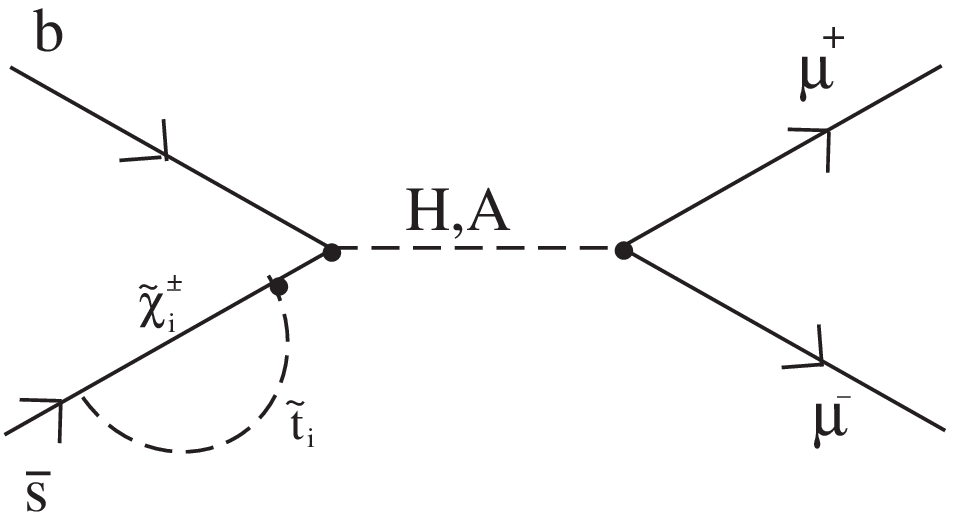}    \end{center}
\vspace{-0.5cm}
\caption{The diagrams contributing to $B_s\to \mu\mu$ decay in the SM and in the MSSM.} \label{bmu}
\end{figure}

\begin{figure}[ht]\vspace{0.3cm}
\begin{center}
 \leavevmode
  \epsfysize=3.5cm    \epsfxsize=3.9cm\epsffile{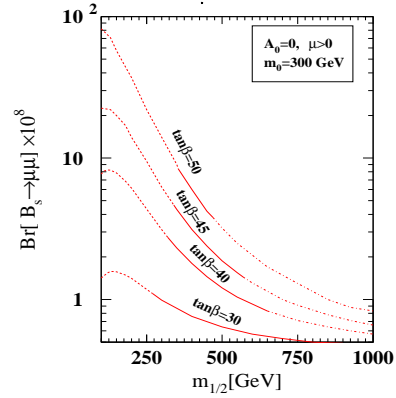}
    \end{center}
\vspace{0.2cm}
\caption{The values of the branching ratio  $B_s\to \mu\mu$ decay  in the MSSM.} \label{br}
\end{figure}

The second decay is $B_s\to\mu^+\mu^-$. In the SM it is given by the diagrams shown in Fig.\ref{bmu}.
The branching ratio is $BR^{SM}(B_s\to\mu^+\mu^-)=3.5\cdot  10^{-9}$, while the recent experiment
gives only the lower bound $BR^{Ex}(B_s\to\mu^+\mu^-)<4.5 \cdot 10^{-8}$\cite{bmu}.  In the MSSM one has several diagrams but the main contribution enhanced by $(\tan\beta)^6$ (!) comes from the one  shown in the bottom of Fig.\ref{bmu}. It is proportional to~\cite{bmususy}
\begin{eqnarray*}
&&BR^{SUSY}(b_s\to\mu\mu) \propto \tan^6\beta  \frac{m_b^2m_t^2 m_\mu^2 \mu^2}{M_W^4m_A^4}\times \\&&
\times \left(\frac{\tilde m^2_{t_1}\log\frac{m^2_{t_1}}{\mu}}{mu^2-m^2_{t_1}}-\frac{\tilde m^2_{t_2}\log\frac{m^2_{t_2}}{\mu}}{mu^2-m^2_{t_2}}\right)^2\end{eqnarray*}
As a result for large $\tan\beta$ one comes in a contradiction with experiment. The values of the branching ratio for various  parameters are shown in Fig.\ref{br} ~\cite{bmususy} and the restrictions on the parameter space in Fig.\ref{constr}

\begin{figure}[ht]\vspace{-0.1cm}
 \leavevmode
 \epsfysize=2.cm  \epsfxsize=2.cm\epsffile{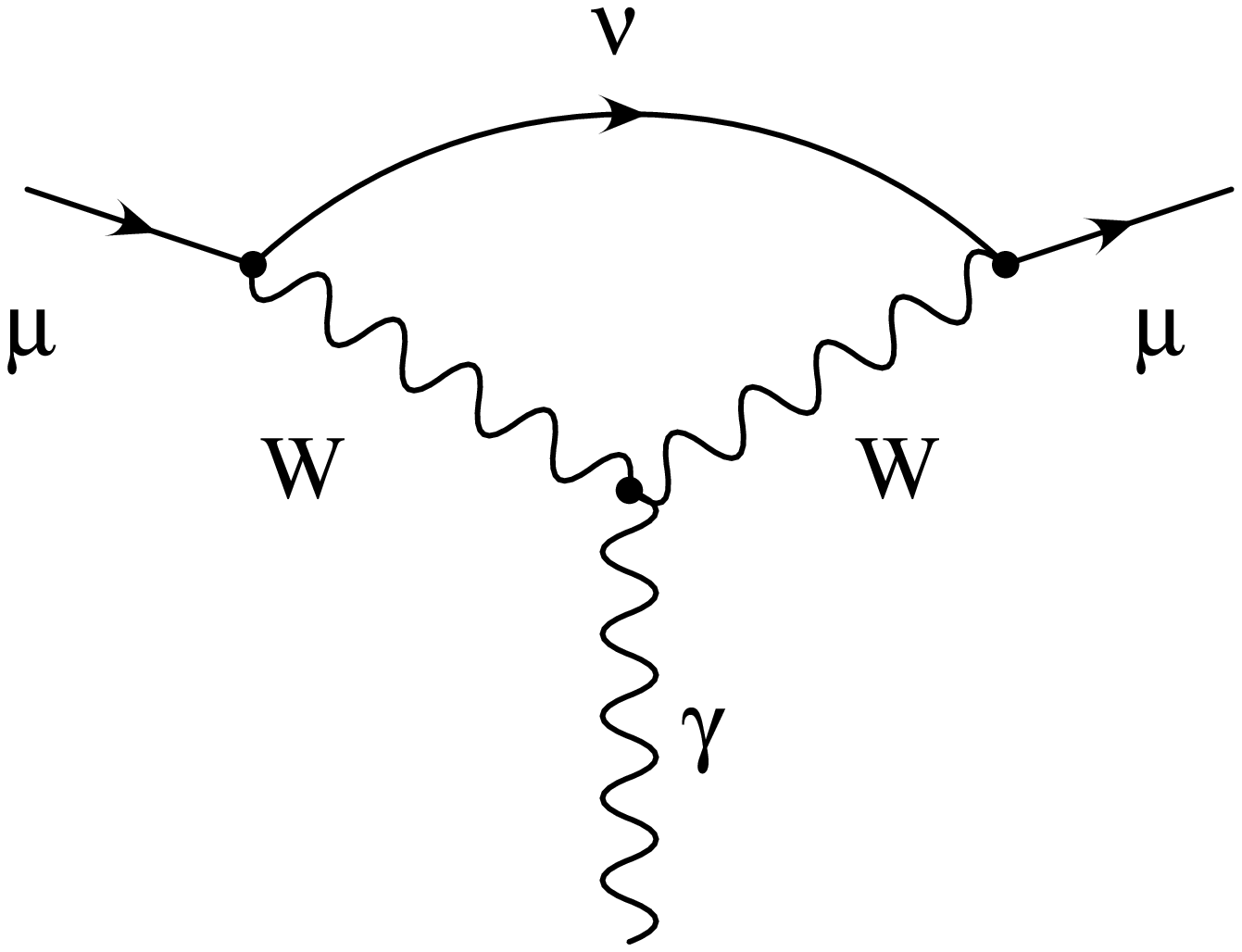}
  \epsfysize=2.cm  \epsffile{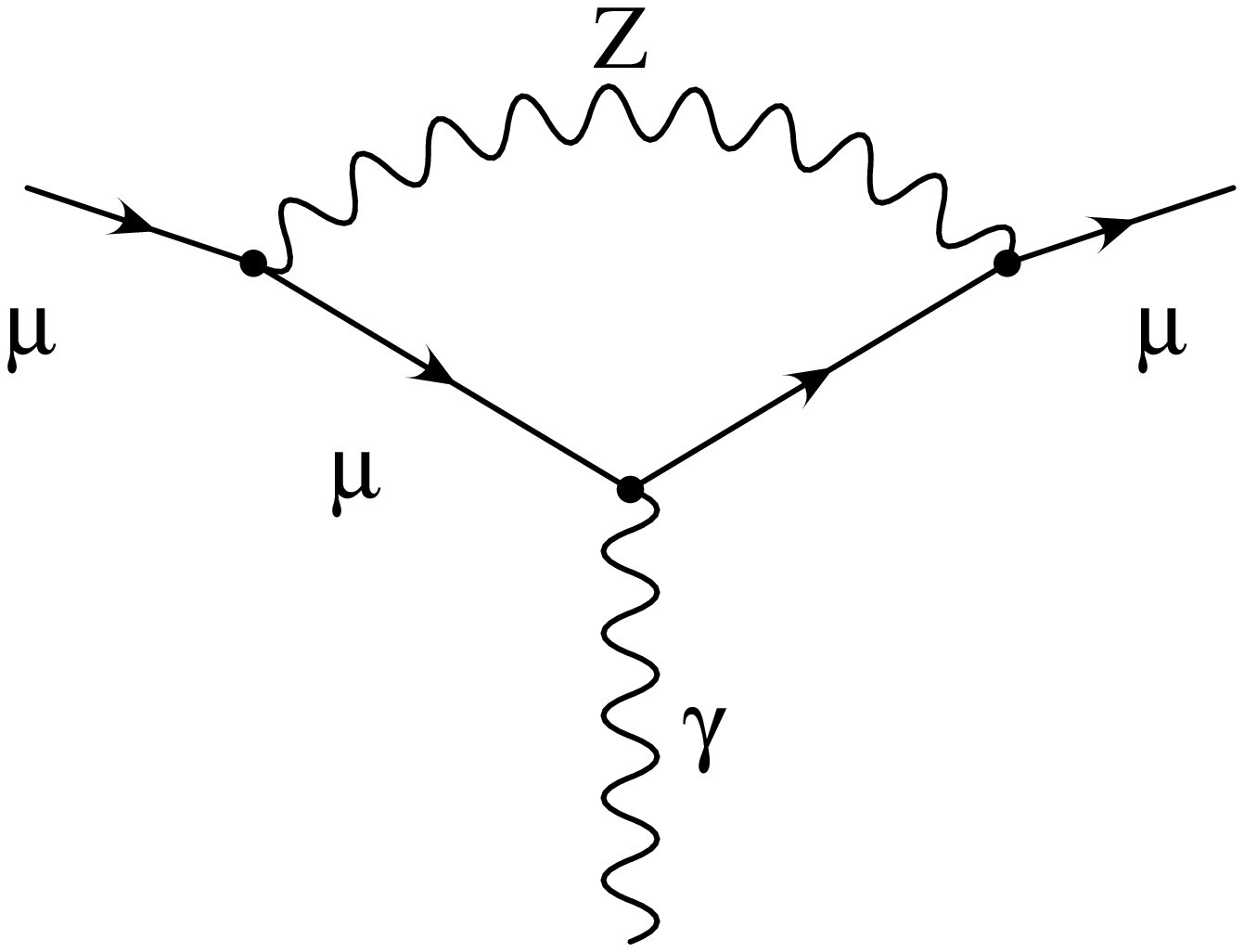}
  \epsfysize=2.cm  \epsffile{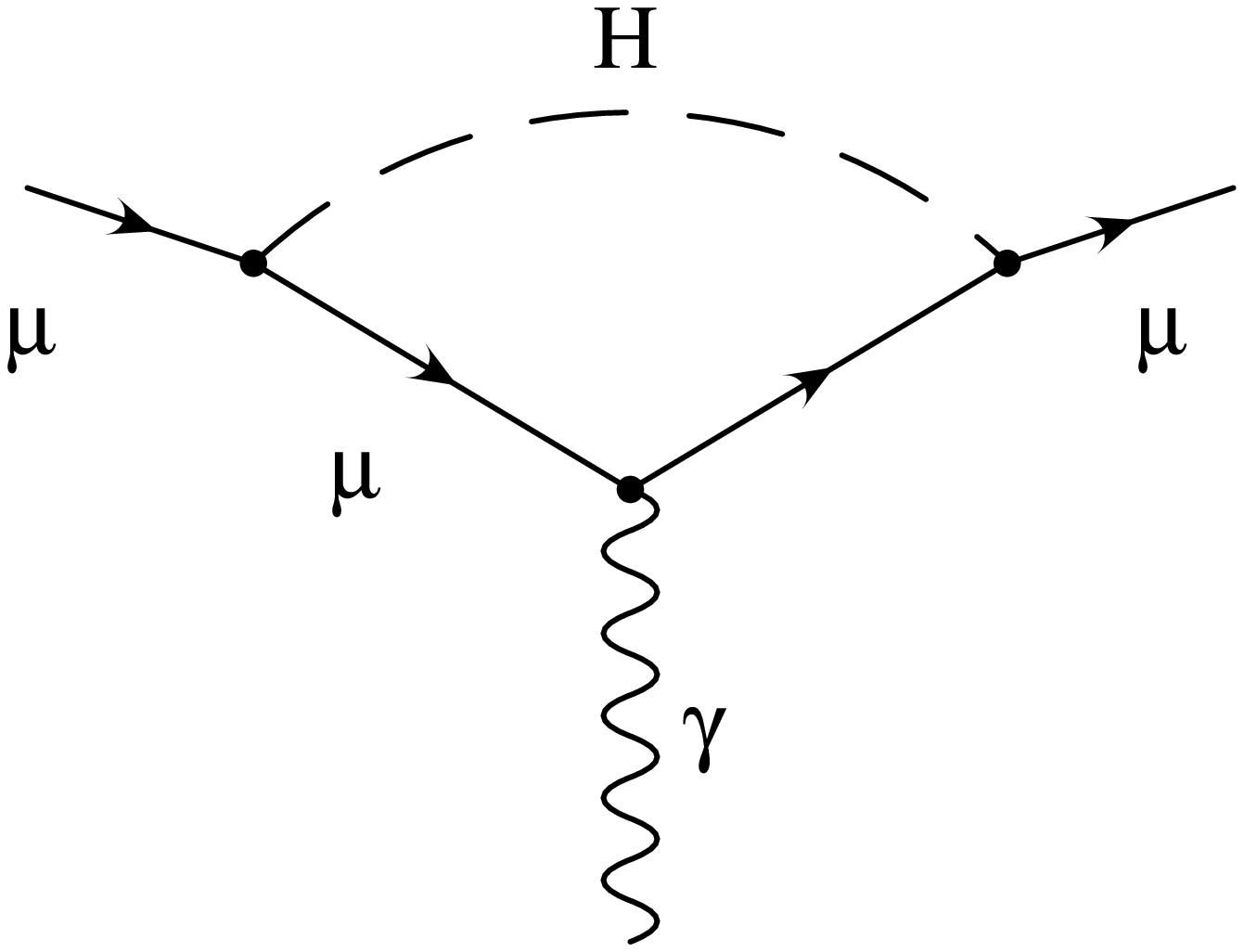}

   \epsfysize=2.3cm\epsffile{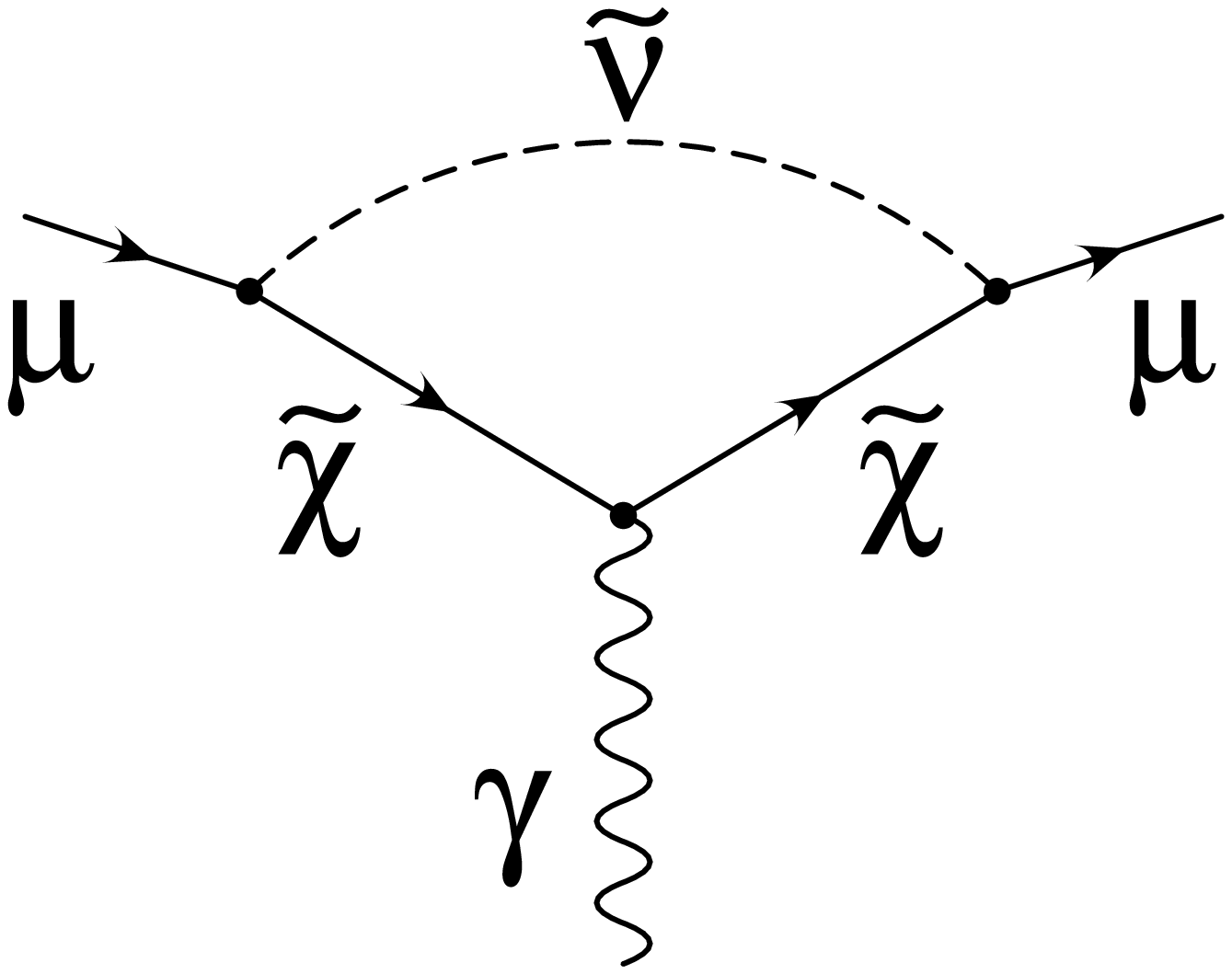}
  \epsfysize=2.3cm \epsffile{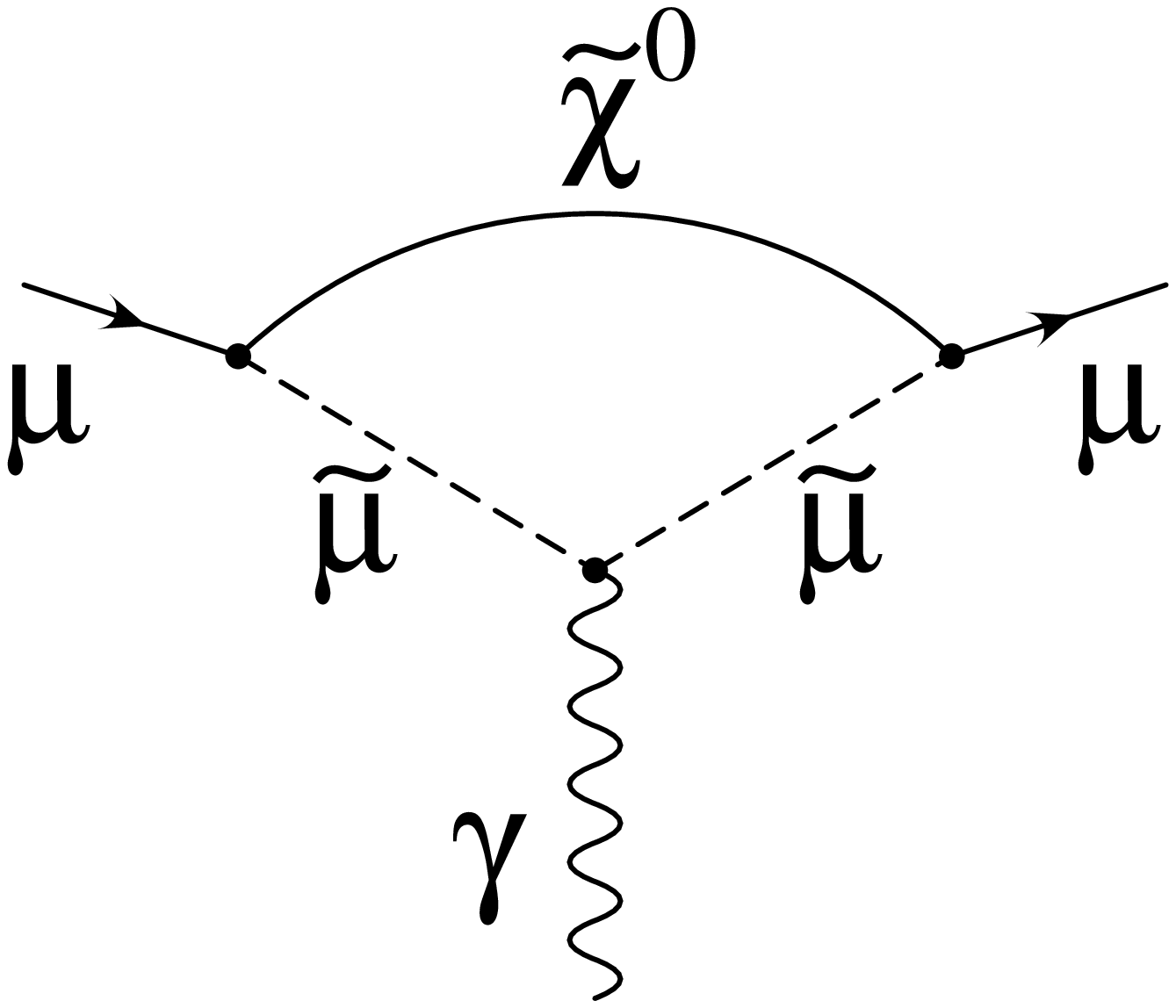}   
\vspace{-0.1cm}
\caption{The diagrams contributing to $a_\mu$ in the SM and in the MSSM.} \label{anom}
\end{figure}

$\bullet$  Anomalous magnetic moment of muon.
Recent measurement of the anomalous magnetic moment
indicates small deviation from the SM of the order of 2.5 $\sigma$\cite{am}:
\begin{eqnarray*}
a_\mu^{exp}&=&11\ 659\ 202\ (14)(6)\cdot  10^{-10} \\
a_\mu^{SM}&=&11\ 659\ 159.6\ (6.7) \cdot 10^{-10} \\
\Delta a_\mu&=&a_\mu^{exp}-a_\mu^{theor}=(27\pm10)\cdot 10^{-10},
\end{eqnarray*}
where the SM contribution comes from
\begin{eqnarray*}
a_\mu^{QED}&=&11\ 658\ 470.56\ (0.29)\cdot 10^{-10} \\
a_\mu^{Weak}&=&15.1\ (0.4) \cdot10^{-10} \\
a_\mu^{hadron}&=&673.9\ (6.7) \cdot 10^{-10},
\end{eqnarray*}
so that the accuracy of the experiment finally reaches  the order of the weak contribution.
The corresponding diagrams are shown in Fig.\ref{anom}.
\begin{figure}[ht]\vspace{-0.4cm}
\begin{center}
 \leavevmode\epsfxsize=5cm
  \epsfysize=5.5cm \epsffile{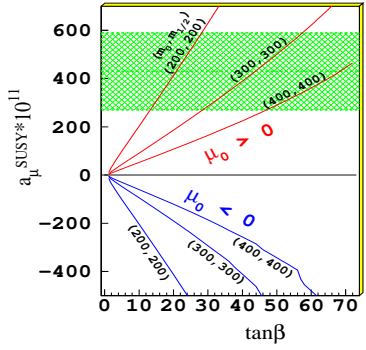}
    \end{center}
\vspace{-1.1cm}
\caption{The dependence of  $a_\mu^{SUSY}$ versus $\tan\beta$  for various
values of the SUSY breaking parameters $m_0$ and $m_{1/2}$.
The horizontal band shows the discrepancy between the experimental data and
the SM estimate.\label{value}}
\end{figure}
The deficiency may be easily filled with SUSY contribution coming from the diagrams shown
 in the bottom of Fig.\ref{anom}. They are similar to  that of the weak interactions after replacing
the vector bosons by charginos and neutralinos.

The total contribution to $a_\mu$ can be approximated by~\cite{CM}
\begin{eqnarray*}  &&\hspace*{-0.7cm}
|a_\mu^{SUSY}| \simeq \!\frac{\alpha(M_Z)}{8\pi \sin^2\theta_W}
\frac{m_\mu^2\tan\beta}{m_{SUSY}^2}\left(\!1\!\!-\!\!\frac{4\alpha}{\pi}
\ln\frac{m_{SUSY}}{m_\mu}\!\right)\\
&& \simeq 140 \cdot 10^{-11}
 \left(\frac{100 \ GeV}{m_{SUSY}}\right)^2 \tan\beta,
\end{eqnarray*}
where $m_\mu$ is the muon mass, $m_{SUSY}$ is an average mass of supersymmetric
particles in the loop (essentially the chargino mass).
It is proportional to $\mu$ and $\tan\beta$ as shown in Fig.\ref{value}. This
requires positive sign of $\mu$ that kills a half of the parameter space of the
MSSM~\cite{Anom}.

If SUSY particles are light enough they  give the desired contribution to  the anomalous magnetic moment.
However, if they are too light the contribution exceeds the gap between the experiment and the SM. For too heavy particles the contribution is too small. This defines the allowed regions as shown in Fig.\ref{constr}.

$\bullet$
The requirement that the lightest supersymmetric particle (LSP) is neutral also restricts the parameter space. This constraint is a consequence of R-parity conservation.
The regions excluded by the  LSP constraint are shown in Fig.\ref{constr}.

Summarizing all together we have the allowed region in parameter space as shown at the last plots
in Fig.\ref{constr} ~\cite{Anom,BSK}. Some requirements are complimentary  being essential for smaller or larger values of $\tan\beta$. One can see that a) all requirements are consistent and b) they still leave a lot of
freedom for the choice of parameters. Analogous analysis has been performed in a number of papers~\cite{regions} with similar results.

\begin{figure}[ht]\vspace{-0.2cm}
 \begin{center}
 \leavevmode
  \epsfysize=5.cm \epsffile{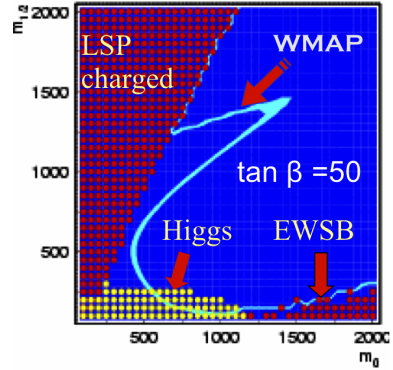}
    \end{center}
\vspace{-0.9cm}
\caption{The light (lbue) band is the region allowed by the WMAP data.
The excluded regions where the LSP is stau (red up left
corner), where the radiative electroweak symmetry breaking mechanism does not work (red
low right corner), and where the Higgs boson is too light (yellow lower left corner) are
shown with dots.
 \label{DMcon}}
\end{figure}

$\bullet$ Astrophysical constraints. One can also impose the constraint that comes from astrophysics.
The accuracy of measurement of the amount of the Dark Matter in the Universe defines with high precision
the cross-section of DM annihilation. This in its turn requires the adjustment of parameters. We consider this problem in more detail in the last section. As a result one finds that this constraint is fulfilled in
a narrow band in $m_0,m_{1/2}$ plane for any fixed value of $\tan\beta$ as shown in Fig.\ref{DMcon}~\cite{BKZ}.
This plot corresponds to $\tan\beta=50$. With decreasing  values of $\tan\beta$ the curve moves to the left and the
funnel disappears.

In the narrow allowed region one fulfills all the constraints simultaneously and has the suitable amount of the dark matter.
Phenomenology essentially depends on the region of parameter space and has direct influence on the strategy of SUSY searches. Each point in this
plane  corresponds to a fixed set of parameters and allows one to
calculate the spectrum, the cross-sections, etc.
\clearpage

\subsection{Experimental signatures at  $e^+e^-$ colliders}
Experiments are finally beginning to push into a significant
region of supersymmetry parameter space. We know the sparticles
and their couplings, but we do not know their masses and mixings.
Given the mass spectrum one can calculate the cross-sections and
consider the possibilities of observing  new particles at modern
accelerators. Otherwise, one can get  restrictions on unknown
parameters.

 We start with $e^+e^-$ colliders. In the leading order creation of superpartners is
given by the diagrams shown in Fig.\ref{creation} above. For a
given center of mass energy the cross-sections depend on the mass
of created particles and vanish at the kinematic  boundary.
Experimental signatures are defined
by the decay modes which vary with the mass spectrum. The main
ones are summarized below.
$$\begin{array}{lll}
\mbox{\underline{Production}}&\mbox{\underline{Decay
Modes}}&~~ \mbox{\underline{Signatures}} \\ && \\ \bullet
~\tilde{l}_{L,R}\tilde{l}_{L,R}~~ &\tilde{l}^\pm_R \to l^\pm
\tilde{\chi}^0_i& \mbox{acompl
pair of}
\\ &  \tilde{l}^\pm_L \to l^\pm \tilde{\chi}^0_i  &
\mbox{charged lept} \!+\! \Big/ \hspace{-0.3cm E_T} \\ \bullet
~\tilde{\nu}\tilde{\nu}& \tilde{\nu}\to l^\pm \tilde{\chi}^0_1
 & \Big/ \hspace{-0.3cm E_T}\\
\bullet  ~\tilde{\chi}^\pm_1\tilde{\chi}^\pm_1 &\tilde{\chi}^\pm_1 \to
 \tilde{\chi}^0_1 l^\pm \nu &
 \mbox{isol lept} \!\!+\!\! 2 \mbox{jets}\!\! +\!\! \Big/ \hspace{-0.3cm E_T} \\
&  \tilde{\chi}^\pm_1 \to \tilde{\chi}^0_2 f \bar f' &\mbox{pair
of acompl}\\
 &  \tilde{\chi}^\pm_1 \to l \tilde{\nu}_l &\mbox{leptons} + \Big/ \hspace{-0.3cm E_T}
\\
& ~~~~~~~  \to l\nu_l\tilde{\chi}^0_1 & \\
& \tilde{\chi}^\pm_1 \to \nu_l \tilde{l}&\mbox{4 jets} + \Big/ \hspace{-0.3cm E_T}
\\ & ~~~~~~~ \to \nu_l
l\tilde{\chi}^0_1 & \\
 \bullet  ~\tilde{\chi}^0_i\tilde{\chi}^0_j &
\tilde{\chi}^0_i \to \tilde{\chi}^0_1 X& X=\nu_l \bar \nu_l \ \mbox{invisible} \\ &&
~~= \gamma,2l,\mbox{2 jets} \\ && 2l \!+\! \Big/ \hspace{-0.3cm E_T},
l\!+\!2j \!+\! \Big/ \hspace{-0.3cm E_T} \\
\bullet ~\tilde{t}_i\tilde{t}_j & \tilde{t}_1 \to c
\tilde{\chi}^0_1 & \mbox{2 jets}\!+\! \Big/ \hspace{-0.3cm E_T} \\ &
\tilde{t}_1 \to b \tilde{\chi}^\pm_1& \mbox{2b jets}\!+\! \mbox{2lept}\!\! +\!\! \Big/
\hspace{-0.3cm E_T} \\
& ~~~~~  \to b f\bar
f'\tilde{\chi}^0_1 & \\
&& \mbox{2 b jets}\!+\!\mbox{lept} \!+\! \Big/
\hspace{-0.3cm E_T}
\\
\bullet  ~\tilde{b}_i\tilde{b}_j & \tilde{b}_i \to b
\tilde{\chi}^0_1 & \mbox{2 b jets}+ \Big/ \hspace{-0.3cm E_T} \\ &
\tilde{b}_i \to b \tilde{\chi}^0_2
& \mbox{2 b jets}\!\!+\!\! \mbox{2 lept}\! \!+\!\! \Big/ \hspace{-0.3cm E_T} \\
&~~~~~~  \to b f\bar f'\tilde{\chi}^0_1 &
 \mbox{2 b jets}\!\!+\!\! \mbox{2 jets}\!\!+\!\! \Big/ \hspace{-0.3cm E_T}
 \end{array} $$
A characteristic feature of all possible signatures is the missing
energy and transverse momenta, which is a trade mark of a new
physics.

 Numerous attempts to find  superpartners at LEP II
gave no positive result thus imposing the lower bounds on their
masses~\cite{LEPSUSY}. Typical LEP II limits on the masses of
superpartners are
\begin{eqnarray*}
\hspace{-0.7cm}&& m_{\chi^0_1} > 40 \ GeV, \ m_{\tilde e}>105\ GeV, \
 m_{\tilde t}> 90\ GeV \\
\hspace{-0.7cm} &&m_{\chi^\pm_1} > 100 \ GeV, \ m_{\tilde \mu}>100\ GeV, \
 m_{\tilde b}> 80\ GeV \\
 \hspace{-0.7cm}&& m_{\tilde \tau}>80\ GeV
\end{eqnarray*}

\subsection{Experimental signatures at hadron colliders}

 Experimental
SUSY signatures at the Tevatron and LHC are similar.
The strategy of SUSY search  is based on an assumption that the masses of
superpartners indeed are in the region of  1~TeV so that they might be created on the
mass shell with the cross section big enough to distinguish them from the background of
the ordinary particles. Calculation of the background in the framework of the Standard
Model thus becomes essential since the secondary particles in all the cases will be the
same.

There are many possibilities to create superpartners at hadron colliders. Besides the
usual annihilation channel there are numerous processes of gluon fusion, quark-antiquark
and quark-gluon scattering. The maximal cross sections of the order of a few picobarn can
be achieved in the process of gluon fusion.

As a rule all superpartners are short lived and decay into the ordinary particles and the
lightest superparticle.  The main decay modes of superpartners which serve as  the
manifestation of SUSY  are
 $$\begin{array}{lll}
\mbox{\underline{Production}}&\mbox{\underline{Decay
Modes}}&~~ \mbox{\underline{Signatures}} \\ && \\ \bullet
~\tilde{g}\tilde{g}, \tilde{q}\tilde{q},
\tilde{g}\tilde{q}
&\begin{array}{l} \tilde{g}
\to q\bar q \tilde{\chi}^0_1   \\
 ~~~~~ q\bar q' \tilde{\chi}^\pm_1  \\
 ~~~~~ g\tilde{\chi}^0_1 \end{array}
  &
\begin{array}{c} \Big/ \hspace{-0.3cm E_T} + \mbox{multijets}\\
 (+\mbox{leptons}) \end{array} \\
& \begin{array}{l}\tilde{q} \to q \tilde{\chi}^0_i \\
    \tilde{q} \to q' \tilde{\chi}^\pm_i \end{array}
  &\\
 \bullet
~\tilde{\chi}^\pm_1\tilde{\chi}^0_2 &\tilde{\chi}^\pm_1 \to
 \tilde{\chi}^0_1 l^\pm \nu &
 \mbox{Trilepton} + \Big/ \hspace{-0.3cm E_T} \\
 & ~~~~\ \tilde{\chi}^0_2 \to
 \tilde{\chi}^0_1 ll & 
 \end{array}$$
$$\begin{array}{lll}
 &  \tilde{\chi}^\pm_1 \to \tilde{\chi}^0_1 q \bar q'&\mbox{Dilept}\! \!+\!\!\mbox{ jet} \!\!+\!\! \Big/ \hspace{-0.3cm E_T}\\
 & ~~~~ \tilde{\chi}^0_2 \to
\tilde{\chi}^0_1 ll &\\
 \bullet  ~\tilde{\chi}^+_1\tilde{\chi}^-_1 &
\tilde{\chi}^+_1 \to l \tilde{\chi}^0_1 l^\pm \nu &
\mbox{Dilepton} + \Big/ \hspace{-0.3cm E_T} \\ \bullet
~\tilde{\chi}^0_i\tilde{\chi}^0_i & \tilde{\chi}^0_i \to
\tilde{\chi}^0_1 X &
\Big/ \hspace{-0.3cm E_T} \!\!+\!\! \mbox{Dilept+jets}\\
& ~~~~  \tilde{\chi}^0_i  \to \tilde{\chi}^0_1 X'  & \\
\bullet  ~\tilde{t}_1\tilde{t}_1 & \tilde{t}_1 \to c
\tilde{\chi}^0_1 & \mbox{2 acollin jets}\!\!+\!\! \Big/ \hspace{-0.3cm
E_T} \\ & \tilde{t}_1 \to b \tilde{\chi}^\pm_1 &
 \mbox{sing lept}\!\! +\!\! \Big/ \hspace{-0.3cm E_T} \!\! +\!\! b's\\
 & ~~~~  \tilde{\chi}^\pm_1\to \tilde{\chi}^0_1 q\bar q'  & \\
 &\tilde{t}_1 \to b \tilde{\chi}^\pm_1
& \mbox{Dilept} \!\!+\!\! \Big/ \hspace{-0.3cm E_T}\!\! +\!\! b's\\
& ~~~ \tilde{\chi}^\pm_1 \to \tilde{\chi}^0_1
l^\pm \nu &\\
\bullet
~\tilde{l}\tilde{l},\tilde{l}\tilde{\nu},\tilde{\nu}\tilde{\nu}
&
 \tilde{l}^\pm \to l\pm \tilde{\chi}^0_i& \mbox{Dilepton}+ \Big/ \hspace{-0.3cm E_T} \\
 & ~~ \tilde{l}^\pm \to \nu_l \tilde{\chi}^\pm_i  &  \mbox{Single lept} +
/ \hspace{-0.25cm E_T} \\
& \tilde{\nu} \to \nu \tilde{\chi}^0_1& / \hspace{-0.25cm E_T}
 \end{array}$$

Noe again the typical events with missing energy and transverse momentum that is the main
difference from the background processes of the Standard Model. Contrary to  $e^+e^-$ colliders, at hadron
machines the background is extremely rich and essential. The missing energy is
carried away by the heavy particle with the mass of the order of 100~GeV that is
essentially different from the processes with neutrino in the final state.
\begin{table}[htb]
\begin{tabular}{|c|p{1.7cm}|}
\hline Process & final states \\
 \hline
\includegraphics[width=50mm,height=35mm]{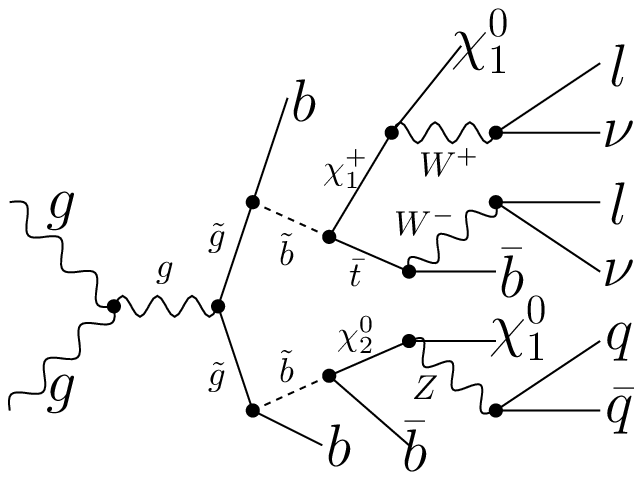}
& \vspace*{-35mm}
\begin{minipage}[t]{1.8cm}
$\begin{array}{c}
2\ell \\ 2\nu \\ 6j \\ \Big/\hspace{-0.3cm E_T}
\end{array}$
\end{minipage} \\ \hline
\includegraphics[width=50mm,height=35mm]{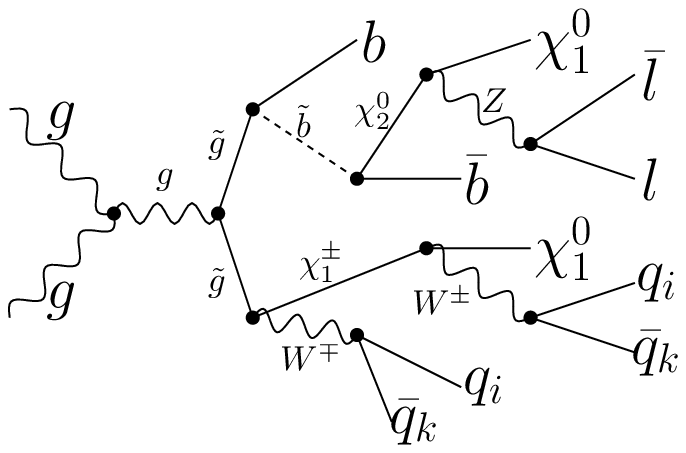}
& \vspace*{-30mm}
\begin{minipage}[t]{1.8cm}
$\begin{array}{c}
2\ell \\ 6j \\ \Big/ \hspace{-0.3cm E_T}
\end{array}$
\end{minipage} \\ \hline
\includegraphics[width=50mm,height=35mm]{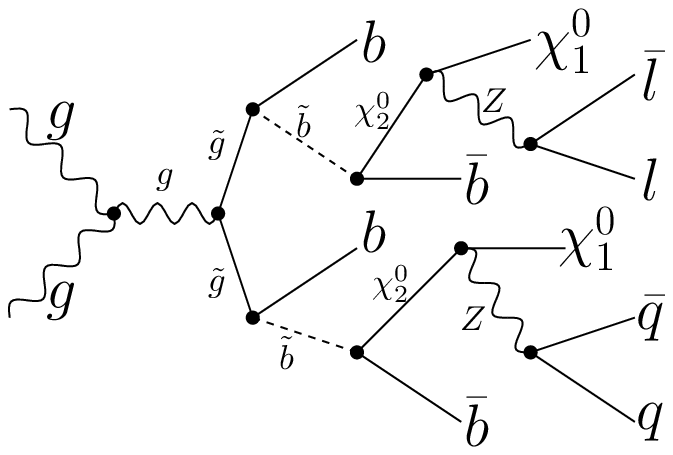}
& \vspace*{-30mm}
\begin{minipage}[t]{1.8cm}
$\begin{array}{c}
2\ell \\ 6j \\ \Big/ \hspace{-0.3cm E_T}
\end{array}$
\end{minipage} \\ \hline
\includegraphics[width=50mm,height=35mm]{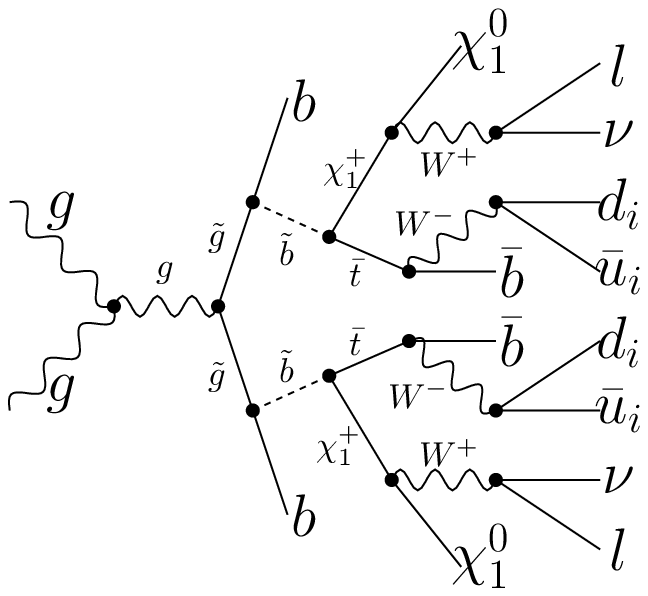}
& \vspace*{-35mm}
\begin{minipage}[t]{1.8cm}
$\begin{array}{c}
2\ell \\ 2\nu \\ 8j \\ \Big/\hspace{-0.3cm E_T}
\end{array}$
\end{minipage} \\ \hline
\includegraphics[width=50mm,height=35mm]{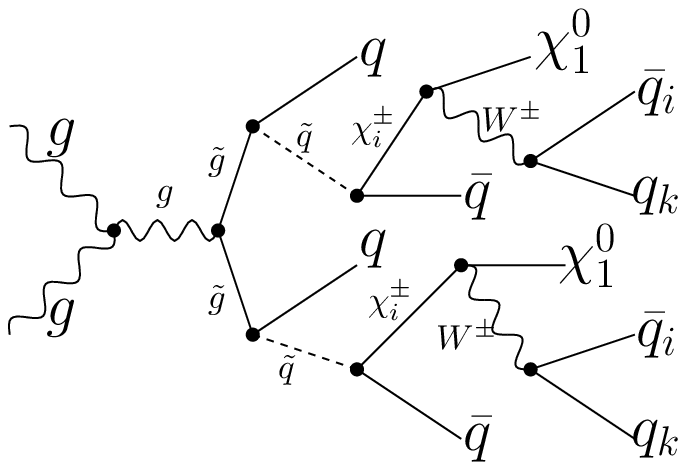}
& \vspace*{-25mm}
\begin{minipage}[t]{1.8cm}
$\begin{array}{c}
8j \\ \Big/ \hspace{-0.3cm E_T}
\end{array}$
\end{minipage} \\ \hline
\end{tabular}
\bigskip
\caption{Creation of the pair of gluino with further cascade decay} \label{tab:n}
\end{table}
In hadron collisions the superpartners are always created in pairs and then further
quickly decay creating a cascade with the ordinary quarks (i.e. hadron jets) or leptons
at the final state plus the missing energy. For the case of gluon fusion with creation of gluino
it is presented in Table~\ref{tab:n}.

Chargino and neutralino can also be produced in pairs through the Drell-Yang mechanism
$pp \to \tilde \chi^\pm_1 \tilde \chi^0_2$ and can be detected via their lepton decays
$\tilde \chi^\pm_1 \tilde \chi^0_2 \to \ell\ell\ell+\Big/ \hspace{-0.3cm E_T}$. Hence the
main signal of their creation is the isolated leptons and missing energy
(Table~\ref{tab:j}). The main background in trilepton channel comes from creation of the
standard particles  $WZ/ZZ,t\bar t, Zb\bar b$ è $b\bar b$. There might be also the
supersymmetric background from the cascade decays of squarks and gluino into multilepton
modes.

\begin{table}[htb]
\begin{tabular}{|c|p{1.7cm}|} \hline
Process & final  states \\
\hline
\includegraphics[width=50mm,height=30mm]{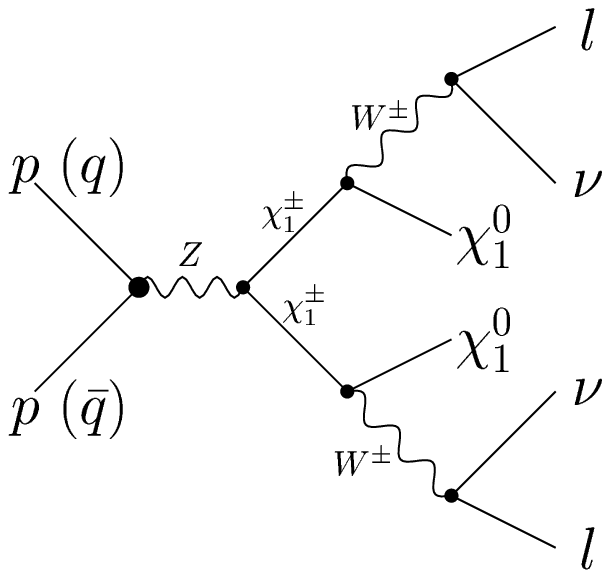}
& \vspace*{-30mm}
\begin{minipage}[t]{1.8cm}
$\begin{array}{c}
2\ell \\ 2\nu \\ \Big/\hspace{-0.3cm E_T}
\end{array}$
\end{minipage} \\ \hline
\includegraphics[width=50mm,height=30mm]{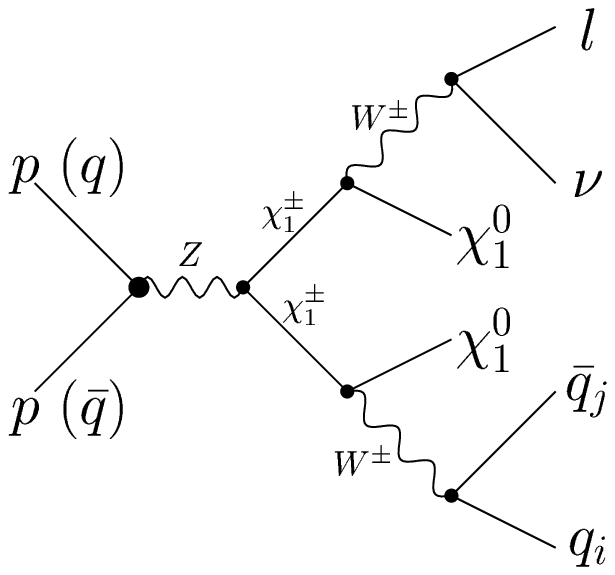}
& \vspace*{-35mm}
\begin{minipage}[t]{1.8cm}
$\begin{array}{c}
\ell \\ \nu \\ 2j \\ \Big/\hspace{-0.3cm E_T}
\end{array}$
\end{minipage} \\ \hline
\includegraphics[width=50mm,height=30mm]{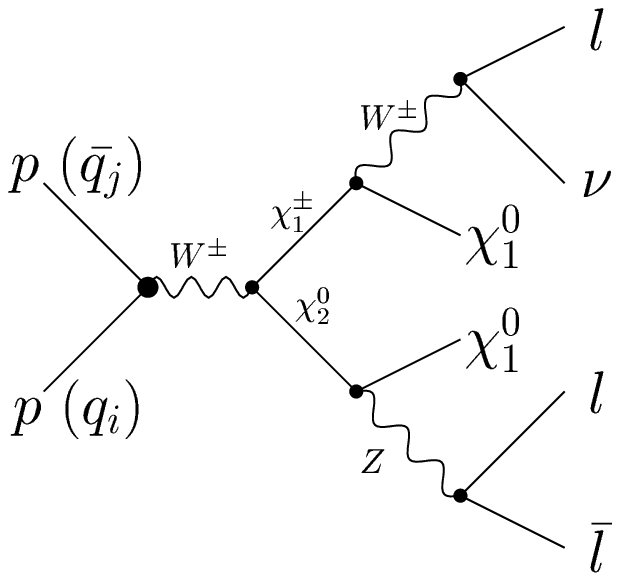}
& \vspace*{-30mm}
\begin{minipage}[t]{1.8cm}
$\begin{array}{c}
3\ell \\ \nu \\ \Big/\hspace{-0.3cm E_T}
\end{array}$
\end{minipage} \\ \hline
\includegraphics[width=50mm,height=35mm]{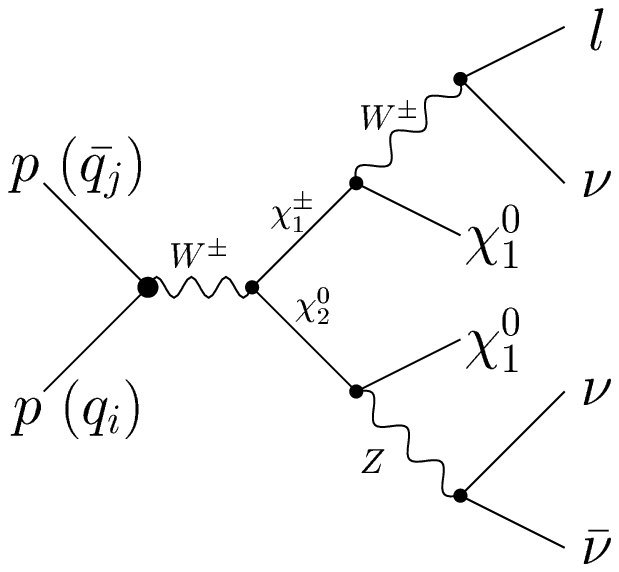}
& \vspace*{-30mm}
\begin{minipage}[t]{1.8cm}
$\begin{array}{c}
\ell \\ 3\nu \\ \Big/\hspace{-0.3cm E_T}
\end{array}$
\end{minipage} \\ \hline
\includegraphics[width=50mm,height=35mm]{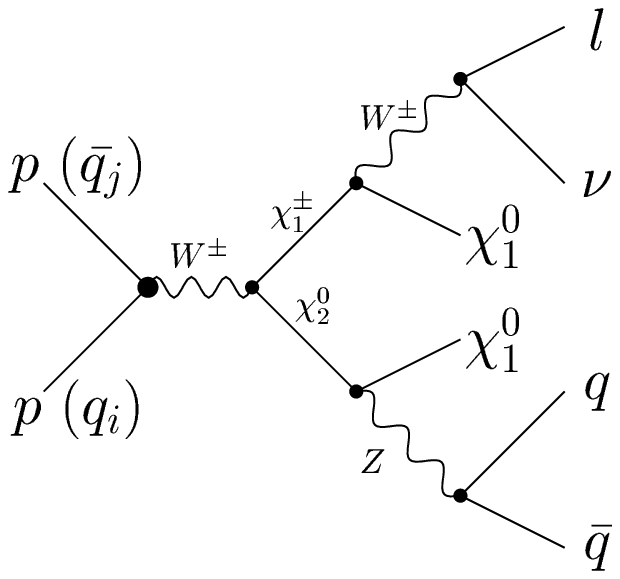}
& \vspace*{-35mm}
\begin{minipage}[t]{1.8cm}
$\begin{array}{c}
\ell \\ \nu \\ 2j \\ \Big/\hspace{-0.3cm E_T}
\end{array}$
\end{minipage} \\ \hline
\end{tabular}\bigskip
\caption{Creation of the lightest chargino and the second neutralino with further cascade
decay.} \label{tab:j}\end{table}
%

\begin{table}[htb]
\begin{tabular}{|c|p{1.7cm}|} \hline
Process & final  states \\
\hline
\includegraphics[width=50mm,height=30mm]{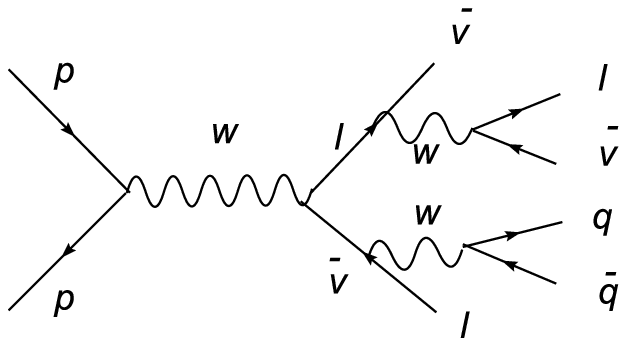}
& \vspace*{-30mm}
\begin{minipage}[t]{1.8cm}
$\begin{array}{c} \\
2\ell \\ 2 j \\ \Big/\hspace{-0.3cm E_T}
\end{array}$
\end{minipage} \\ \hline
\includegraphics[width=50mm,height=30mm]{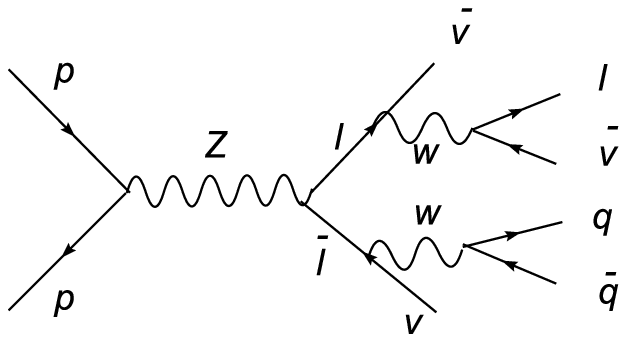}
& \vspace*{-35mm}
\begin{minipage}[t]{1.8cm}
$\begin{array}{c} \\ \\
\ell  \\ 2j \\ \Big/\hspace{-0.3cm E_T}
\end{array}$
\end{minipage} \\ \hline
\end{tabular}\bigskip
\caption{The background processes at hadron colliders (weak interactions).} \label{tab:b1}\end{table}
%

\begin{table}[htb]
\begin{tabular}{|c|p{1.7cm}|} \hline
Process & final  states \\
\hline
\includegraphics[width=50mm,height=35mm]{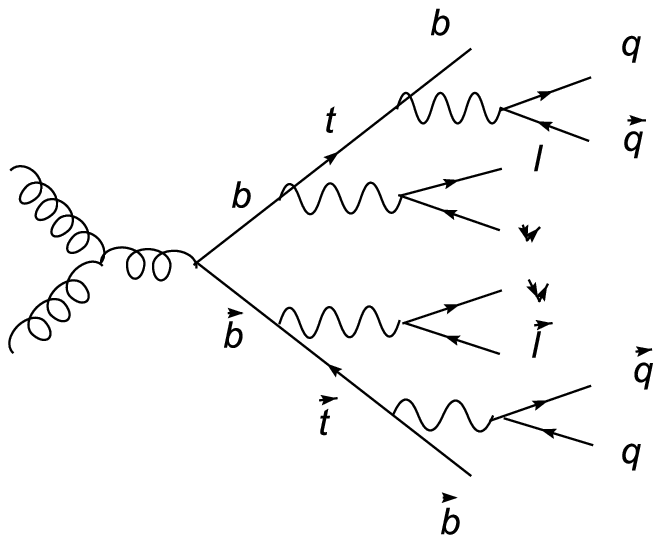}
& \vspace*{-30mm}
\begin{minipage}[t]{1.8cm}
$\begin{array}{c}\\
2\ell \\ 6j \\ \Big/\hspace{-0.3cm E_T}
\end{array}$
\end{minipage} \\ \hline
\includegraphics[width=50mm,height=35mm]{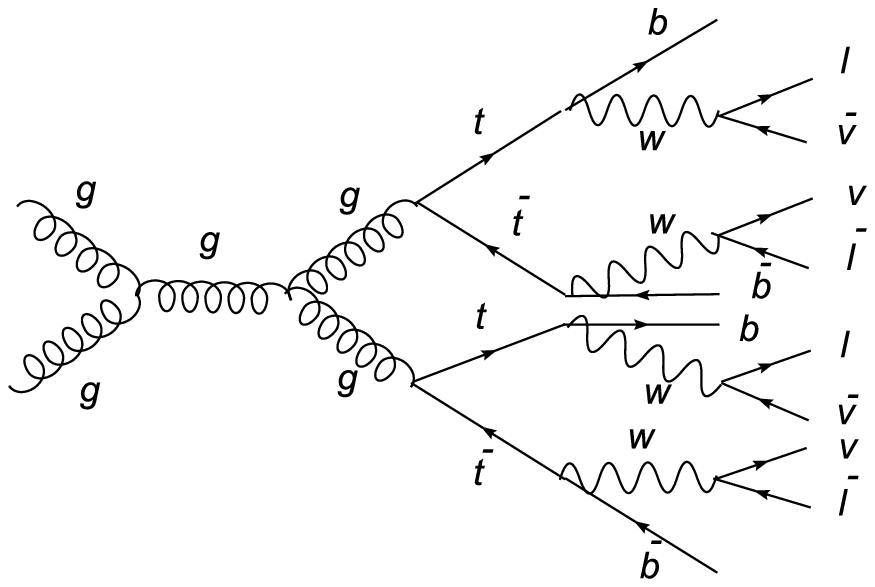}
& \vspace*{-35mm}
\begin{minipage}[t]{1.8cm}
$\begin{array}{c}\\
4\ell \\  4j \\ \Big/\hspace{-0.3cm E_T}
\end{array}$
\end{minipage} \\ \hline
\end{tabular}\bigskip
\caption{The background processes at hadron colliders (strong interactions).} \label{tab:b2}
\end{table}
Numerous  SUSY searches have been  already performed at the
Tevatron.  Pair-produced squarks
and gluinos have at least two large-$E_T$ jets associated with
large missing energy.  The final state with lepton(s) is possible
due to leptonic decays of the $\tilde \chi^\pm_1$ and/or $\tilde
\chi^0_2$.

 In the trilepton channel the Tevatron  is sensitive up to
$m_{1/2} \leq 250$ GeV if $m_0\ \leq 200$ GeV  and  up to $m_{1/2}
\leq 200$ GeV if $m_0\ \geq 500$ GeV. The existing  limits on the squark and gluino masses at the Tevatron are~\cite{tevatron} :\
$m_{\tilde q} \geq 300 \ GeV, \ \ m_{\tilde g} \geq 195\ GeV.$

\begin{figure}[htb]
\begin{center}
\includegraphics[width=0.40\textwidth]{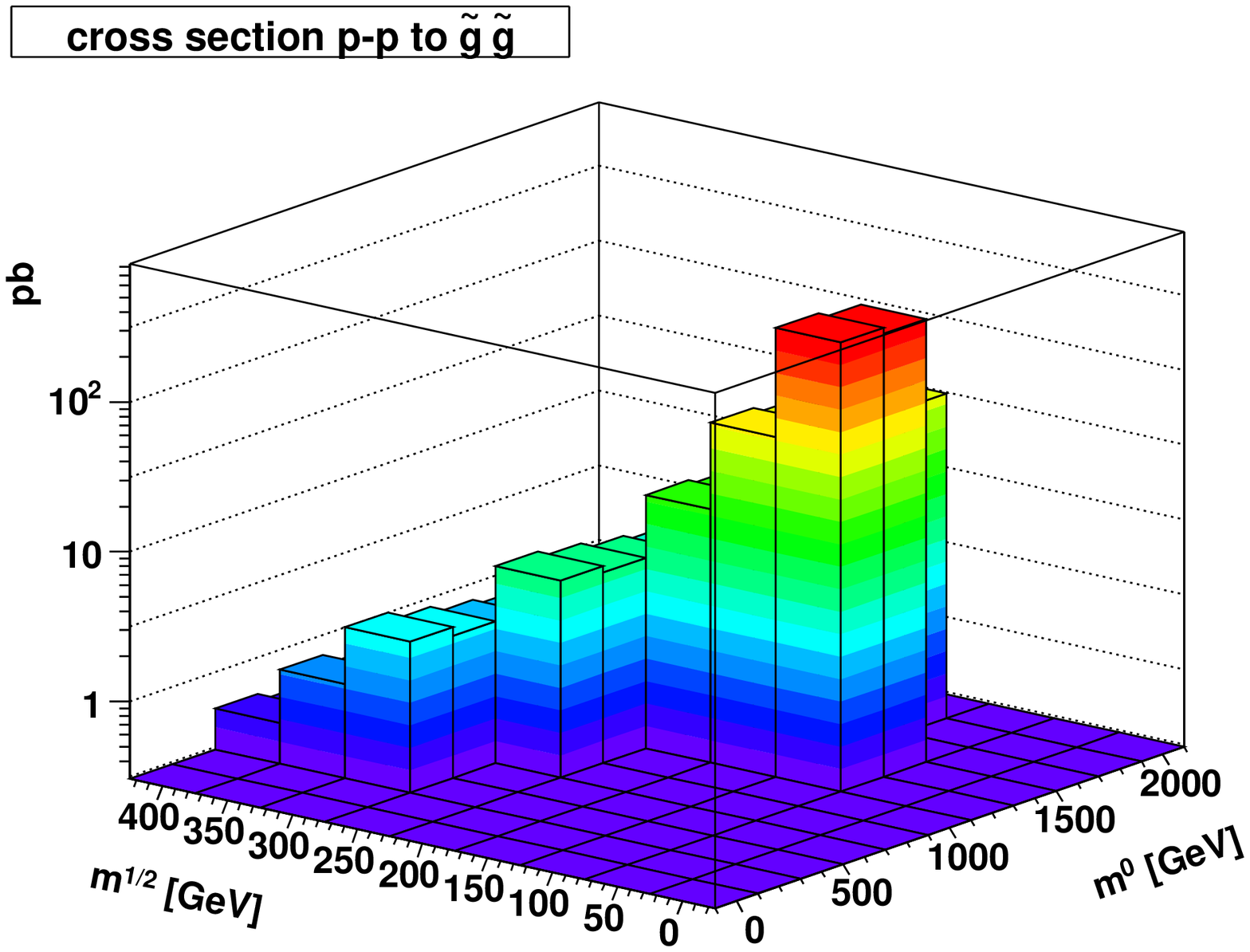}
\includegraphics[width=0.40\textwidth]{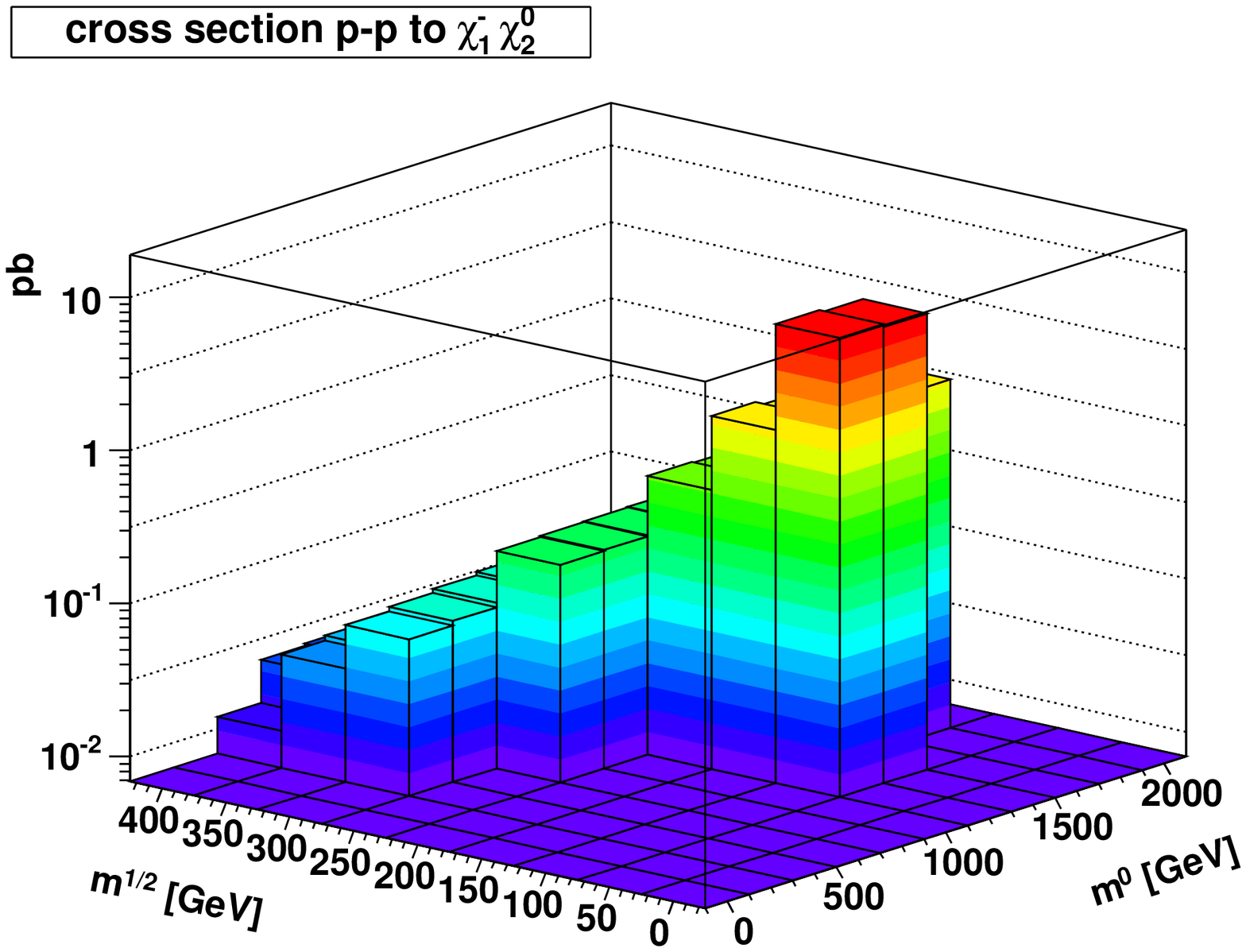}
\includegraphics[width=0.40\textwidth]{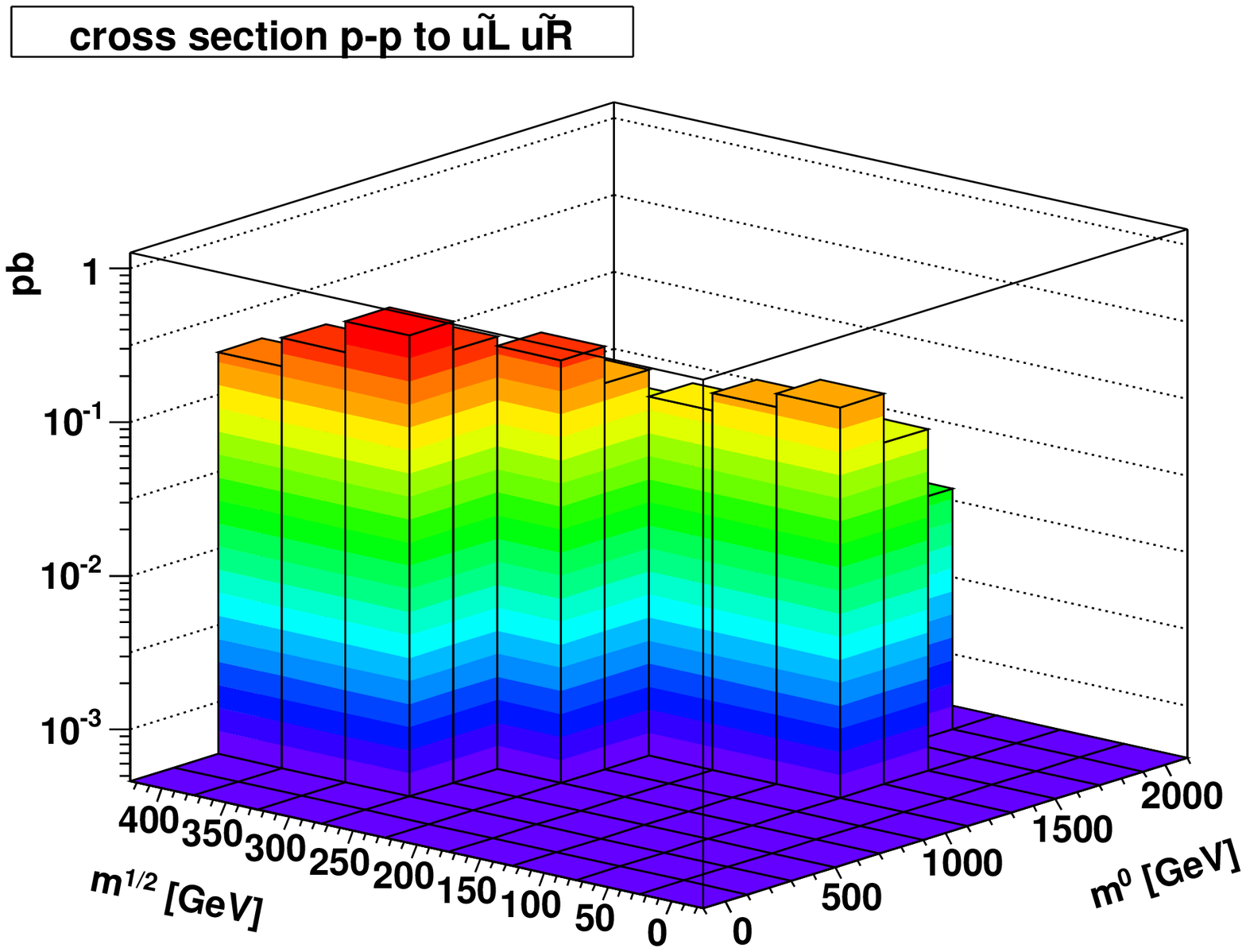}
\includegraphics[width=0.40\textwidth]{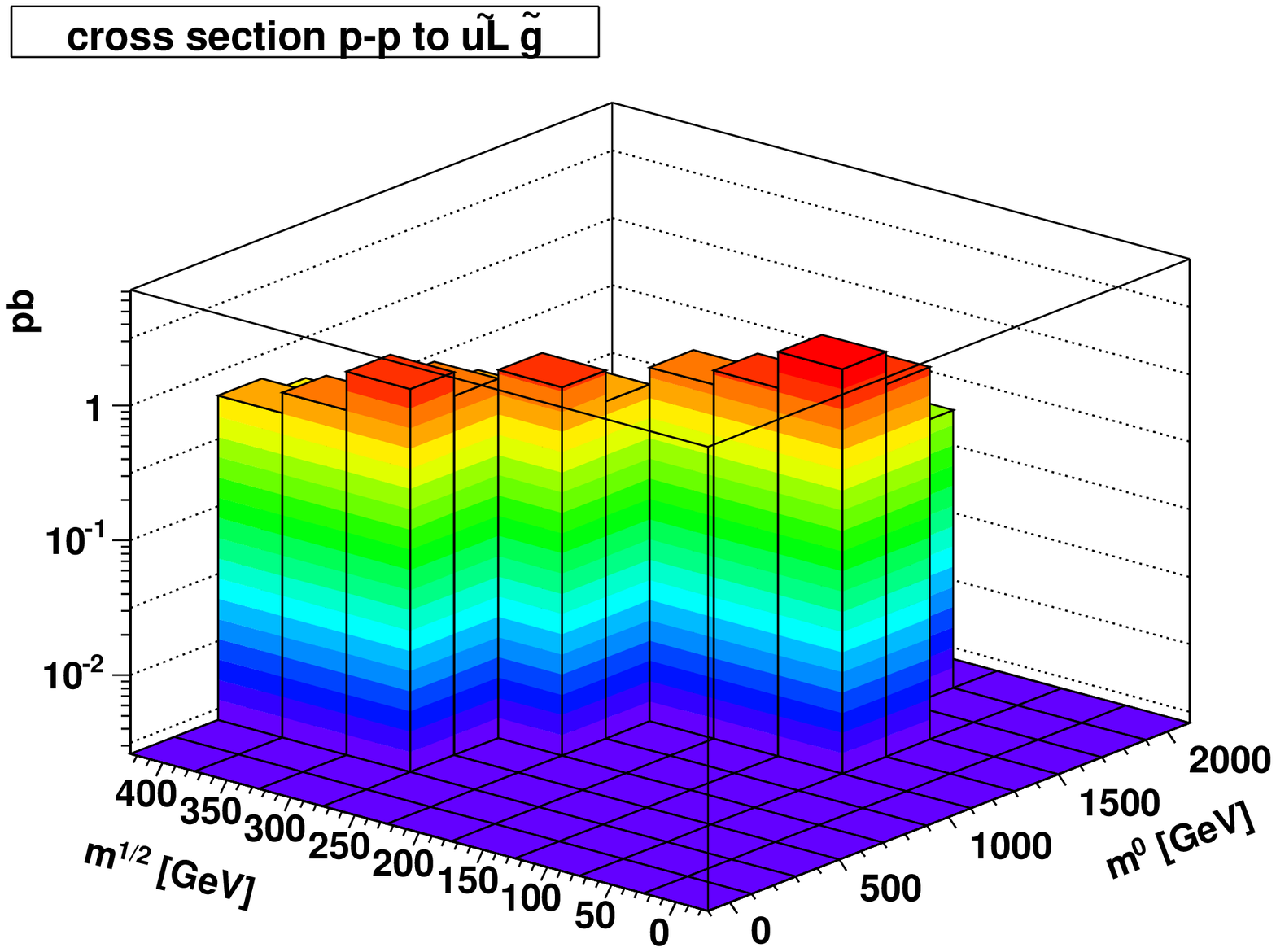}
\end{center}\vspace{-1.5cm}
\caption{The cross sections of superpartners creation as functions of $m_{1/2}$ and
$m_{0}$ for $\tan\beta=51$, $A_0=0$ and positive sign of $\mu$.} \label{fig:Mss}
\end{figure}

The LHC has an advantage of higher energy and bigger luminosity. The cross sections for various superpartner production at the LHC in  $m_0,m_{1/2}$ plane are shown in
Fig.~\ref{fig:Mss}. One can see that  the biggest cross-section reaching 100 pb in the maximum is achieved for gluino production. And though it strongly depends on the gluon mass, with a planned luminosity of LHC one may have a  reliable detection. It should be mentioned, however, that  being produced in collisions the superpartners follow the cascade decay chain and the cross section at the final stage  is essentially smaller being multiplied by the  branching ratios  of the corresponding processes. The resulting cross-sections for particular final states are in the fb region. They are  higher for   hadron final states where one has jets with missing energy and lower for  lepton ones which  are  cleaner for detection.  The cross-sections for chargino production are almost one order of magnitude lower reaching 10 pb in the maximum and those for squark production are  below 1 pb. In some regions of parameter space with
light neutralino and chargino  the production cross-sections   can reach
those of the strongly interacting particles~\cite{unexpLHC}.

\begin{figure}[htb]
\begin{center}
\includegraphics[width=5.cm]{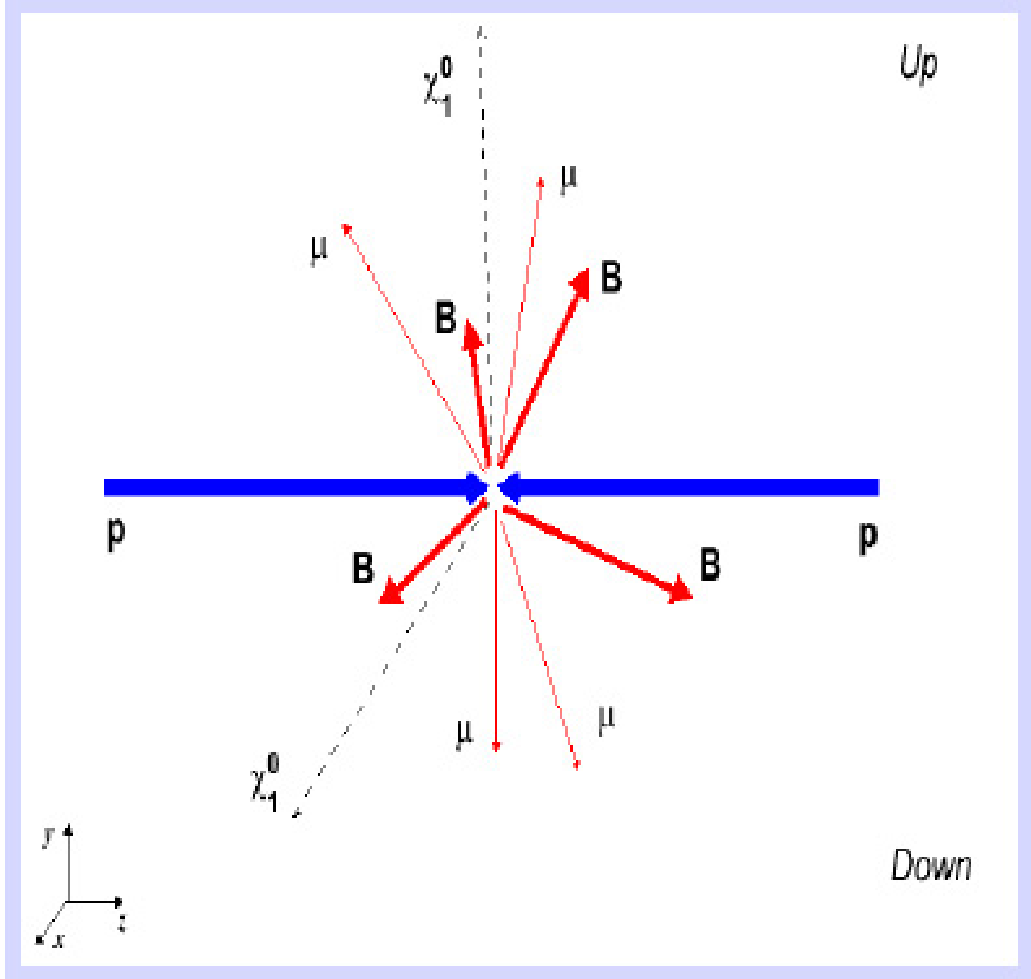}
\vspace{0.4cm}

\includegraphics[width=6.cm]{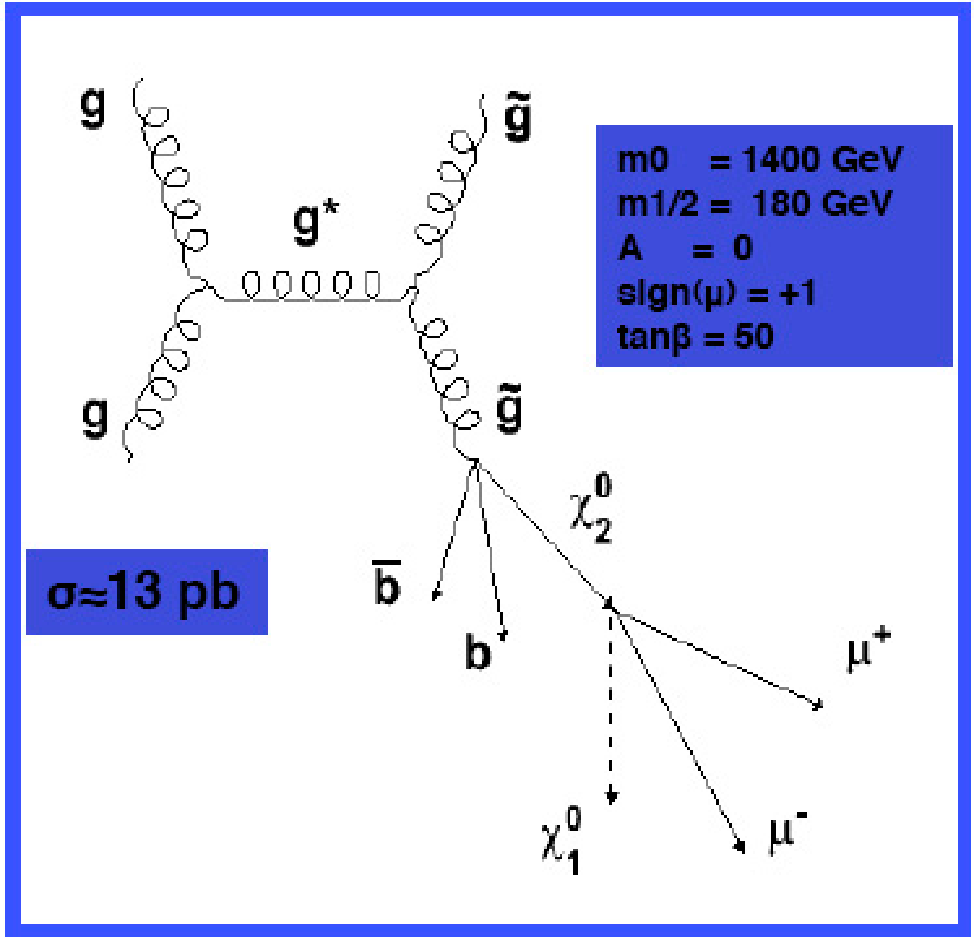}
\end{center}\vspace{-0.7cm}
\caption{Gluino production at the LHC accompanied with cascade decays} \label{gluino}
\end{figure}

In most of the cases the superpartners are very short lived and decay practically at the collision
point without leaving a secondary vertex.  One then has hadron jets (mostly b-jets) and leptons
flying  outside. The typical process of gluino production is presented in Fig.\ref{gluino} where the cascade decay of one of the gluinos is shown~\cite{LNP}. For a given choice of soft SUSY breaking parameters  the cross-section at the first stage reaches 13 pb  but with the 4-lepton + 4-jet final state is reduced to a few fb.
To distinguish this reaction from the background one has to perform the analysis of the missing energy and consider the peculiarities in the invariant mass distribution of the muon pair,  free pass of B-mesons, etc~\cite{LNP}.

\clearpage

\subsection{The long-lived superpartners}

Within the framework of  Constrained MSSM with gravity mediated soft supersymmetry breaking mechanism there exists an interesting possibility to get long-lived next-to-lightest supersymmetric particles (staus or stops or charginos). Their production cross-sections crucially depend on a single parameter, the mass of the superparticle, and for light staus can reach a few \% of pb.
This might be within the LHC reach. The stop production cross-section can achieve even hundreds pb.  Decays of long-lived superpartners would have an unusual signature if heavy charged particles decay with a considerable delay producing secondary vertices inside the detector, or even escape the detector. This possibility can be realized  at the boundary regions of allowed parameter space where one can have a mass degeneracy between the LSP and the NLSP. The life time of the NLSP is inversely proportional  to the mass degeneracy.
\begin{figure}[htb]
\begin{center}
\includegraphics[width=0.35\textwidth]{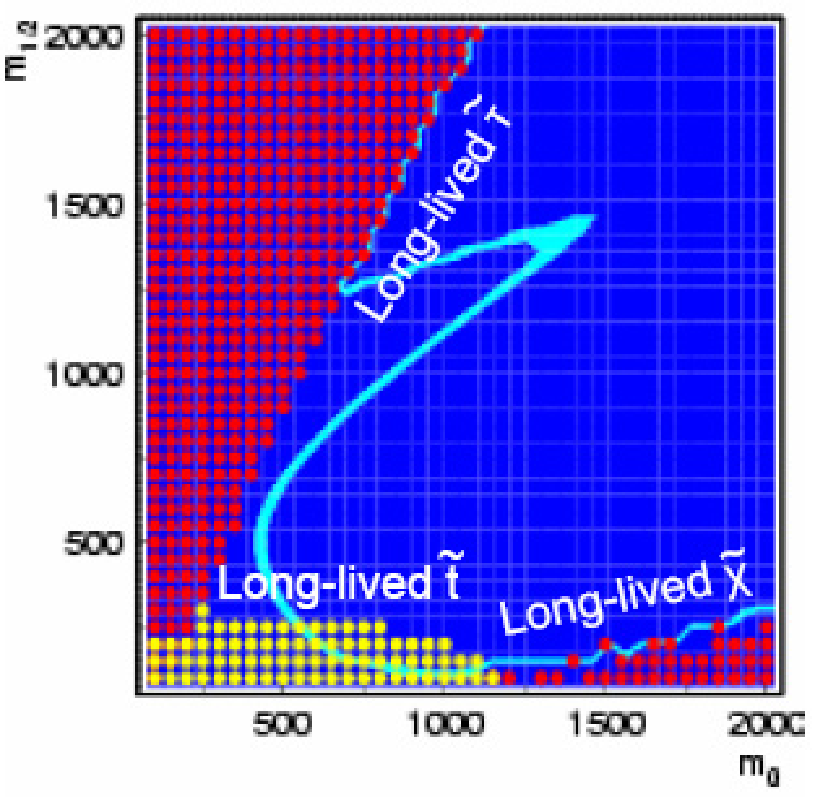}
\end{center}\vspace{-1.3cm}
\caption{The regions of the parameters space of mSUGRA where the long-lived  sparticles might exist \label{long}}
\end{figure}

We show in Fig.\ref{long} the regions where this might happen. The first region is the so-called
co-annihilation region which is interesting from the point of view of existence of long-lived
charged sleptons. Their life-time may be large enough to be produced in proton-proton
collisions and to fly away from the detector area or to decay inside the detector at a
considerable distance from the collision point. Clearly that such an event can not be
unnoticed. However, to realize this possibility one need a fine-tuning of  parameters
of the model~\cite{manuel}.
The second region is the border of the bulk region where light long-lived stops can exist. It appears only for large negative trilinear soft supersymmetry breaking parameter $A_0$. On the border of this region, in full analogy with the stau co-annihilation region, the top squark becomes the LSP and near this border one might get the long-lived stops. Stops can  form the so-called R-hadrons (bound states of SUSY particles) if their lifetime is bigger than the hadronisation time.

The last interesting  region of parameter space  is a narrow band along the line where the radiative electroweak symmetry breaking fails. On the border of this region the Higgs mixing parameter $\mu$, which is determined from the requirement of electroweak symmetry breaking via radiative corrections, tends to zero. This leads to existence of light and degenerate states: the second chargino and two neutralinos, all of them being essentially higgsinos.

Experimental Higgs and chargino mass limits as well as WMAP relic density limit can be easily satisfied in these scenarios. However, the strong fine-tuning is required.  The discussed scenarios differ from the GMSB scenario~\cite{llgmsb} with the gravitino as the LSP, and next-to-lightest supersymmetric particles typically live longer.

Searches for long-lived particles were already made by	LEP	collaborations~\cite{llsearches}. It is also of great interest at the moment since the first physics results of the coming LHC are expected in the nearest future. Light long-lived sparticles could be produced already during first months of its operation~\cite{lllhc}.
Since staus and stops are relatively light in this scenario, their production cross-sections are quite large and may achieve a few per cent of pb for stau production, and tens or even hundreds of pb for light stops,  $\tilde m_t < 150$ GeV. The cross-section then quickly falls down when the mass of stop is increased. However, even for very large values of $|A|$ when stops become heavier than several hundreds GeV, the production cross-section is of order of few per cent of pb, which is enough for detection with the high LHC luminosity.

\subsection{The reach of the LHC}

\begin{figure}[htb]
\begin{center}
\includegraphics[width=0.35\textwidth]{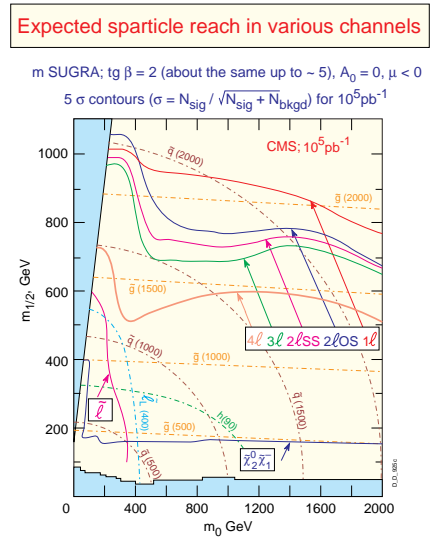}
\end{center}\vspace{-1.0cm}
\caption{Expected range of reach for superpartners in various channels at
LHC~\cite{LHC}.} \label{LHCreach}
\end{figure}
\begin{figure}[htb]
\begin{center}
\includegraphics[width=0.3\textwidth,height=6.0cm]{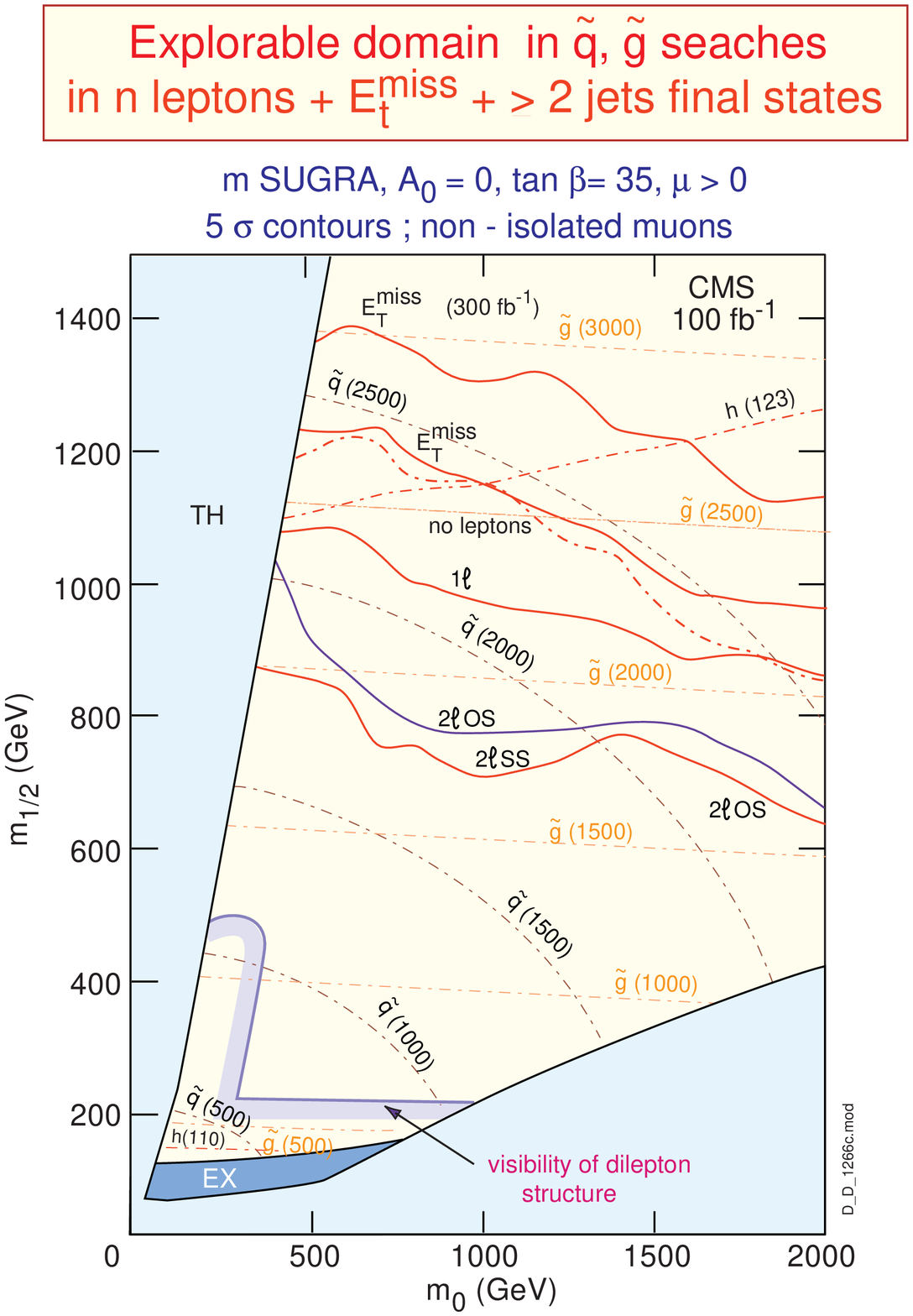}
\end{center}\vspace{-0.7cm}
\caption{Expected domain of searches for  squarks and gluions  at
LHC~\cite{LHC}.} \label{LHCchannel}
\end{figure}
The LHC hadron collider is the ultimate machine for new physics at
the TeV scale. Its c.m. energy is planned to be 14 TeV with very
high luminosity up to a few hundred fb$^{-1}$.  The LHC is
supposed to cover a wide range of parameters of the MSSM (see
Figs. below) and discover the superpartners with the masses below
2 TeV~\cite{Kras}. This will be a crucial test for the MSSM and
the low energy supersymmetry. The LHC potential to discover
supersymmetry is widely discussed in the
literature~\cite{Kras}-\cite{LHCSUSY}.

To present the region of reach for the LHC in different channels of sparticle production it
is useful to take  the same plane of soft SUSY breaking parameters $m_0$ and $m_{1/2}$. In this
case one usually assumes certain luminosity which will be presumably achieved during the
accelerator operation.

Thus, for instance, in Fig.~\ref{LHCreach} it is shown the regions of reach in different
channels. The lines of a constant squark mass form the arch curves, and those for gluino
are almost horisontal. The curved lines show the reach bounds in different channel of
creation of secondary particles. The theoretical curves are obtained within the MSSM for
a certain choice of the other soft SUSY breaking parameters.
 As one can see, for the fortunate circumstances the wide range of the parameter
space  up to the masses of the order of 2~Tev will be examined.

The other example is shown in Fig.~\ref{LHCchannel} where the regions of reach for squarks
and gluino  are shown for various luminosities. One can see that for the maximal
luminosity the discovery range for squarks and gluino reaches 3~TeV for the center of
mass energy of 14~TeV and even higher for the double energy.

  The
LHC will be able to discover SUSY with squark and gluino masses up
to $2\div 2.5$ TeV for the luminosity $L_{tot}=100\ fb^{-1}$.  The most powerful signature for squark
and gluino detection are multijet events; however, the discovery
potential depends on relation between the LSP, squark, and gluino
masses, and decreases with the increase of the LSP mass.

\begin{figure}[htb]
\hspace*{-0.3cm}
\includegraphics[width=0.5\textwidth, height=6cm]{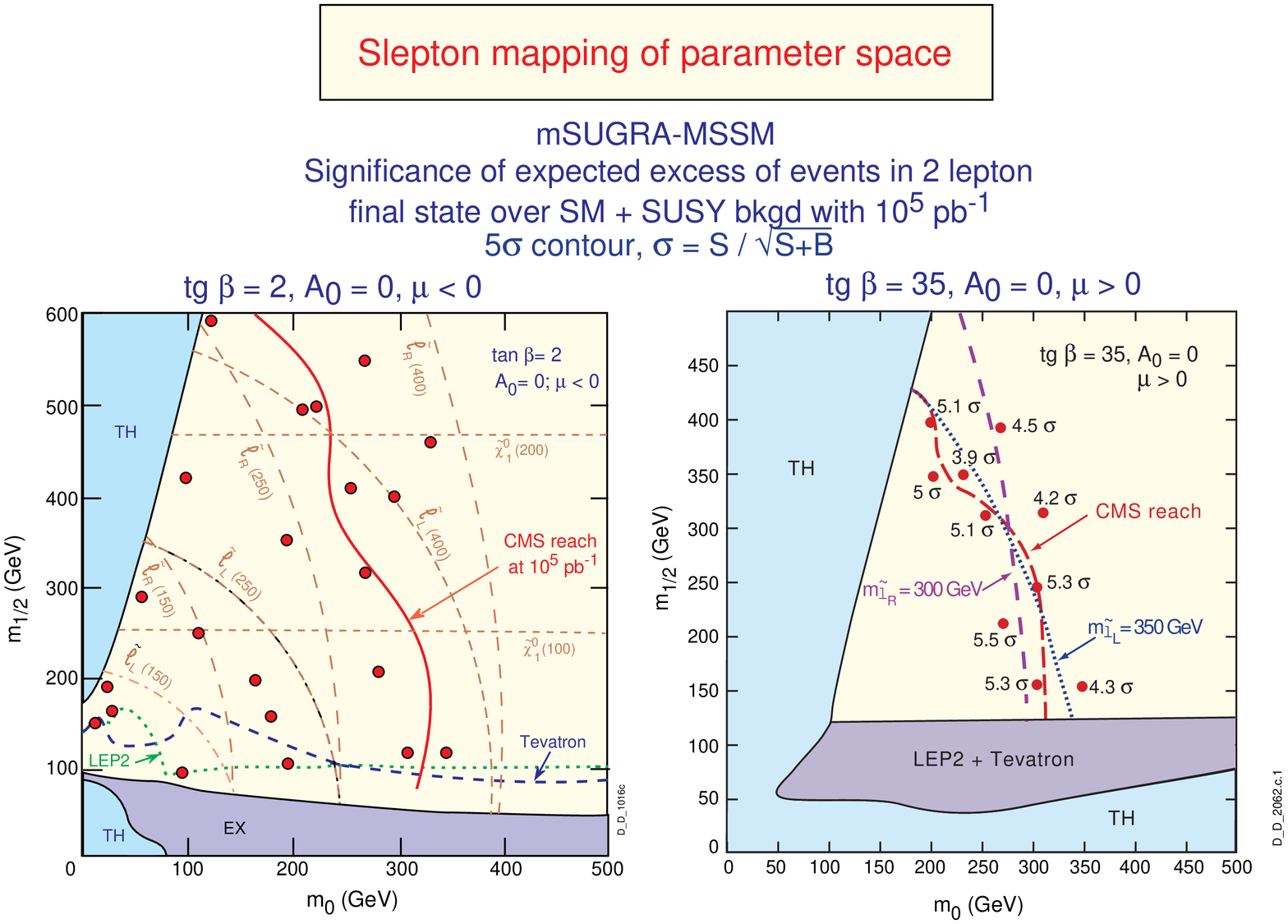}
\vspace{-0.7cm}
\caption{Expected range of reach for sleptons at LHC~\cite{LHC}.} \label{fig:LHC3}
\end{figure}

The same is true for the sleptons as shown in Fig.~\ref{fig:LHC3}. The slepton pairs can
be created via the Drell-Yang mechanism  $pp\to\gamma^*/Z^*\to\tilde \ell^+\tilde \ell^-$
and can be detected through the lepton decays $\tilde \ell \to \ell+\tilde \chi^0_1$. The
typical signal used for slepton detection is the dilepton pair with the missing energy
without hadron jets. For the luminosity of $L_{tot}=100$~fb$^{-1}$ the LHC will be able
to discover sleptons with the masses up to 400~GeV~\cite{Kras}. The discovery reach for
sleptons in various channels is shown in Fig.\ref{fig:LHC3}.

\subsection{The lightest superparticle}

One of the crucial questions is the properties of the lightest
superparticle. Different SUSY breaking scenarios lead to different
experimental signatures and different LSP.

$\bullet$ Gravity mediation

In this case, the LSP is the lightest neutralino
$\tilde{\chi}^0_1$, which is almost 90\% photino for a low
$\tan\beta$ solution and contains more higgsino admixture for high
$\tan\beta$. The usual signature for LSP is missing energy;
$\tilde{\chi}^0_1$ is stable and is the best candidate for the
cold dark matter in the Universe. Typical processes, where the LSP
is created, end up with jets + $\Big/ \hspace{-0.3cm}E_T$, or
leptons + $\Big/ \hspace{-0.3cm}E_T$, or both jets + leptons +
$/ \hspace{-0.25cm}E_T$.

$\bullet$ Gauge mediation

In this case the LSP is the  gravitino $\tilde G$ which also leads
to missing energy. The actual question here is what the NLSP, the
next lightest particle, is. There are two possibilities:

i) $\tilde{\chi}^0_1$ is the NLSP. Then the decay modes are: \
 $\tilde{\chi}^0_1 \to \gamma \tilde G, \ h \tilde G, \ Z \tilde
 G.$\
 As a result, one has two hard photons + $/
 \hspace{-0.25cm}E_T$, or jets + $/ \hspace{-0.25cm}E_T$.

ii) $\tilde l_R$ is the NLSP. Then the decay mode is  $\tilde l_R
\to \tau \tilde G$ and the signature is a charged lepton and the
missing energy.

$\bullet$ Anomaly mediation

In this case, one also has two possibilities:

i) $\tilde{\chi}^0_1$ is the LSP and wino-like. It is almost
degenerate with the NLSP.

ii) $\tilde \nu_L$ is the LSP. Then it appears in the decay of
chargino $\tilde \chi^+ \to \tilde \nu l$ and the signature is the
charged lepton and the missing energy.

$\bullet$ R-parity violation

In this case, the LSP  is no longer stable and decays into the SM
particles. It may be charged (or even colored) and may lead to
rare decays like neutrinoless double $\beta$-decay, etc.

Experimental limits on the LSP mass follow from non-observation of
the corresponding events. Modern lower limit is around 40 GeV .

\section{Supersymmetric Dark Matter}

\subsection{The problem of the  dark matter in the Universe}
As was already mentioned the shining matter does not compose all the matter in the Universe.
According to the latest precise data~\cite{WMAP} the matter content of the
Universe is the following:
\begin{eqnarray*}
&&\Omega_{total}=1.02 \pm 0.02 ,\\
&&\Omega_{vacuum}=0.73 \pm 0.04, \\
&&\Omega_{matter}=0.23 \pm 0.04,\\
&& \Omega_{baryon}=0.044 \pm 0.004,
\end{eqnarray*}
so that the dark matter prevails the usual matter by factor of 6.

Besides the rotation curves of stars the dark matter manifests itself in the observation of gravitational lensing effects~\cite{lensing} and the large structure formation. It is believed that the dark matter played the crucial role in formation of large structures like clusters of galaxies and  the usual matter just fell down in a potential well attracted by  gravitational interaction afterwards. The dark matter can not make compact objects like the usual matter since it does not take part in strong interaction and can not lose energy by photon emission since it is neutral. For this reason the dark matter can be trapped in much larger scale structures like galaxies.

In general one may assume two possibilities: either the dark matter interacts  only gravitationally, or it participates also in  the weak interaction.  The latter case is preferable since then one may hope to detect it via the methods of particle physics. What makes us to believe that the dark matter is probably the Weakly Interacting Massive Particle (WIMP)? This is because the cross-section of DM annihilation which can be figured out of the amount of the DM in the Universe is close to a typical weak interaction cross-section. Indeed,  let us assume that all the DM is made of  particles  of a single type. Then the amount of the DM can be calculated from the Boltzman equatio~\cite{Kolb,Rub}
\begin{equation}
\frac{dn_\chi}{dt}+ 3 H n_\chi = - < \sigma v > ( n^2_\chi- n^2_{\chi,eq}),	
\end{equation}
 where $H = \dot{R}/ R$ is the Hubble constant and $n_{\chi,eq}$ is the equilibrium concentration.
The relic abundance is expressed in terms of $n_\chi$ as
\begin{equation}
\Omega_\chi h^2 =\frac{m_\chi n_\chi}{\rho_c}\approx \frac{2\cdot 10^{27}\ cm^3\ sec^{-1}}{<\sigma v>}.
\end{equation}
Having in mind that  $\Omega_\chi h^2 \approx 0.113\pm 0.009$ and $v\sim 300$ km/sec one gets
\begin{equation}
\sigma\approx 10^{-34}\ cm^2 = 100\ pb,
\end{equation}
which is a typical EW cross-section.

\subsection{Detection of the Dark matter}

There are two methods of the DM detection: direct and indirect. In direct detection one assumes that the particles of DM come to Earth and interact with the nuclei of a target. In underground experiments one can hope to observe such events measuring the recoil energy. There are several experiments of this type: DAMA, Zeplin, CDMS and Edelweiss. Among them only DAMA collaboration claims to observe a positive outcome in annual modulation of the signal with the fitted DM particle mass around 50 GeV~\cite{DAMA}.
\begin{figure}[htb]\vspace{-0.2cm}
\begin{center}
  \includegraphics[width=.38\textwidth]{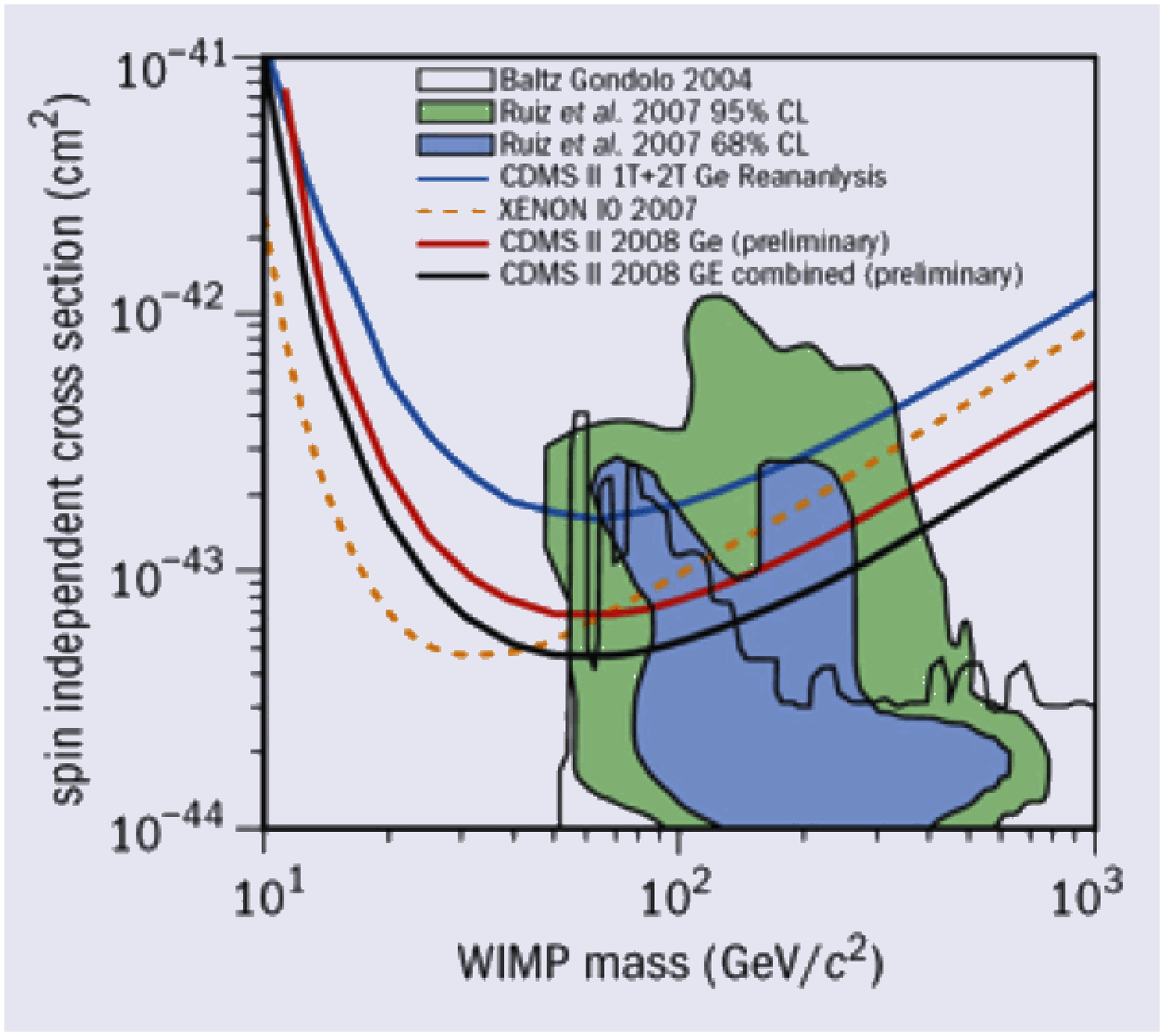}
  \includegraphics[width=.38\textwidth]{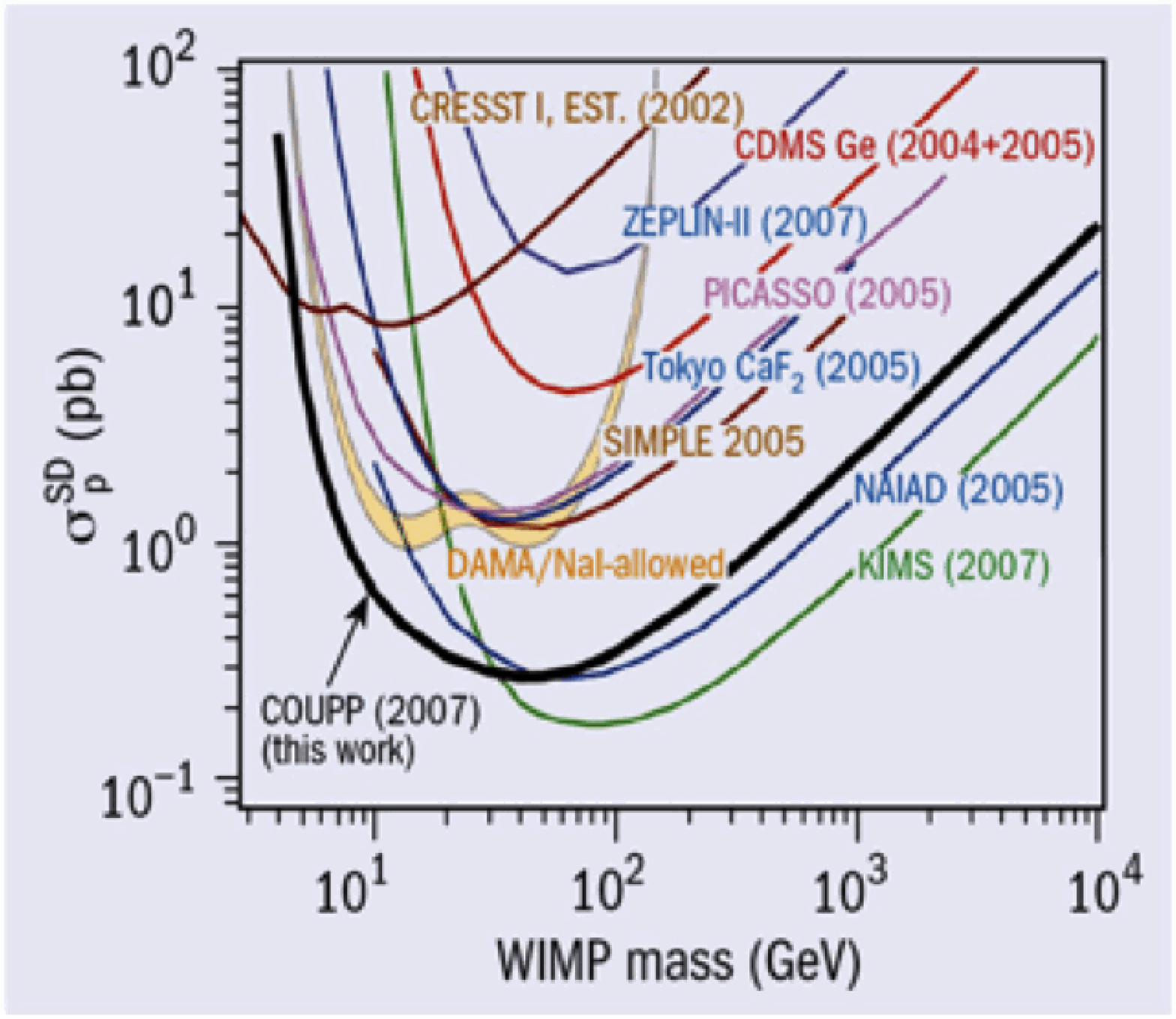}\vspace{-0.8cm}

    \caption{The exclusion plots from direct DM detection experiments. Spin-independent case (top) from Chicagoland Observatory for Underground Particle Physics (COUPP) and spind-dependent case (bottom)  from Cryogenic Dark Matter Search (CDMS)
  }
\label{direct}
  \end{center}
\end{figure}

\vspace{-0.2cm}All the other experiments do not see it though CDMS collaboration recently announced about a  few events of a desired type~\cite{CDMS}. The reason of this disagreement might be in different methodology and the targets used since the cross-section depends on a spin of a target nucleus.  The collected statistics is also essentially different, DAMA has accumulated by far more data and this is  the only collaboration which studies the modulation of the signal that may be crucial for reducing the background.

The typical exclusion plots for spin-independent and spin-dependent cross-sections are shown in Fig.\ref{direct}
where one can see DAMA allowed region overlapping with the other exclusion ones.
Still today we have no convincing evidence for direct DM detection or exclusion.

Indirect detection of the DM is aimed to the registration of a signal from the DM annihilation in the form of additional gamma rays and charged particles (antiprotons and positrons) in cosmic rays.  These particles should have the energetic spectrum reflecting their origin from annihilation of heavy massive particles which is different from the background coming from the known sources.  Hence one may expect the appearance of a "shoulder``  in the cosmic ray spectrum.  There are several experiments of this kind: EGRET (diffuse gamma rays) which is followed by FERMI; HEAT and AMS1 (positrons) which is followed by PAMELA; BESS (antiprotons) which will be followed by PAMELA and AMS2. All these experiments see some deviation from the background though the uncertainties are large and the background is not known very accurately especially for charged particles.

From this point of view the most  detailed information was obtained by EGRET cosmic telescope~\cite{egret1} which orbited the Earth for 9 years and measured the spectrum and intensity of  diffuse gamma rays over the whole celestial sphere with the  4 degree bins. The form of the spectrum was measured in the region of 0.1-10 GeV.
It allows one to  perform the independent analysis in different directions of the celestial sphere.
Gamma rays have the advantage that they point back to the source and do not suffer energy losses, so they
are the ideal candidates to trace the dark matter density. The charged components interact with Galactic matter and are deflected by the Galactic magnetic field, so they do not point back to the source.

\begin{figure*}[t]\vspace{-0.0cm}
\begin{center}
  \includegraphics[width=.80\textwidth,height=7cm]{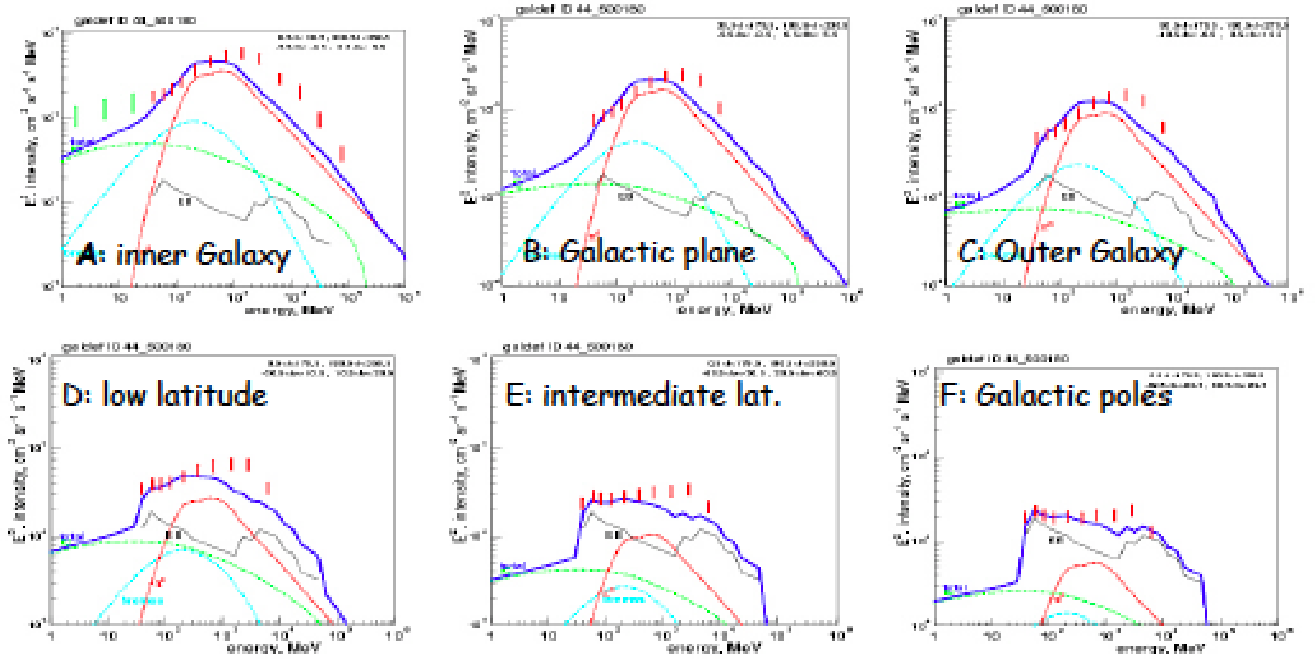}
  \vspace{-0.2cm}
   \caption{Excess in diffuse gamma rays as measured by EGRET  in various regions in the sky. The solid(blue) line is the background as calculated by the GALPROP  code.  Discrete slashes represent EGRET data. Also shown are the contributions of the known background sources  \label{excess}}
\vspace{0.7cm}
  \includegraphics[width=.80\textwidth]{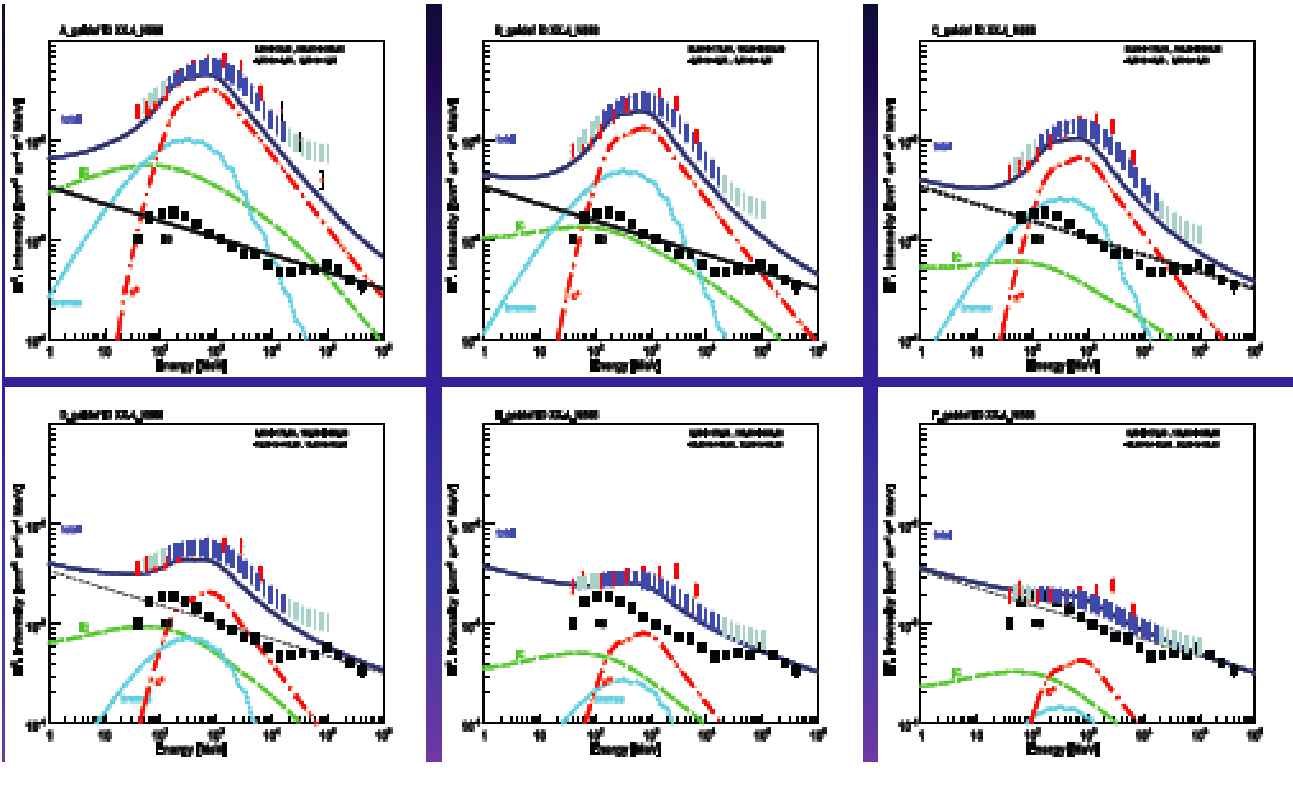}\vspace{-0.2cm}
   \caption{Excess in diffuse gamma rays as measured by FERMI  - dark (blue) slashes in comparison with EGRET -
   light (red) slashes in the same regions in the sky\label{excessnew}}
  \end{center}
\end{figure*}

 The diffuse component shows a clear excess for the energy above 1 GeV  about a
factor two over the expected background from known nuclear
interactions, inverse Compton scattering and bremsstrahlung  as shown in Fig.\ref{excess}~\cite{egret2}.
Different  plots  correspond to  different regions in the sky: A - inner galaxy, B - outer disk, C - outer galaxy, D- low attitude, E- intermediate latitude, F -galactic poles.

\begin{figure}[h]\vspace{-0.2cm}
\begin{center}
  \includegraphics[height=.235\textheight]{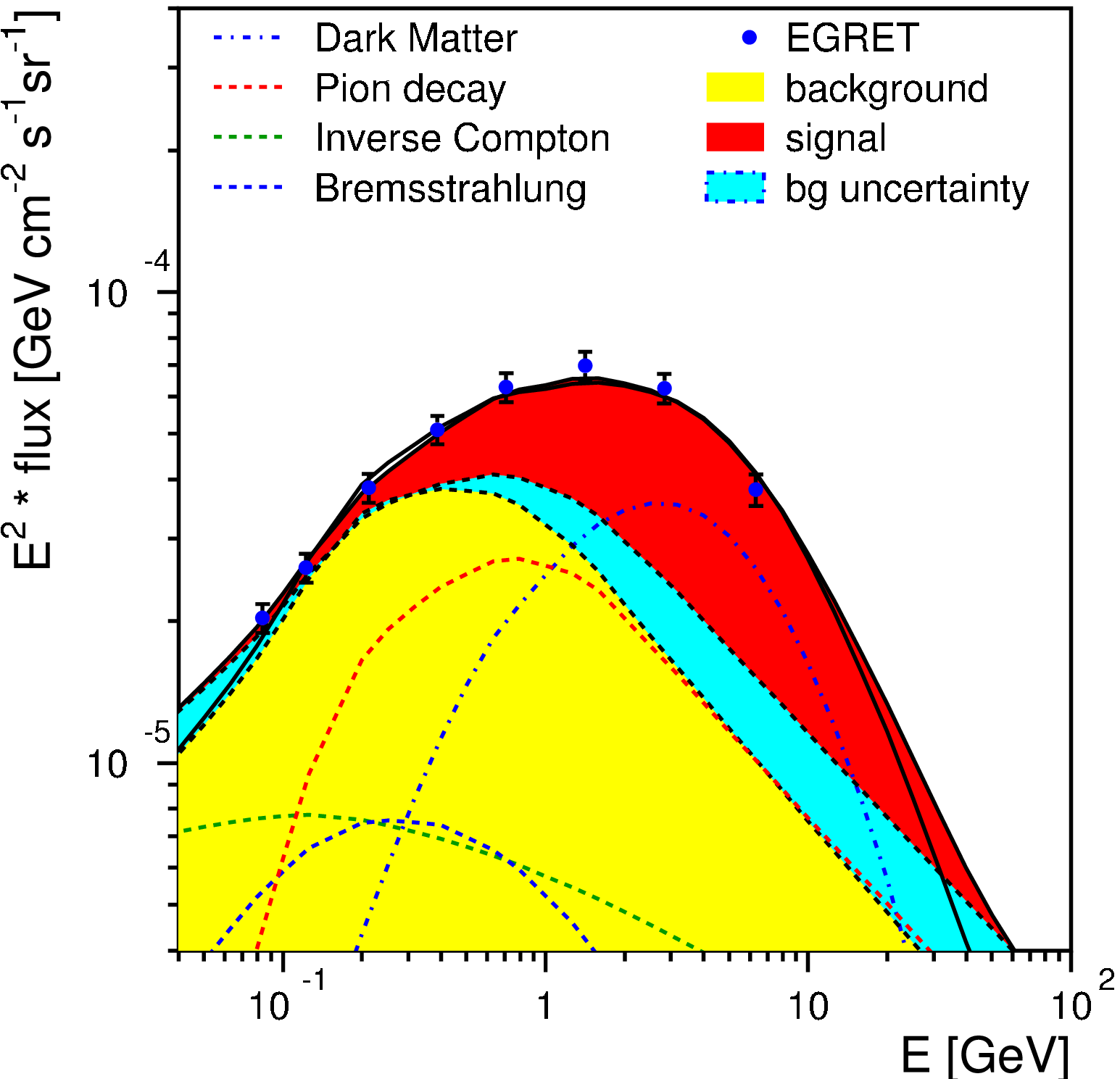}
  \includegraphics[height=.26\textheight]{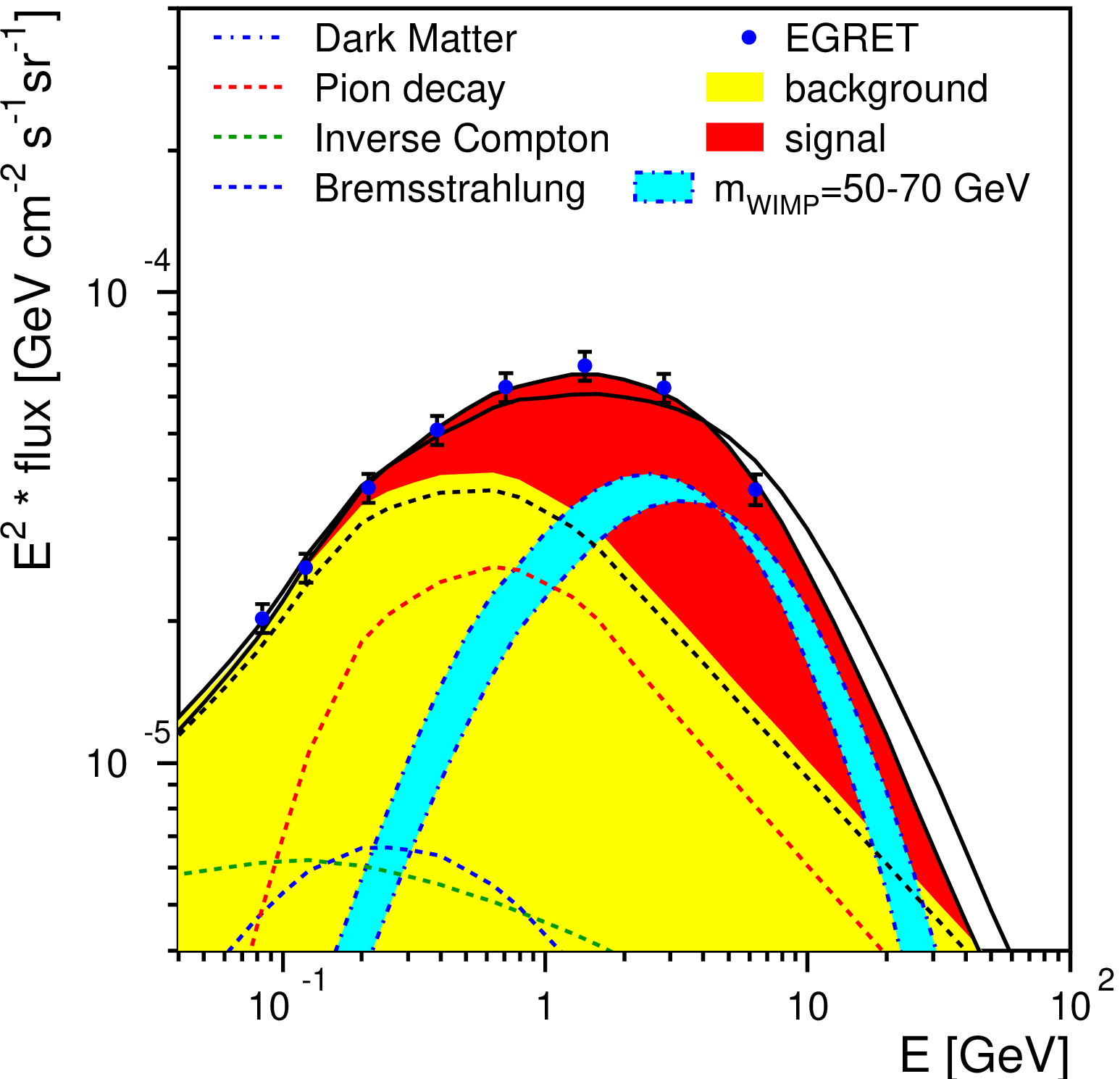}\vspace{-0.6cm}
  \caption{The spectrum of diffuse gamma rays measured by EGRET in the direction of the galaxy center and the fit to the data. The light shaded (yellow)
    areas indicate the background using the shape of the
    conventional GALPROP model~\cite{GALPROP}, while the dark shaded (red)
    areas are the signal contribution from DMA for a 60 GeV WIMP mass. }  \label{var}
  \end{center}
\end{figure}

As one can see the excess of a signal above the background is isotropic in celestial sphere that suggests the common source which might be the DM. It was shown that the observed excess in the spectrum of diffuse gamma reays, if taken seriously,  has all the features of the decay of $\pi^0$ mesons produced by monoenergetic quarks coming from the DM annihilation. Fitting the background together with the signal  from the DM annihilation one can get remarkable agreement for all directions if the mass of the DM particle is around 60 GeV. A detailed picture  for the region of the sky in the direction of the galactic center is shown in Fig.\ref{var}.  Here one can see the allowed background variations and the variations of the DM particle mass used for fitting the data.  Possible background variations are not enough to ex-
\clearpage
\noindent plain EGRET data while the variation of the WIMP mass  within 50-70 GeV does not contradict these data.

It is instructive to compare EGRET data with recently released FERMI data which are much more precise.  This comparison is shown in Fi.\ref{excessnew}~\cite{deBoer}. One can see that the new data are not in contradiction with the old ones. The excess is still visible though is smaller compared to EGRET. On has to admit, however, that the interpretation of the data in favour of background modification is also possible. So, taking the optimistic point of view, one can interpret these data as a signal from the DM annihilation, otherwise everything is sinked in the error bars.

\begin{figure}[htb]\vspace{-0.4cm}
\begin{center}
  \includegraphics[width=.35\textwidth]{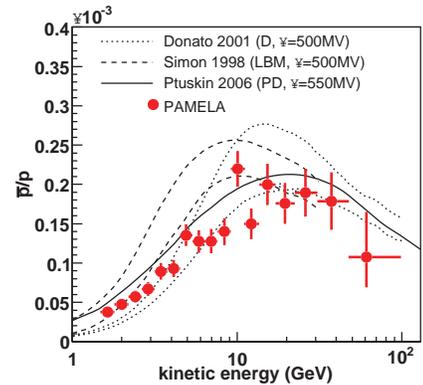}\vspace{-0.5cm}
   \caption{Antiproton/proton ratio as measured by PAMELA. No excess is found \label{antipr}}
  \end{center}
\end{figure}

\begin{figure}[htb]\vspace{-0.1cm}
\begin{center}
  \includegraphics[width=.33\textwidth]{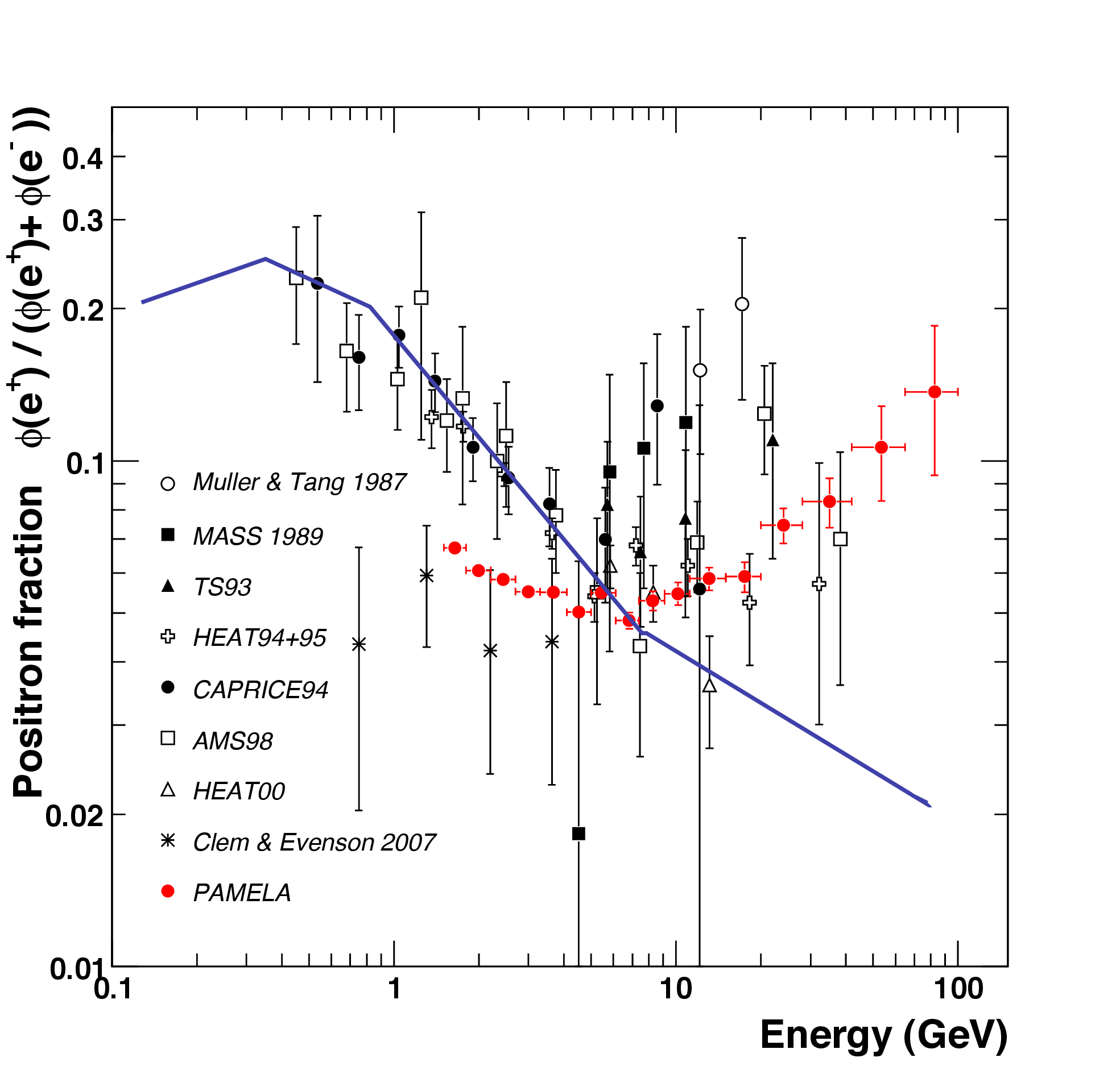}\vspace{-0.5cm}
   \caption{Positron fraction as measured by PAMELA in comparison with the background\label{pamela}}
  \end{center}
\end{figure}

The experimental data with the charge particles looks more contradictory. We present the antiproton  and positron data in Figs. \ref{antipr}~\cite{antipr} and \ref{pamela}~\cite{pamela}, respectively. While there is no excess observed in antiproton data, the positron spectrum measured by PAMELA is quite unusual. It strongly contradicts the expectations from the GALPROP. Possible interpretations of PAMELA  data include:
background from hadronic showers with large electromagnetic component; astrophysical sources like pulsars, positron acceleration in SNR, locality of sources;  leptophilic DM annihilation, very heavy ($\sim$ 1 TeV) WIMPs, etc.  The situation is still to be clarified.

One should mention here, that interpreting the excess in diffuse gamma rays data as the WIMP annihilation one has to enhance the intensity of a signal by factor of 10-100 that is usually  achieved by   assumption of clumpiness of the DM. This almost obvious property of the DM has no experimental confirmation so far. The same enhancement, however,  is not needed for antiprotons where one seems to have an agreement  with the data. This contradiction might be attributed  to different behaviour of charged particles  in galactic magnetic fields.

\subsection{Supersymmetric Interpretation of the Dark Matter}

Supersymmetry offers several candidates for the  role of the cold dark matter.  If one looks at the particle content of the MSSM from the point of view of a heavy neutral particle, one finds several such particles, namely: a  superpartner of the  photon (photino $\tilde \gamma$), a superpartner of the  Z-boson (zino $\tilde z$),  a superpartner of neutrino  (sneutrino $\tilde \nu$) and superpartners of the Higgs bosons  (higgsino $\tilde H$).
The DM particle can be the lightest of them, the LSP.  The others decay on the LSP and the SM particles, while the LSP is stable and can survive since the Big Bang. As a rule the lightest  supersymmetric particle is the so-called neutralino, the spin 1/2 particle which is the combination of photino, zino and two neutral higgsinos  and is the eigenstate of the mass matrix
$$
|\tilde \chi^0_1\rangle =N_1|B_0\rangle
+N_2|W^3_0\rangle +N_3|H_1\rangle +N_4|H_2\rangle .$$

 Thus, supersymmetry actually predicts the existence of the dark matter.  Moreover, we have shown above that one can choose the parameters of  a soft supersymmetry  breaking in such a way that one gets the right amount of the DM. This requirement  serves as a constraint for these parameters and is consistent with the requirements coming
 from particle physics.

\begin{figure}[htb]
\begin{center}
\hspace*{0.3cm}  \includegraphics[width=.35\textwidth]{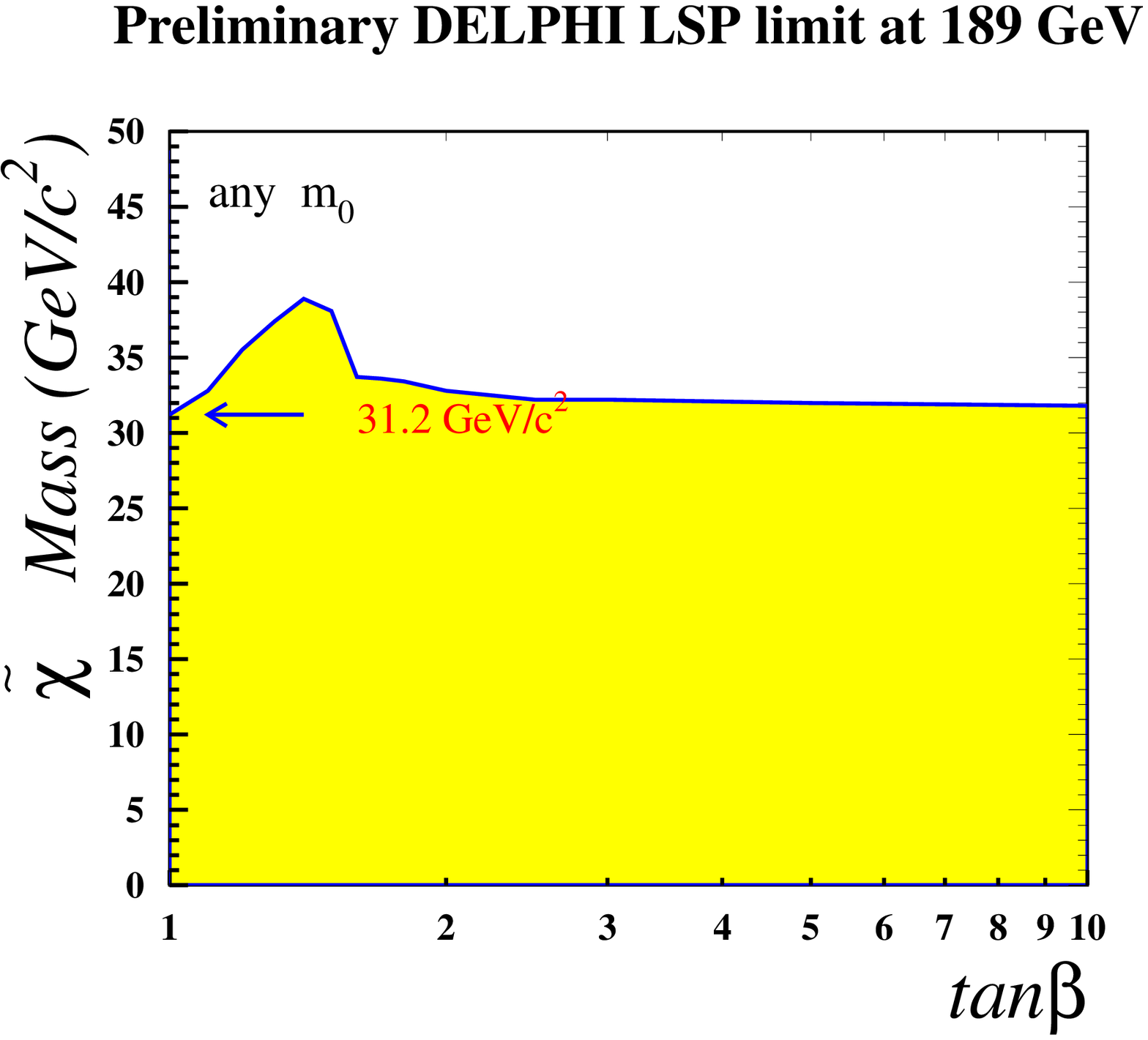}
 \includegraphics[width=.35\textwidth]{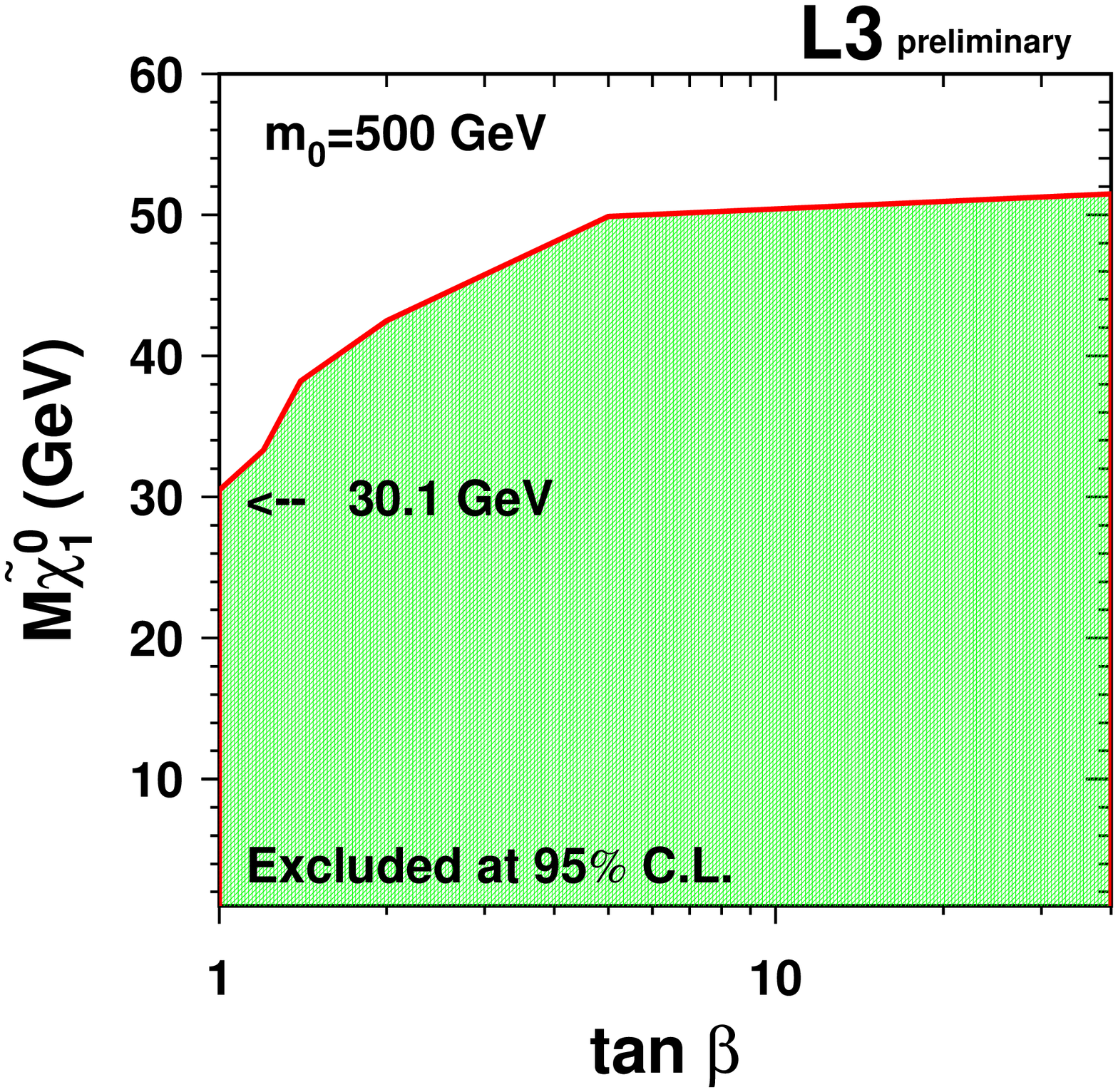}
    \vspace{-0.4cm}
  \caption{Exclusion limit on the LSP mass from Delphi coll.
  and L3 coll. (LEP)~\cite{LEP}
  \label{delphi}}\end{center}
\end{figure}

 The search for the LSP was one of the tasks of  LEP. They were supposed to be produced as a result of chargino decays and be detected via missing transverse energy and momentum. Negative results  defined the limit on the LSP mass as shown in Fig.\ref{delphi}.

\begin{figure}[htb]
  \includegraphics[width=.45\textwidth]{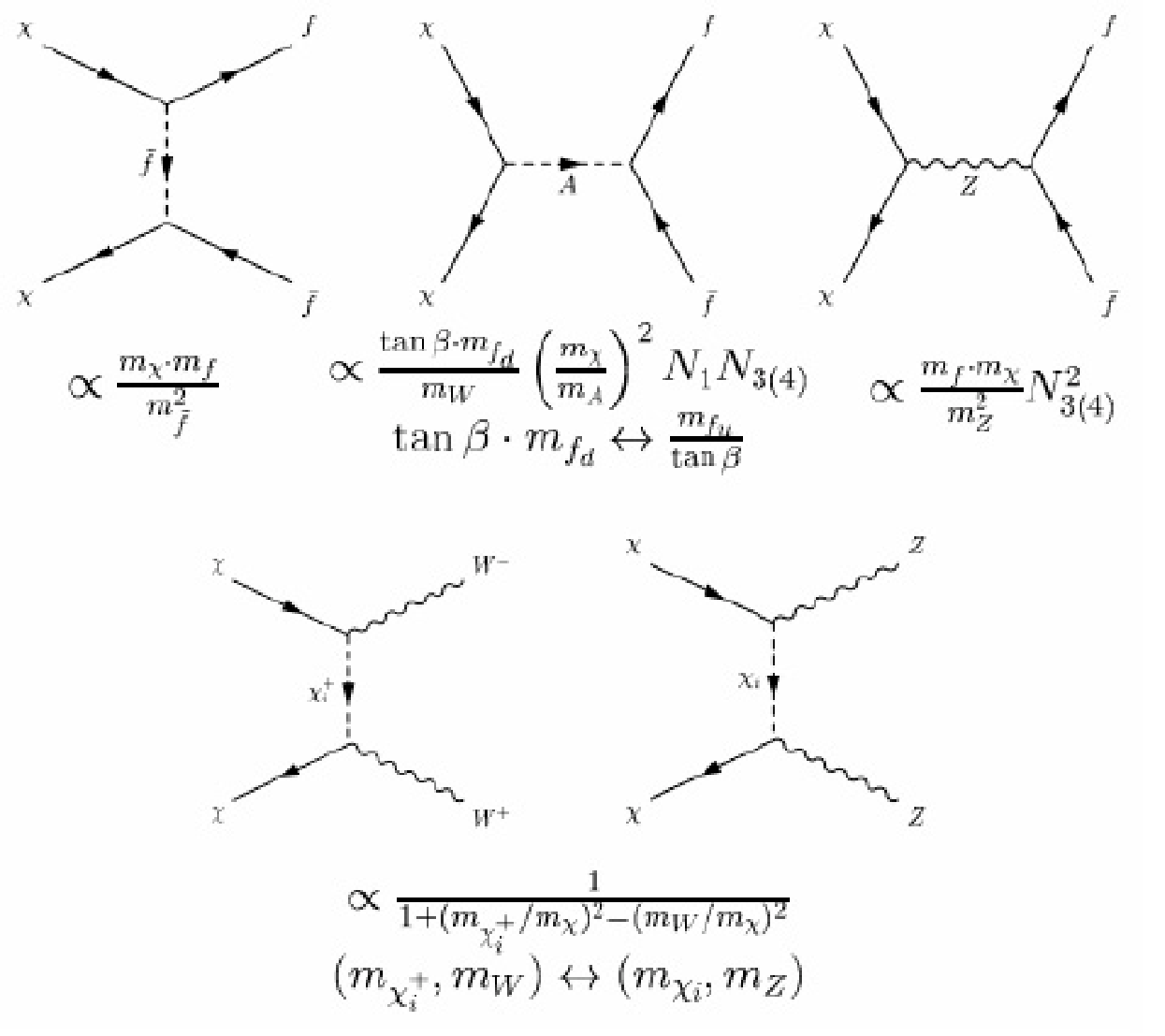}\vspace{-0.3cm}
  \caption{The dominant annihilation diagrams for the lightest neutralino in the MSSM
  \label{annihilation}}
\end{figure}

The DM particles which form the halo of the galaxy annihilate  to produce
the ordinary particles in the cosmic rays. Identifying them with the LSP from a supersymmetric
 model one can calculate the annihilation rate and to study the secondary particle spectrum.
 The dominant annihilation diagrams of the lightest supersymmetric
particle (LSP) neutralino are shown in Fig.\ref{annihilation}.  The usual final states are either the quark-antiquark pairs or the W and Z  bosons.  Since the cross sections are proportional to the final state fermion mass,
 the heavy fermion final states, i.e. third generation quarks and leptons, are
expected to be dominant. The W- and Z-final states from t-channel chargino and
neutralino exchange have usually a smaller cross section.

The dominant contribution comes from A-boson exchange: $\chi+\chi \to A \to b\bar b$. The sum of the diagrams should yield $<\sigma v>=2\cdot  10^{-26}\ cm^3/sec$ to get the correct relic density.

\begin{figure}[htb]\vspace{-0.3cm}
  \includegraphics[width=.45\textwidth]{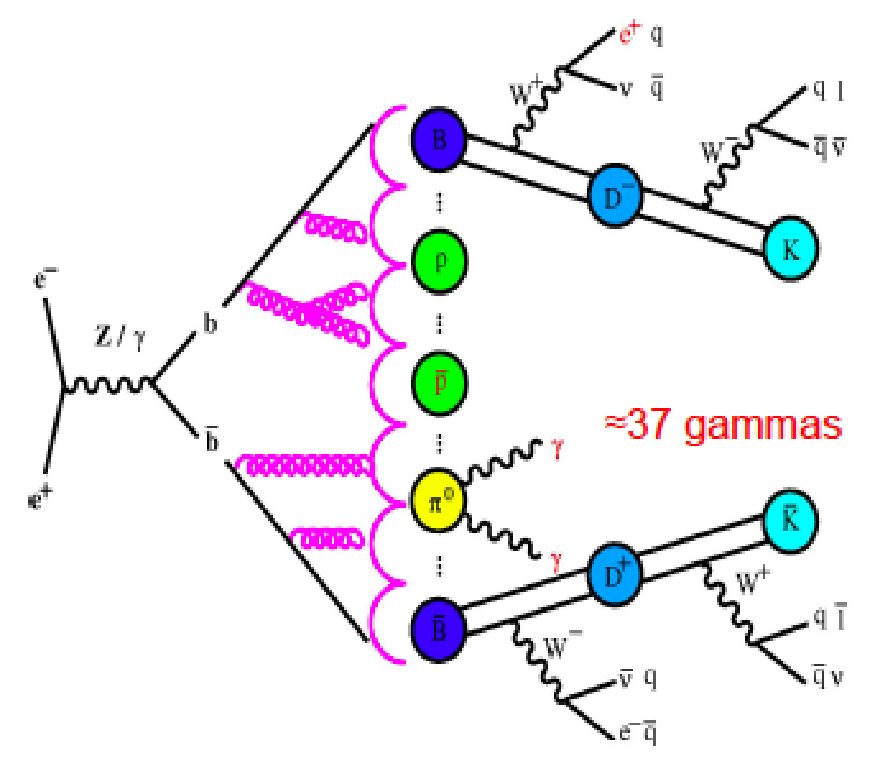}
  \caption{The final states in the process of $e^+e^-$-annihilation at colliders in the SM
  \label{SMfinal}}
\end{figure}

The spectral shape of the secondary particles  from DMA is well known from the
fragmentation of mono-energetic quarks studied at electron-positron
colliders, like LEP at CERN, which has been operating up to
centre-of-mass energies of about 200 GeV, i.e. it corresponds to
the neutralino masses up to 100 GeV (see Fig.\ref{SMfinal}). The different
quark flavours all yield similar gamma spectra at high energies.  Hence, the specrta of positrons, gammas and
antiprotons is known.  The relative ammount of $\gamma, p^-$ and $e^+$ is also known. One expects around 37 gammas per collision.

The gamma rays from the DM annihilation can be distinguished
from the background  by their completely different spectral
shape: the background originates mainly from cosmic rays
hitting the gas of the disc and producing abundantly $\pi^0$ mesons,
which decay into two photons. The initial cosmic ray spectrum is a steep
power law spectrum, which yields a much softer gamma ray spectrum
than the fragmentation of the hard mono-energetic quarks from the DM annihilation.
The spectral shape of the gamma rays from the background is well
known from fixed target experiments given the known cosmic ray spectrum.

\subsection{SUSY interpretation of EGRET excess}

If one takes the EGRET excess in diffuse gamma rays seriously then one can try to identify the DM particle responsible for this excess with the LSP. The mass of this hypothetical WIMP as it follows from EGRET data is in the range of 50 to 100 GeV and is fully compatible with the neutralino. Since in the MSSM all the couplings are known one can calculate the annihilation rate given by the diagrams in Fig.\ref{annihilation}. The only unknown parameters are the SUSY masses (and mixings) which one can choose to fit the data.

\begin{figure}[htb]
  \includegraphics[height=.307\textheight]{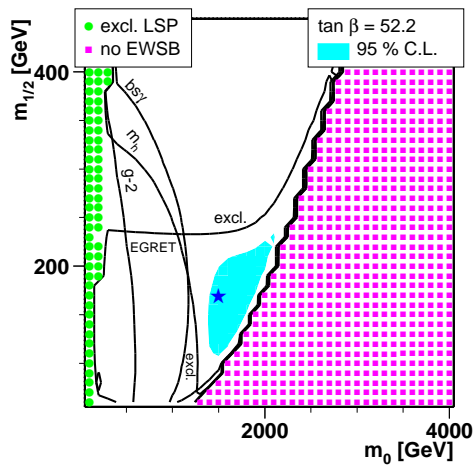}
  \caption{ The allowed region of parameter space with account of the EGRET data on diffuse gamma rays.
  The star indicates the best fir value.
  \label{msugra}}
\end{figure}

Combining various requirements on  soft SUSY parameters together with the assumed EGRET energy range for the mass of neutralino one gets an essentially constrained allowed region shown in Fig.\ref{msugra}~\cite{egret3}
One can see that the "EGRET" region of parameter space  corresponds to high values of $\tan\beta$, low $m_1/2$ and high $m_0$. This is the range of the so-called focus point region where  chrarginos and neutralinos, whose mass is governed by the value of  $m_1/2$, are light and squarks and sleptons, whose mass is governed by  $m_0$, are heavy.  The lightest neutralino in the this region is 95\% photino being the superpartner of a photon of the cosmic microwave background.
Choosing the point in this allowed region one can calculate the whole mass spectrum of superpartners.
 We present in the Table\ref{t1} the sample mass spectrum corresponding to the best fit point in the "EGRET" region.~\cite{egret3}.
\begin{table}[htb]
\begin{center}
 {\begin{tabular}{|c|c|}
   \hline
  {\bf Particle} & {\bf Mass [GeV]} \\
   \hline
   $\tilde \chi^0_{1,2,3,4}$ & 64, 113, 194, 229 \\
   $\tilde \chi^\pm_{1,2},\tilde{g}$ & 110, 230, 516 \\
   $\tilde u_{1,2}=\tilde c_{1,2}$ & 1519, 1523 \\
   $\tilde d_{1,2}=\tilde s_{1,2}$ & 1522, 1524 \\
   $\tilde t_{1,2}$ & 906, 1046 \\
   $\tilde b_{1,2}$ & 1039, 1152 \\
   $\tilde e_{1,2}=\tilde \mu_{1,2}$ & 1497, 1499 \\
   $\tilde \tau_{1,2}$ & 1035, 1288 \\
   $\tilde \nu_e, \tilde \nu_\mu, \tilde \nu_\tau$ & 1495, 1495, 1286 \\
   $h,H,A,H^\pm$ & 115, 372, 372, 383 \\
   \hline
   Observable & Value \\
   \hline
   $Br(b\to s\gamma)$ &  $3.02\cdot  10^{-4}$\\
   $\Delta a_\mu$ &$1.07 \cdot 10^{-9}$ \\
   $\Omega h^2$ & $0.117$ \\
   \hline
  \end{tabular}}\vspace{0.3cm}
\caption{The mass spectrum of superpartners in the EGRET point: $m_0=1500$ ÃýÂ,
$m_{1/2}=170$ ÃýÂ, $A_0=0$, $\tan\beta=52.2$, $\mu>0$ \label{t1}}
\end{center}
\end{table}

As one can see from the table, in the "EGRET" point one has considerable splitting between the relatively light superpartners of the gauge fields and heavy squarks and sleptons. The masses of neutralinos and charginos are almost at the lower boundary of experimentally allowed range.  The same is true for the lightest Higgs boson.
Experimental lower limit on the SM Higgs boson mass today is 114.7 GeV as follows from the negative results of the search  at LEP. This bound is also true for the MSSM for large $\tan\beta$.

Thus, taking the optimistic point of view that the excess in diffuse gamma rays actually exists and accepting the supersymmetric interpretation of this excess one can {\em simultaneously} give answer to the following questions:\\
$\bullet$ In Cosmology:
What is CDM  made of?\\
$\bullet$ In Astrophysicists:
What is the origin of excess of diffuse Galactic Gamma Rays? \\
$\bullet$ In Particle physicists :
Where are the Super\-symmetric Particles?\\
And the answer is:\\
$\bullet$
DM is made of WIMPs which are SUSY particles distributed in Halo of our Galaxy
with a mass around 70 GeV.

What is important, supersymmetric  interpretation of the DM is testable  since it predicts the mass spectrum which can be directly checked at the LHC in the nearest future.

\section{Conclusion}

\begin{figure}[htb]
\begin{center}
 \includegraphics[width=.45\textwidth]{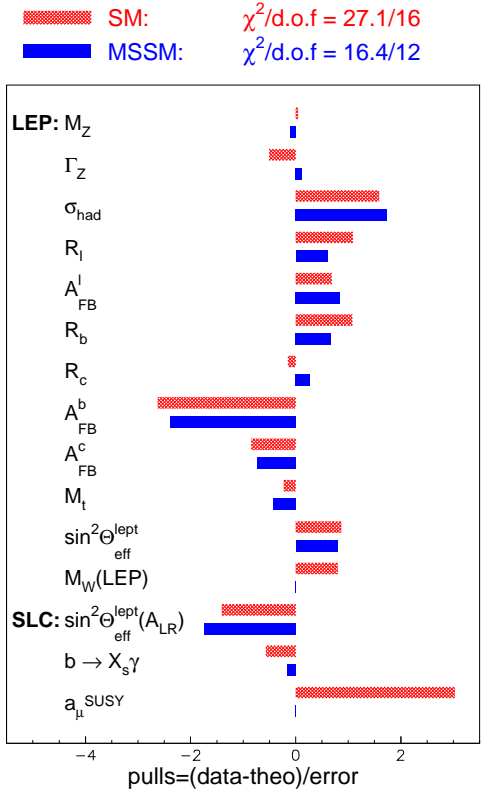}
  \caption{The SM versus the MSSM in comparison with precision experimental data}\label{SMvMSSM}
  \end{center}
\end{figure}

Supersymmetry is now the most popular extension of the Standard
Model. Comparison of the MSSM with precision experimental data for the MSSM
is as good as  for the SM and sometimes even better. For example the branching ratio
$BR(b\to s\gamma)$ and the anomalous magnetic moment of muon are fitted better in the MSSM than in the SM.
The relic density of the DM is not described in the SM but is naturally predicted by the MSSM. One can see this
comparison for main observables in Fig.\ref{SMvMSSM}~\cite{BSK}.

Still today after 30 years since the invention of supersymmetry we have no single
convincing evidence that supersymmetry  is realized in particle physics. It remains very popular in quantum field theory and in string theory due to its exceptional properties but needs experimental justification.

Let us remind the main pros and contras for supersymmetry in particle physics

Pro:\\
$\bullet$  Provides natural framework for unification with gravity\\
$\bullet$  Leads to gauge coupling unification (GUT) \\
$\bullet$  Solves the hierarchy problem  \\
$\bullet$ Is a solid quantum field theory\\
$\bullet$  Provides natural candidate for the WIMP cold DM\\
$\bullet$  Predicts new particles and thus generates new job positions

Contra:   \\
$\bullet$ Does not shed new light on the problem of

\ \ \ $*$  Quark and lepton mass spectrum

\ \ \ $*$ Quark and lepton mixing angles

\ \ \ $*$ the origin of CP violation

\ \ \ $*$ Number of flavours

\ \ \ $*$  Baryon assymetry of the Universe\\
$\bullet$ Doubles the number of particles

Low energy supersymmetry promises us that new physics is round the corner at a
TeV scale to be exploited at colliders and astroparticle experiments of this decade. If our
expectations are correct, very soon we will face new discoveries,
the whole world of supersymmetric particles will show up and the
table of fundamental particles will be enlarged in increasing
rate.  This would be a great step in
understanding the microworld.
The future will show whether we are right in our expectations or not.

\vspace{0.4cm} {\large \bf Acknowledgements} \vspace{0.1cm}

The author would like to express his gratitude to the organizers of the School for  their effort
in creating  a pleasant  atmosphere and support.  This work was partly supported by
RFBR grant \# 08-02-00856 and
Russian MIST grant \# 1027.2008.2. I would also like to thank A.Gladyshev for his help in preparation of the manuscript.

\end{document}